\documentclass{aa}

\usepackage{natbib}
\bibpunct{(}{)}{;}{a}{}{,}             %% natbib format for A&A and ApJ
\usepackage{graphicx}
\usepackage{multirow}
\usepackage[varg]{txfonts}
\usepackage{kotex}

\usepackage[breaklinks=true]{hyperref} %% to avoid \citeads line fills
\hypersetup{colorlinks=true,urlcolor=blue,citecolor=blue,pdfborder= 0 0 0}

\begin{document}
\title{Cosmic gas highways in C-EAGLE simulations}

\author{I. Vurm\inst{1}\fnmsep\thanks{indrek.vurm@ut.ee}
        \and
        J. Nevalainen\inst{1}
        \and
        S.~E. Hong\inst{2,3}
        \and Y.~M. Bah\'{e}\inst{4}
        \and C.~Dalla Vecchia\inst{5,6}
        \and P. Hein\"{a}m\"{a}ki\inst{7}
   }

\institute{Tartu Observatory, University of Tartu, 61602 T{\~{o}}ravere, Tartumaa, Estonia 
\and
Korea Astronomy and Space Science Institute, 776 Daedeok-daero, Yuseong-gu, 34055 Daejeon, Republic of Korea
\and
Astronomy Campus, University of Science \& Technology, 776 Daedeok-daero, Yuseong-gu, 34055 Daejeon, Republic of Korea
\and
Leiden Observatory, Leiden University, PO Box 9513, 2300 RA Leiden, The Netherlands
\and
Instituto de Astrof\'{i}sica de Canarias, C/V\'{i}a L\'{a}ctea s/n, E-38205 La Laguna, Tenerife, Spain
\and
Departamento de Astrof\'{i}sica, Universidad de La Laguna, Av.~Astrof\'{i}sico Francisco S\'{a}nchez s/n, E-38206 La Laguna, Tenerife, Spain
\and
Tuorla Observatory, Department of Physics and Astronomy, Vesilinnantie 5, University of Turku, 20014 Turku,
Finland
}

%\date{Received ; accepted}
\date{}

  \abstract
  % context heading (optional)
  % {} leave it empty if necessary  
   {A substantial fraction of the cosmic baryons is expected to hide in the form of diffuse warm-hot intergalactic medium (WHIM) at X-ray temperatures ($T = 10^5-10^7$~K). Due to the expected low density of the WHIM, it has been very difficult to detect so far. A statistically significant sample of credible detection of the WHIM phase might help solve the problem of the missing cosmic baryons. While the bulk of the cosmic gas is approximately at rest inside the filaments of the Cosmic Web, the fraction of gas located close to galaxy clusters is falling towards them with substantial velocities.
   The infalling gas is influenced by the increasing density in the cluster vicinity and eventually undergoes a termination shock, which may boost its X-ray signal. Thus, the cluster outskirts are potential locations for improved detectability of the missing baryons.}
   {The primary goal of this work is to identify optimal locations of the enhanced X-ray emission and absorption, arising from the interaction of infalling filamentary gas with the cluster material. Our further goal is to improve our understanding of the various physical processes affecting the WHIM as it falls towards clusters of galaxies along the cosmic filaments. We aim to utilise this information for planning future X-ray observations of the WHIM in cluster outskirts.}
   {We applied the DisPerSE filament finder to the galaxy distribution in the surroundings of a single Coma-like ($M_{200} \sim 10^{15.4}~{\rm M}_{\odot}$) simulated C-EAGLE cluster of galaxies. We characterised the distribution of the thermodynamic properties of the gas in such filaments and provided a physical interpretation for the results. This analysis serves as a proof of method to be applied to the full C-EAGLE sample in a future work.}
   {We captured a large fraction ($\sim$ 50\%) of the hot ($T > 10^{5.5}$~K) gas falling towards the cluster in the detected filaments in the cluster outskirts. The gas in the filaments is in approximate free-fall all the way down to the radial distance of $\sim 2 \ r_{200}$ from the cluster. At smaller radii, the filament gas begins to slow down due to the increasing pressure of the ambient gas; approximately half of the filament gas nevertheless penetrates into the cluster before being decelerated. The deceleration is accompanied by the conversion of gas bulk kinetic energy into heat. As a result, the density and  temperature of the gas in the filaments increase from the general Cosmic Web level of $\rho \sim 10\rho_{\rm av}$ (where $\rho_{\rm av}$ is the cosmic mean baryon density) and $T = 10^5-10^6$~K at $r \sim 4 \ r_{200}$ towards $\rho \sim 100\rho_{\rm av}$ and $T = 10^7-10^8$~K  at the virial boundary of the very massive cluster studied in this paper.}
   {The detection of the cosmic filaments of galaxies around clusters may provide a practical observational avenue for locating the densest and hottest phase of the missing baryons.}

\maketitle

\section{Introduction}
\label{Introduction}
In cold dark matter (DM) theory, cosmological structures form by gravitational collapse of primordial density fluctuations and grow hierarchically from smaller to larger structures by mergers. Large galaxy clusters have undergone several such merger events, hence the plasma that makes up the diffuse intracluster medium in a large cluster has likely undergone several strong shocks and been heated to temperatures $T = 10^7 - 10^8$~K within the virial radius. The intracluster gas is virialised and in approximate hydrostatic equilibrium. The high temperature and density of the hot plasma in the inner regions of massive clusters render it relatively easily detectable with modern X-ray instruments. 

However, in the cluster outskirts, the intracluster plasma becomes too tenuous to be routinely detectable with current facilities. The desired boost for the X-ray signal may be provided locally by the physical processes/conditions arising from cluster-filament interactions near the cluster boundary, whereby the filaments which connect the clusters to the Cosmic Web feed the latter with gas and galaxies
\citep[``gas highways'', e.g.,][]{Cautun2014,Kraljic2018,Rost2021,Power20}. In this work, we mainly focus on the gas falling along the filaments towards clusters in order to devise observational strategies for detecting the gas in X-rays.

In the outer filaments ($r \gtrsim 4 r_{200}$), the main driver of gas dynamics is the gravity of the dark matter concentration near the filament spines. It accumulates gas from the surroundings and compresses the filament gas to overdensities of $\sim 10$. The strong external shocks due to the mass flowing towards the filaments \citep[the sonic Mach number $\mathcal{M}_{\rm s} \sim 100$;][]{Ryu2003} heat the gas inside filaments into WHIM domain of $T = 10^5-10^7$~K. Additional heating is provided by weaker internal shocks within gas that has previously been shocked during the hierarchical structure formation process and is associated with accretion onto filaments from non-linear structures as well as halo mergers and chaotic flow motions within filaments \citep[e.g.,][]{Ryu2003,Kang2007}.
 
Closer to the cluster ($r \sim 1-3 r_{200}$), its dark matter and gas become important drivers of the filament gas thermodynamics. In these regions, the infalling gas has substantial ram pressure which is competing against the pressure of the cluster gas. Depending on the properties of the gases, the infalling gas may slow down or even stop close to the hot and dense cluster, producing features in the velocity structure \citep[e.g.,][]{Rost2021}.
When the gas falling along the filaments encounters denser environments close to the cluster, a fraction of the energy associated with ordered gas motion along the filament is converted to thermal energy via shocks. This heats the infalling material at the cluster outskirts. Also, the slowing down of the continuously accreting mass over cosmic timescales will result in the increase of the gas density at the cluster outskirts. It is thus expected that between the outer filaments and the virial region of the cluster there is a transition region with distinct thermodynamic properties. We propose a concept of the cluster-filament interface to describe this transition region.

\citet{Rost2021} used the ThreeHundred simulation to show that the gas particles falling towards galaxy clusters experience different dynamics depending on whether they are within or outside the cosmic filaments. While the filaments act as gas highways, penetrating deeper into the cluster, the particles outside filaments slow down and some even turn back. Our aim in this work is to understand in more detail the physics of the falling gas, inside and outside the filaments, and its interaction with the central cluster by analysing C-EAGLE simulations.
 
\citet{Cornuault2018} considered the interaction of an incoming collimated gas stream with haloes of masses ranging from galaxies to clusters. They suggest that the persistence of the accretion shock, its lifetime, and the possible disruption of the filament depend critically on the relative timescales of various processes in the problem. They delineate different regimes depending primarily on the cooling ability of the post-shock filament gas. The massive cluster CE-29 considered in the present paper falls firmly into the regime in which the postshock gas is unable to cool over the dynamical time (``hot mode accretion"), whereby the filament material remains hot and likely mixes with the ambient halo material. In the picture of \citet{Cornuault2018}, maintaining a hot post-shock gas phase suggests that the accretion shock is persistent, i.e. has a lifetime exceeding the dynamical (free-fall) time associated with the halo. This regime also avoids some of the complexities related to rapid cooling, such as excessive temporal and spatial gradients, highly multiphase media as well as some of the associated hydrodynamical instabilities and turbulence that current cosmological simulations such as C-EAGLE fail to fully capture. This lends some confidence that the shocks identified in the present numerical context have some bearing on the physical reality of filament-cluster interactions.

Several works have been focusing on the infalling substructures in the outskirts of the simulated galaxy clusters. \citet{Mostoghiu} studied the stripping of the gas from the infalling haloes. \citet{2021A&A...653A.171A} studied simulated galaxy clusters with the aim of deriving information for aiding in X-ray detection of the gas in filaments in the cluster outskirts. They focused on the self-gravitating gas clumps, falling along the filaments, and suggested that assuming 0.3 Solar metal abundance, the clumps are at the verge of being detectable with current X-ray instruments via free-free and line emission in the 0.3--2.0~keV band and will be surely detected by future missions like ATHENA.
%However, the clumps contain a small fraction of the total baryon mass in filaments and cover a similarly small fraction of their volumes.
However, the clumps contain a minority of the total baryon mass in filaments and cover a small fraction of their volumes.
In order to investigate observationally the physics of the gas falling towards the clusters and their contribution to the cosmic baryon budget, one needs a robust sampling of the gas outside the clumps. Our aim is to investigate the prospects of detecting the bulk of the gas in filaments in X-rays. The obvious problem is that the median gas overdensity in filaments is at the level of $\sim 10$ \citep[e.g.,][]{Tuominen2021}, while the clumps have densities higher by a factor of 10--100 \citep{2021A&A...653A.171A}. However, the larger extent of the filaments may provide an order of magnitude longer paths through the gas, compared to the clumps. In order to improve the detection chances, we focus on emission and absorption line detection with high-resolution X-ray spectroscopy, which may provide an advantage over the wide-band emission observed with low-resolution instruments, considering the probably dominating cosmic background in the soft X-rays.

There have been some recent reports on possible X-ray detection of the dense and hot phase of intergalactic gas (WHIM) at the outskirts of galaxy clusters \citep[e.g.,][]{2016ApJ...818..131B,2015Natur.528..105E,2021A&A...647A...2R,2008A&A...482L..29W,Tanimura}. While some of the observational works focus on the bridges between cluster pairs, our aim in this paper is different: the intergalactic gas falling along the filaments from the Cosmic Web towards a cluster.

The primary goal of this work is to identify the optimal locations of the enhanced X-ray emission and absorption due to the interaction between the matter falling along filaments and the galaxy clusters. Our secondary goal is to provide some theoretical insights into the underlying physical processes governing gas dynamics in the cluster vicinity, which ultimately affect the X-ray observables. To this end, we characterise and contrast the basic thermodynamical properties of the gas within and outside filaments in the interface region, and provide a physically motivated picture of filaments acting as gas highways. Our motivation is to help harvest the currently existing X-ray data archives (e.g., XMM-Newton and Chandra) and soon arriving data (eROSITA), as well as to develop observational strategies for future X-ray missions (e.g., ATHENA).

Due to the challenges of X-ray observations of diffuse gas like that in cosmological filaments, it is likely that in the foreseeable future detection is only feasible by stacking a large number of filaments. In fact, recent 
soft X-ray emission detections from cosmic filaments at $\sim$4 $\sigma$ level using ROSAT (SRG/eROSITA) required a stacking of $\sim$  15000 (500) filaments from the Sloan Digital Sky Survey  \citep{Tanimura20,Tanimura}. As the primary goal of this paper is observational, we, therefore, limit  our analysis to a level of detail sufficient for informing detection strategies, while retaining the salient features for physical interpretation. This amounts to approximations such as fixing the filament width, analysing spherically averaged quantities, as well as averaging across all filaments. We will discuss the variation of individual filament properties at the level we deem sufficient from an X-ray observer's point of view, as well as consider in more detail two specific filaments whose properties bracket the above variation.
 
The structure of this paper is as follows. In Section~\ref{Data}, we describe the C-EAGLE data and the overall properties of the massive mock galaxy cluster which we use in this work. We also discuss our methods of defining the filaments around the cluster and sampling the gas. After a brief discussion on the gas properties in the filaments at the outer regions of the simulation volume in Section~\ref{Outer}, we characterise in more detail the properties of the hot gas in the cluster-filament interface in Section~\ref{Overall}. A physical interpretation of the thermodynamic structure of the gas is given in Section~\ref{physics}, while in Section~\ref{Xray} we derive implications of the infalling gas on the cosmic highway for X-ray detection. We summarise our results in Section~\ref{Summary}.

\section{Data}\label{Data}
\subsection{C-EAGLE}

Our study is based on data from the Cluster-EAGLE (C-EAGLE) simulation project \citep{Bahe_et_al_2017,Barnes2017b}. The sample of initial 30 DM halos has been selected from a large (3.2 Gpc)$^3$ simulation \citep{Barnes2017a}, evenly sampled per logarithmic mass interval over the range of $M_{200} = 10^{14} - 10^{15.4}~{\rm M}_{\odot}$ to cover the typical mass range of galaxy clusters. The baryonic content of the clusters has then been produced by hydrodynamical zoom simulations using the EAGLE code \citep{Schaye2015, Crain2015} extending to the radial distance of $10 r_{200}$ from the cluster centre. The initial mass of the smoothed-particle hydrodynamics (SPH) gas particles is $m_{\rm gas} = 1.8\times 10^6~{\rm M}_{\odot}$, and the mass of the DM particle is $m_{\rm DM} = 9.7\times 10^6~{\rm M}_{\odot}$.

\subsection{The cluster CE-29}
\label{C29}

In the current work we focus on CE-29, the most massive ($M_{200} = 10^{15.4}~{\rm M}_{\odot}$) halo in the C-EAGLE sample. We are analysing the data within a box of size ($\sim$ 40~Mpc)$^{3}$ centred on CE-29. Such a highly massive cluster is not representative of the true cluster population, where lower-mass clusters are more numerous. However, our aim in this paper is to get first insights into the physics of the cluster-filament interface and the thermodynamic conditions for the production of X-ray emitting ions. We chose the most massive cluster in order to maximise the amount of X-ray material and to establish methods for studying the hot WHIM. In future work, we will apply these methods to the lower-mass clusters in order to obtain a more realistic view of the expected observational situation.

\begin{figure*}[bt]
   \centering
  \includegraphics[width=0.48\hsize]{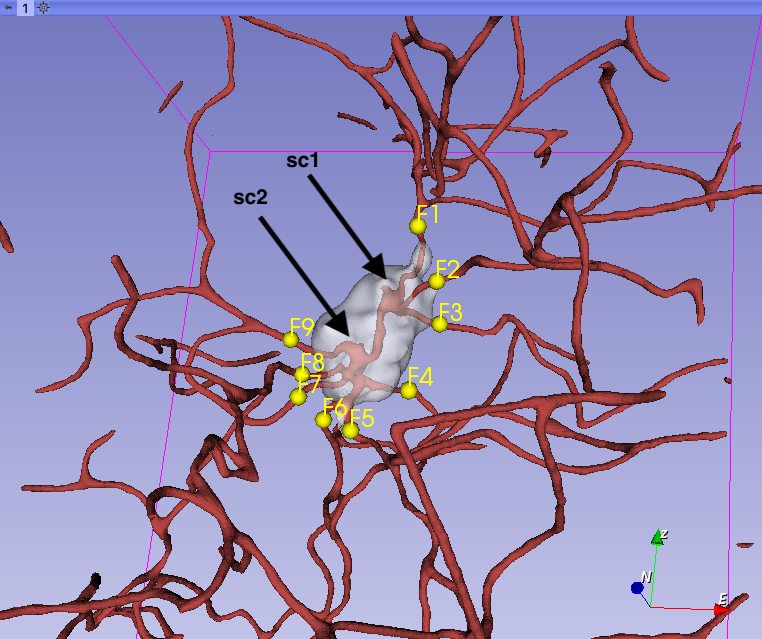}
  \includegraphics[width=0.48\hsize]{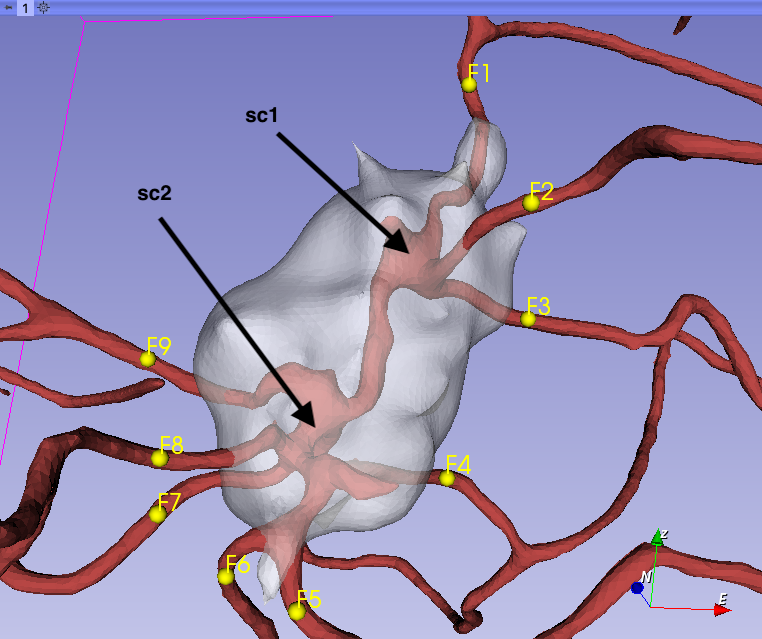}
      \caption{The filament spines as detected by DisPerSE (red lines) are shown together with the virial boundary of the cluster CE-29 (white region) for approximately the full (40~Mpc)$^3$ box (\emph{left}) and zoomed to the central region (\emph{right}).   
      The locations where the filamentary network crosses the virial boundary are labelled with yellow symbols. The centres of the two merging subclusters (sc1 and sc2) are marked with black arrows.}
         \label{vir_fil.fig}
   \end{figure*}

Given the high mass of the cluster CE-29, $M_{200} = 2.4\times 10^{15}~{\rm M}_{\odot}$ \citep[similar to that of Coma, e.g.,][]{2007ApJ...671.1466K}, the virial radius in spherical approximation is quite large, $r_{200} = 2.8$~Mpc. However, visual inspection indicates that the cluster is elongated. The density and velocity structures indicate that two subclusters (sc1 and sc2 in Fig.~\ref{vir_fil.fig}) are merging along the direction of the elongation. Therefore we defined the cluster boundary as an isodensity surface within which $\overline{\rho}_{\rm gas} + \overline{\rho}_{\rm DM} = 200\rho_{\rm crit}$, referred to as the boundary of the virial region below. To suppress local small-scale inhomogeneities, the boundary was calculated on a smoothed density field convolved with a 3D Gaussian kernel with a 0.2~Mpc radius. The result shows that the boundary of the virial region indeed has an elongated shape: the semi-major axis is $\sim 1.5 r_{200}$, the semi-minor axis length is $\sim 0.7 r_{200}$, i.e., the axial ratio is $\sim 2$ (the white region in Fig.~\ref{vir_fil.fig}).

When characterising the radial behaviour of the thermodynamic properties of the whole system below, we use distributions  averaged over spherical shells around the cluster centre. This somewhat complicates the interpretation of the results close to the cluster due to its non-spherical shape. On the other hand, the analysis of individual filaments below takes the shape of the cluster into account and considers the distance from the local cluster boundary (i.e. where the filament crosses) separately for each filament.

\begin{figure*}[bt]
   \centering
   \includegraphics[width=0.48\hsize]{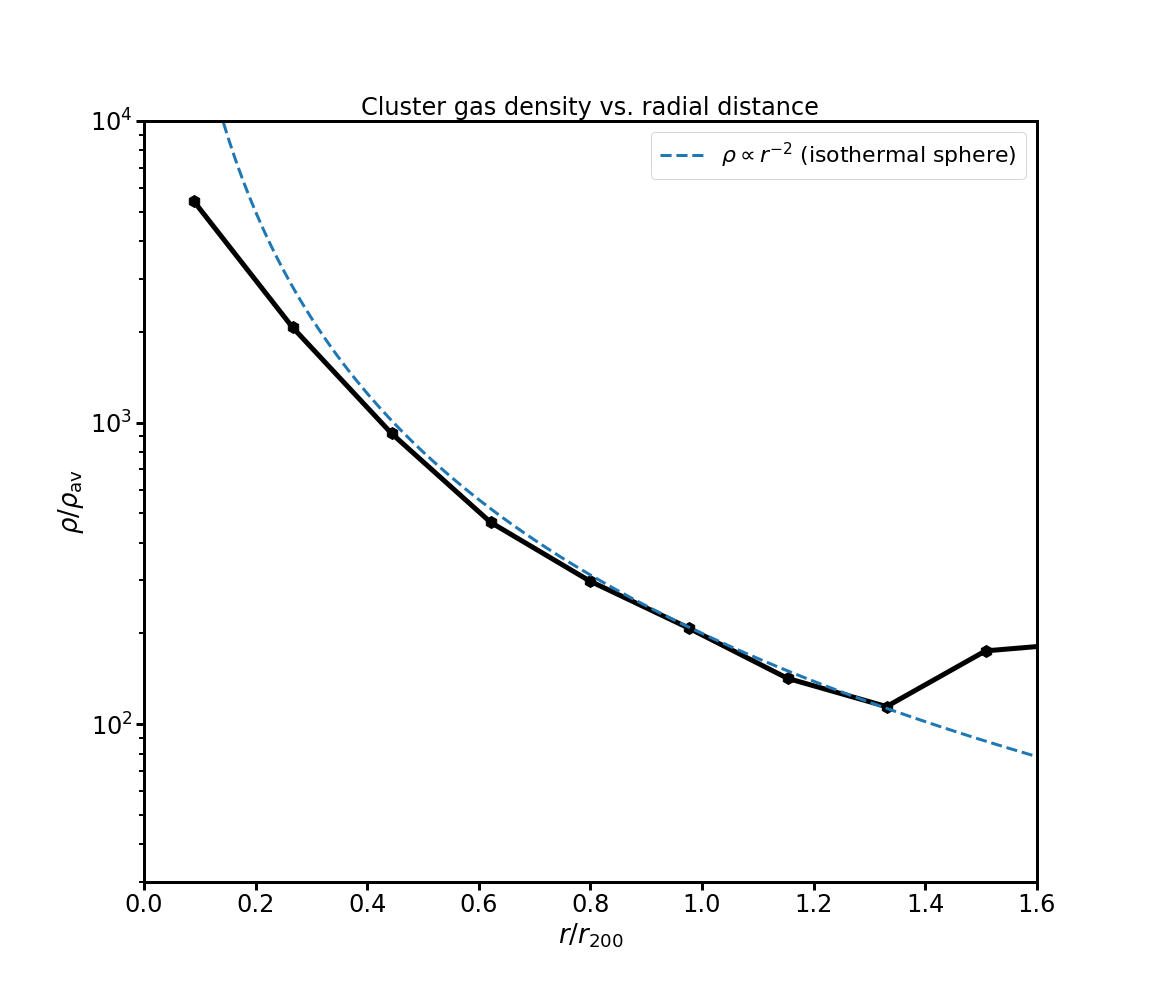}
    \includegraphics[width=0.48\hsize]{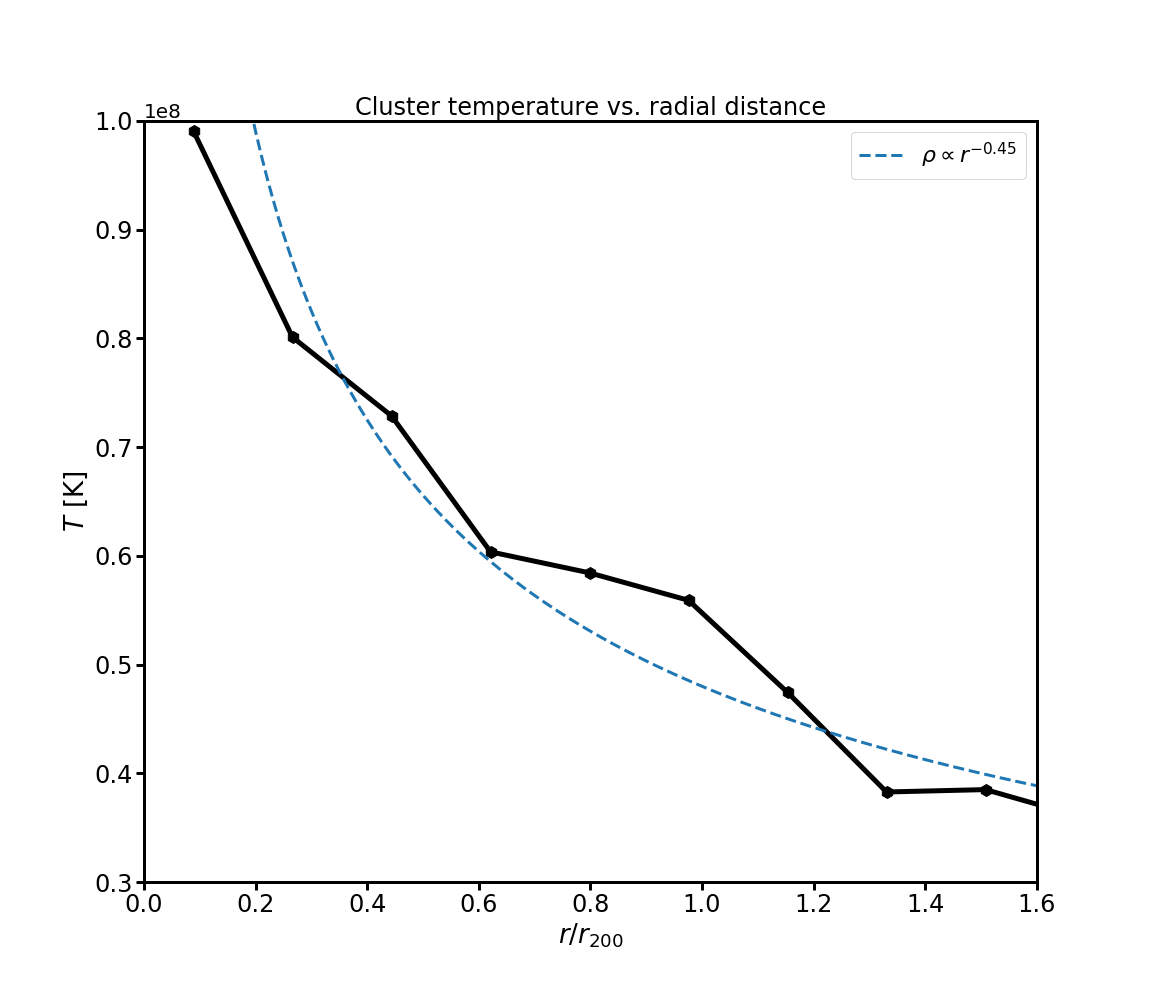}
      \caption{Gas density (\emph{left}) and temperature (\emph{right}) profile within the cluster CE-29. The dashed line in the left panel shows the theoretical density profile for isothermal gas, assuming spherical symmetry and that $\rho_{\rm gas}/\rho_{\rm DM}$ is constant ($\rho \propto r^{-2}$). The dashed line in the right panel shows $\rho \propto r^{-0.45}$ approximation.
              }
         \label{fig:cluster}
   \end{figure*}

The spherically-averaged density and temperature profiles (see Section~\ref{gassampling} for the profile method) of the gas inside the virial boundary decrease with radial distance $r$ from the cluster centre approximately as $\rho \propto r^{-2}$ and $T \propto r^{-0.45}$, respectively, reaching an overdensity of $\sim 100$ and a temperature of a few times $10^7$~K at the virial boundary (see Fig.~\ref{fig:cluster}). 

\subsection{Filaments}
 \label{Filaments}
We applied the Discrete Persistent Structures Extractor (DisPerSE) filament finder \citep{Sousbie11} to C-EAGLE galaxies, instead of gas or dark matter density, because they constitute currently the only robust observable related to these systems. DisPerSE has been shown to trace well the different cosmological environments, including filaments \citep{Sousbie11b}.  It has been widely used for detecting the cosmic web filaments using observational data  \citep[e.g.,][]{Kraljic,Malavasi,Laigle}. The observational angle is fundamental for our purposes since our aim is to locate the enhanced X-ray emission/absorption using galaxies in e.g. SDSS and 4MOST surveys.

\begin{figure}[hbt]
   \centering
   \includegraphics[trim=220 300 150 300, clip, width=1.05\hsize]{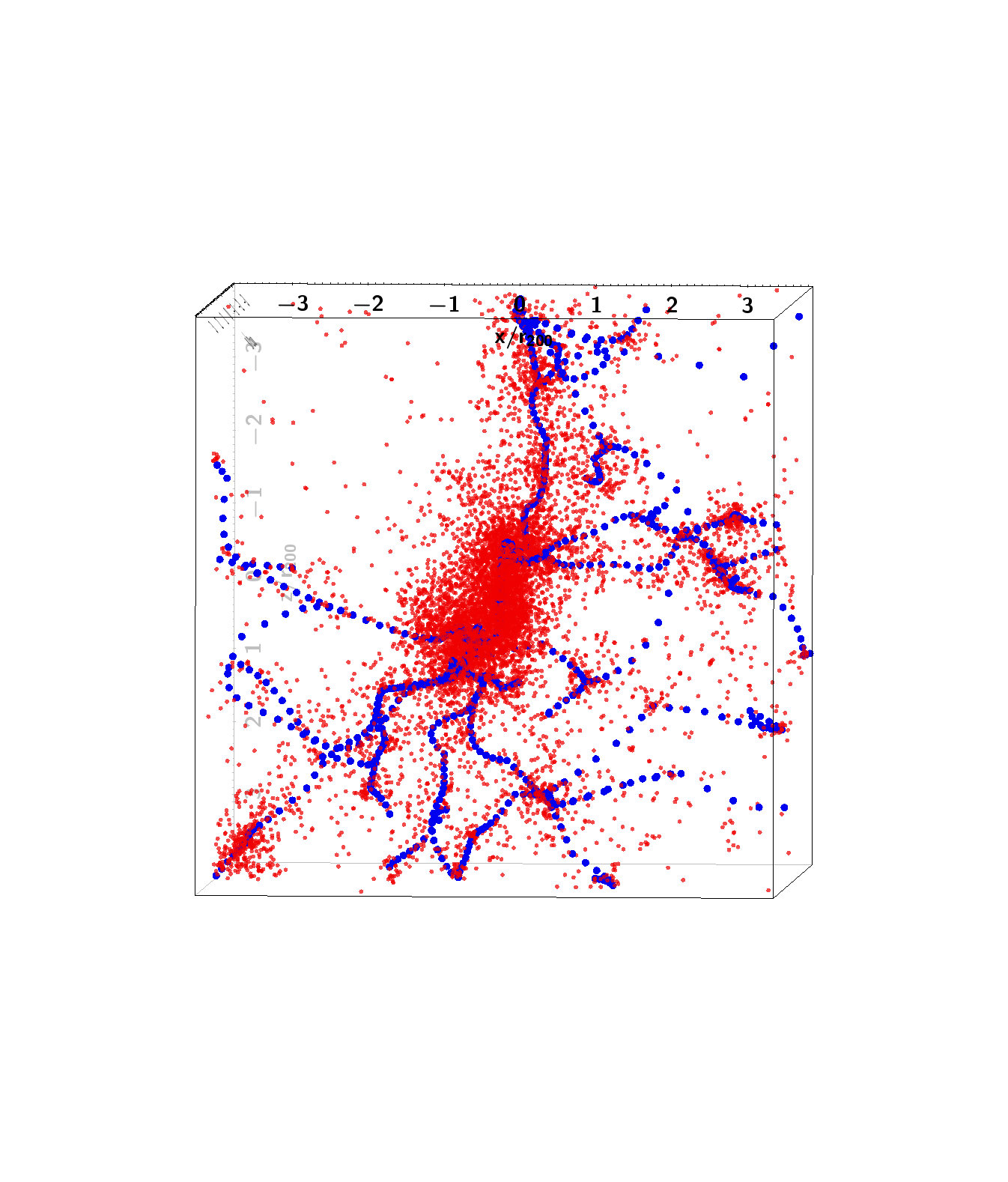}
   \caption{Galaxies (red) and DisPerSE spine points (blue) from a $\sim (8 r_{\rm 200})^3$ region centred on the cluster CE-29. The orientation of the 3D region approximately corresponds to that shown in Figs.~\ref{vir_fil.fig} and \ref{fig:density_filaments}.
   }
    \label{fig:galaxies_filaments}
   \end{figure}

Applying DisPerSE with a default setting (a persistence threshold of $5\sigma$ and a smoothing level $10$, see \citet{Sousbie2013} for details) yields the filamentary network around cluster CE-29 (Figs.~\ref{vir_fil.fig} and \ref{fig:galaxies_filaments}). The network has two centres within the central cluster, approximately located at the centres of the merging subclusters, which in turn are connected with a filament. Thus, we are probably witnessing the merger of two filament networks. The central filament between the two subclusters is approximately aligned with the major axis of the virial boundary of the cluster (see Section \ref{C29}). This is consistent with \citet{Kuchner} who studied a sample of simulated clusters and found a correlation between the alignments of the major axes of the central haloes and the filaments.

The filamentary network is quite complex; there are many locations where one identifiable filament splits into two or more branches. Thus, the filament identity is valid only locally. There are nine locations where the filamentary structure is crossing the virial boundary of the cluster (see Fig.~\ref{vir_fil.fig}). We identify these locations as inner endpoints of filaments F1--3 (F4--9) associated with subcluster sc1 (sc2), respectively. When investigating the individual filaments below (Section~\ref{indiv}), we follow the above filaments only up to the first splitting point. This limits the study of the individual filaments to distances less than $\sim r_{200}$ away from the virial boundary.  
 
\subsection{Gas sampling}
\label{gassampling}
We first divided the computational volume of C-EAGLE into a rectangular grid of (0.2~Mpc)$^3$ cells and computed all physical quantities relevant for our analysis within each cell based on SPH particle data. This allowed a straightforward determination of the median values as well as the distributions of physical quantities within all regions of interest, e.g. within or outside filaments, at a given radial distance from the cluster etc., with minimal computational effort. We verified that the number of SPH particles per cell in filament regions is sufficiently large
($\sim 300$ on average for typical filament overdensity of 10) for robust statistics. We concentrate on the intergalactic baryons and we thus exclude particles defined as halo particles by the EAGLE team.
  
\begin{figure*}[bt]
   \centering
   \includegraphics[trim=20 20 20 60, clip, width=0.48\hsize]{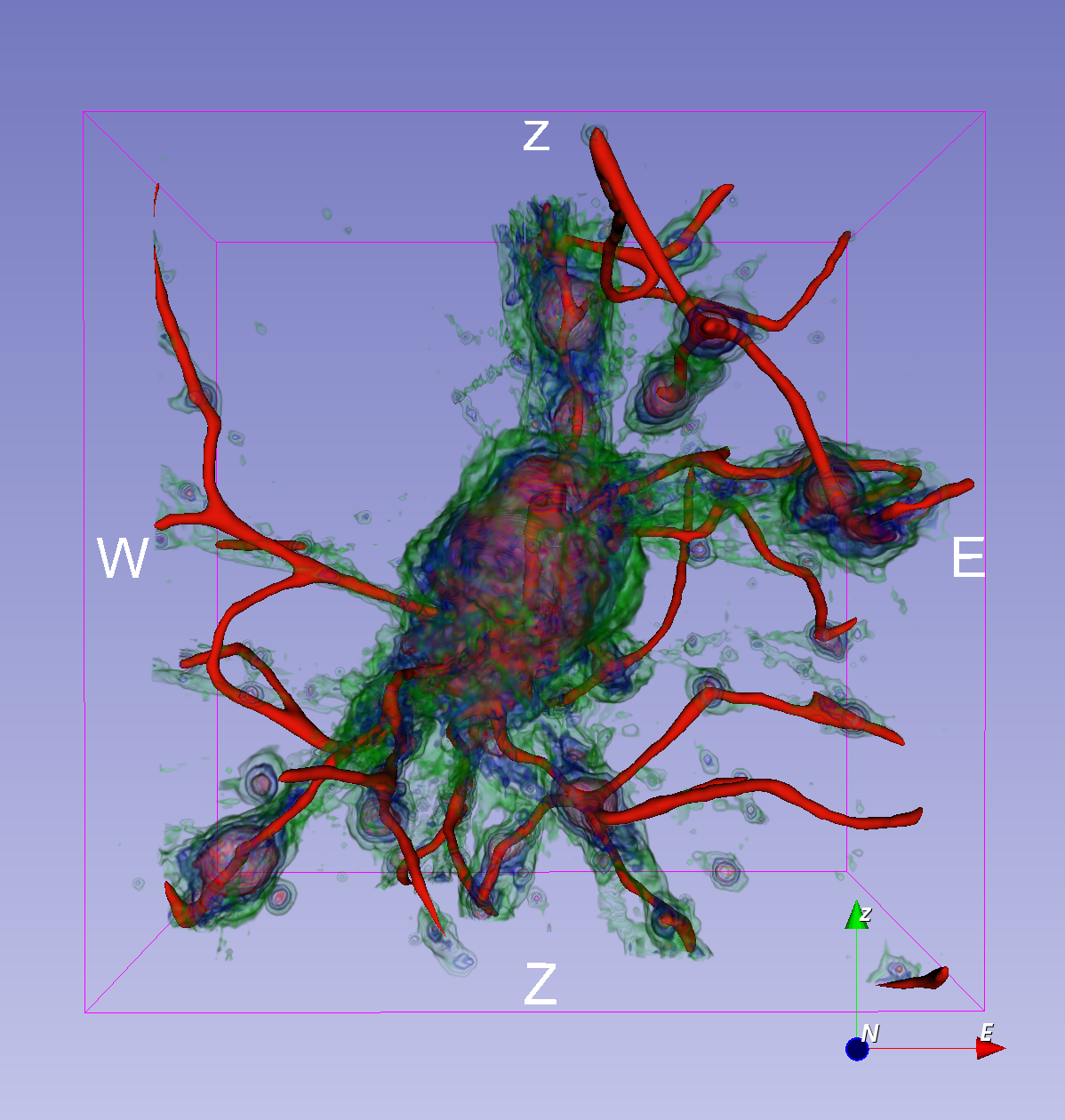}
   \includegraphics[trim=20 20 20 60, clip, width=0.48\hsize]{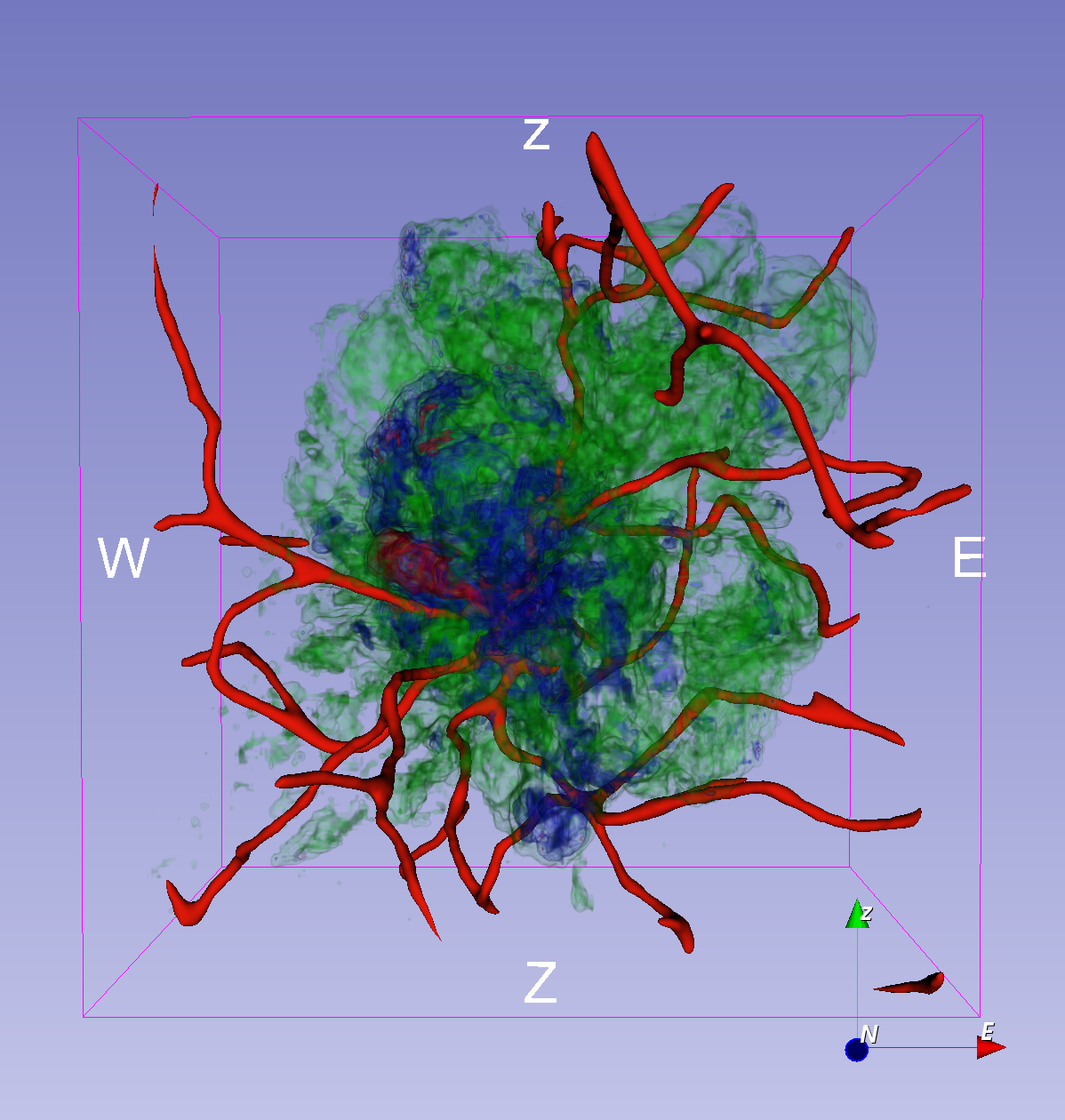}
   \caption{The density (\emph{left}) and temperature (\emph{right}) structure approximately within the cluster-filament interface, i.e., within a $\sim (8r_{200})^3$ region centred on the cluster CE-29. The filaments spines are indicated in red. In the left panel only regions of gas overdensity above 10 are indicated in order to highlight the connection of the densest infalling gas and the filaments (see the electronic version of the journal for video showing a $360^{\circ}$ view). In the right panel, the green, blue and red surfaces approximately correspond to temperatures $2\times 10^7$~K, $6\times 10^7$~K and $2\times 10^8$~K, respectively.
}
    \label{fig:density_filaments}
   \end{figure*}

In each cell, we co-added the SPH particle masses and divided them with the cell volume to obtain the density. For the non-additive properties like temperature and velocity, we computed a mass-weighted SPH particle average  (see Fig.~\ref{fig:density_filaments} for the maps of the density and the temperature).

The DisPerSE procedure does not produce well-defined boundaries for the filaments, which are needed to capture the diffuse baryons within the filament volume. We rely on the results of \citet{Tuominen2021} on main EAGLE filaments, whereby most of the WHIM is located within 1~Mpc distance from the filament spines, while at 2~Mpc, the median gas density profile has already dropped to the cosmic average level. Thus, we defined as filament gas the contents of such (0.2~Mpc)$^3$ boxes which are located within 1~Mpc of any filament spine. While the filaments cover only $\sim 6$\% of the full volume outside the virial boundary of the cluster (out to $6 r_{200}$), the fraction of the diffuse hot ($T \ge 10^{5.5}$~K) baryon mass  located within this volume is $\sim 50$\% of the total in the same temperature range.

Our aim is to investigate quantitatively the physics of the infalling filamentary gas and its interaction with the central cluster. We thus constructed profiles of the thermodynamic properties as a function of the distance from the cluster centre.
We divided the volume into concentric spherical shells of 0.5~Mpc width centred at the cluster centre and extracted the distribution of the properties of such (0.2~Mpc)$^3$ boxes which are located within such a shell and within a filament. The median of this distribution was then assigned as a representative of the baryon properties in that shell. From this data, we constructed the median radial profiles of the baryon properties in the filaments as a function of the nominal $r_{200}$ (see Section~\ref{Overall} for the results).

\section{Outer filaments}
\label{Outer}

\begin{figure*}[bt]
   \centering
 \includegraphics[width=0.48\hsize]{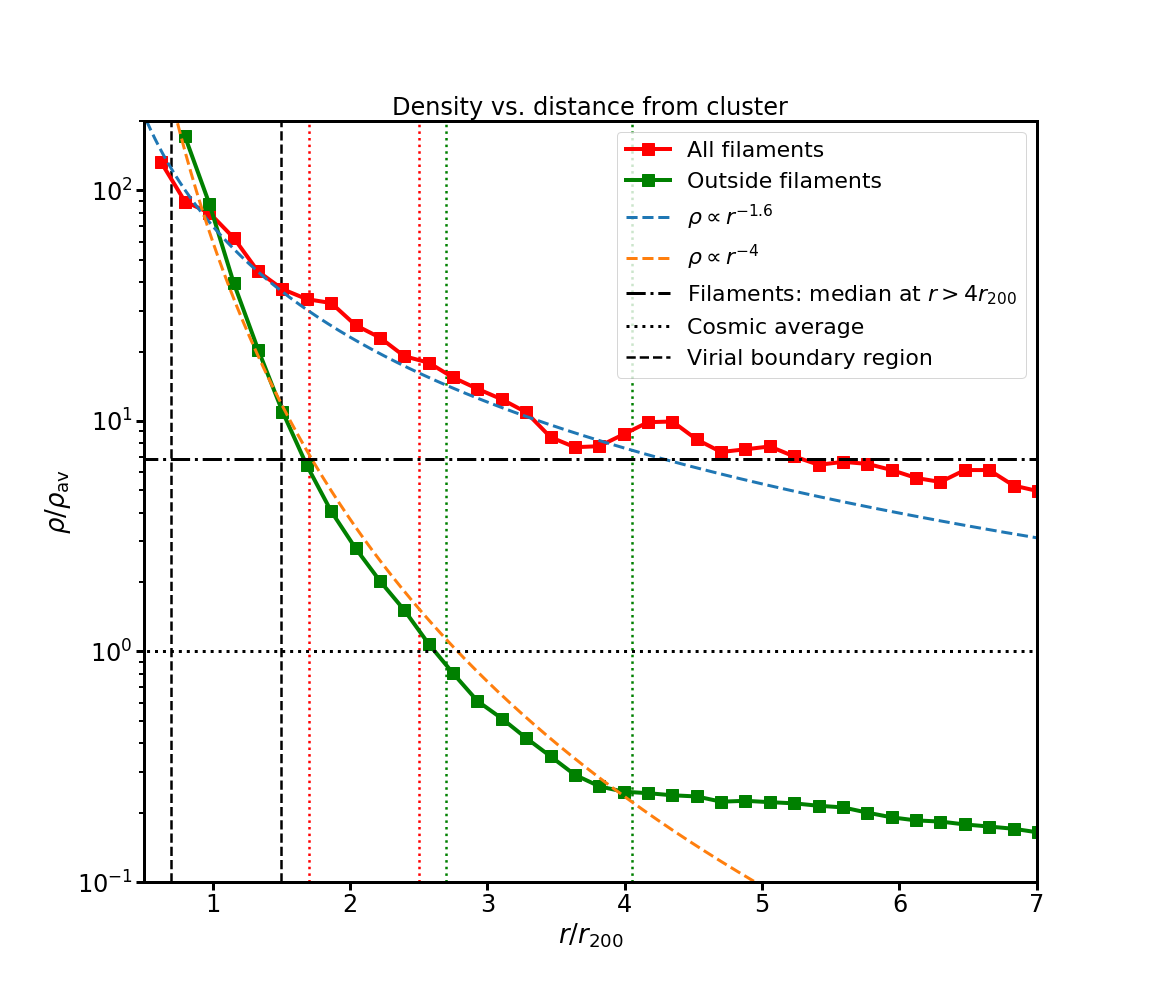}
 \includegraphics[width=0.48\hsize]{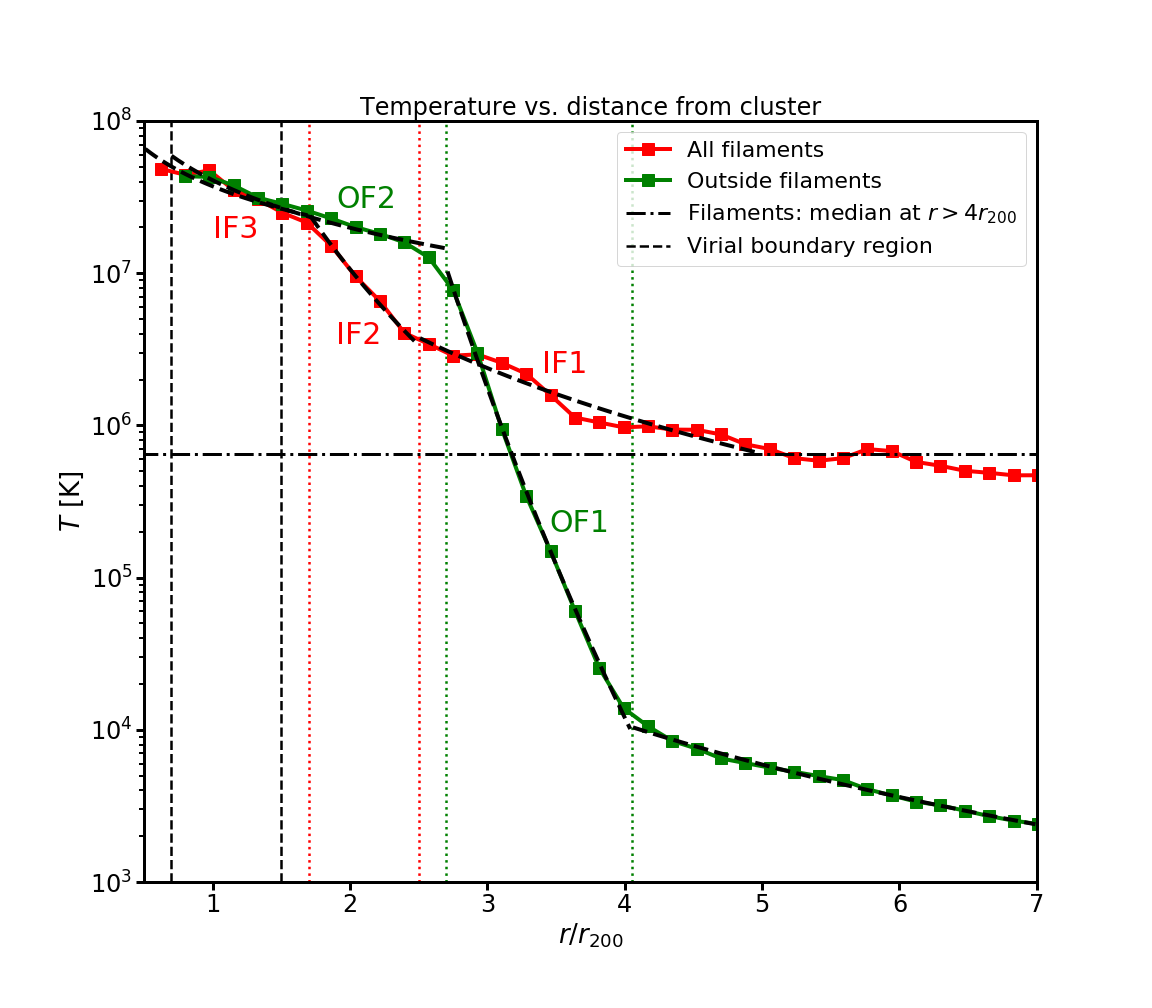}
   \includegraphics[width=0.48\hsize]{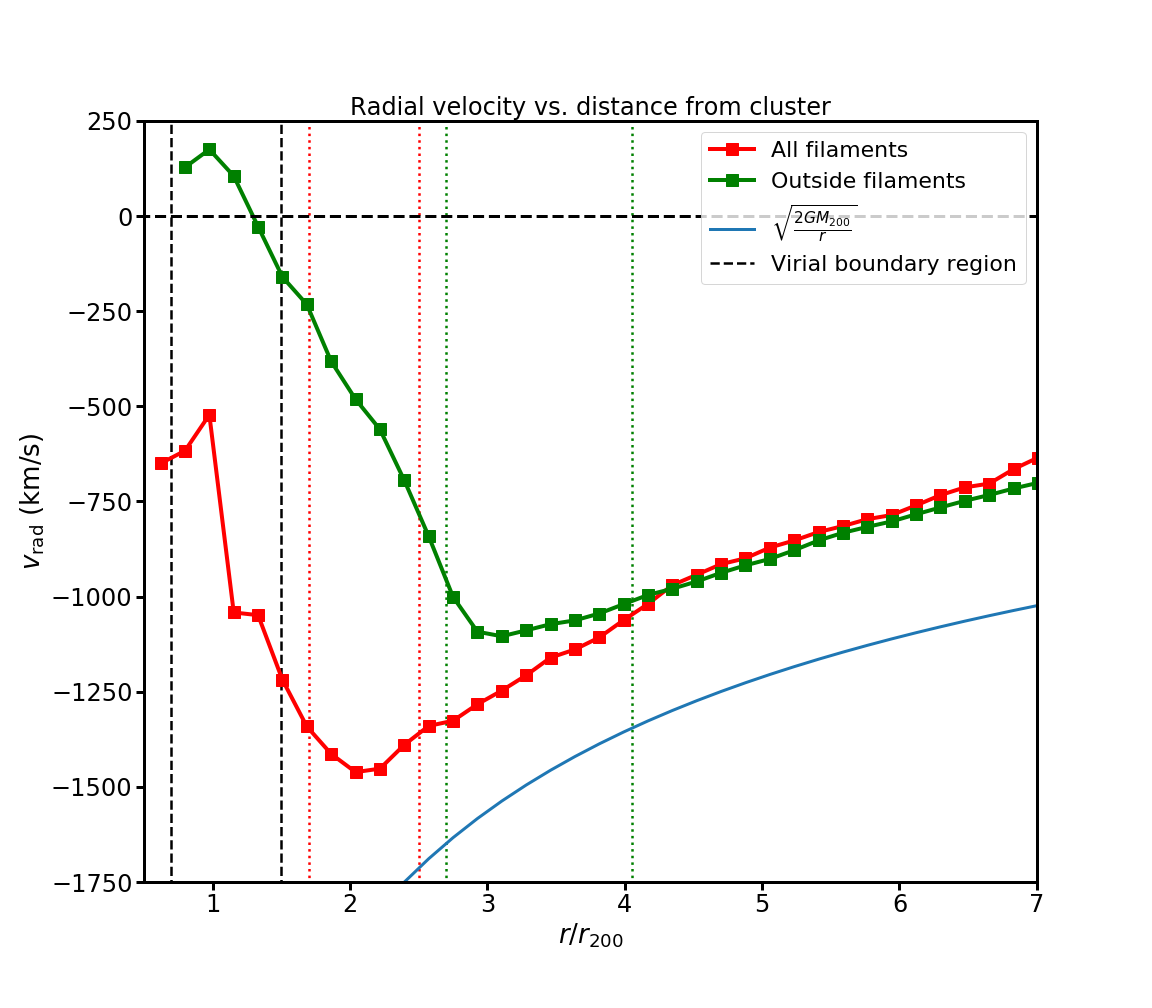}
   \includegraphics[width=0.48\hsize]{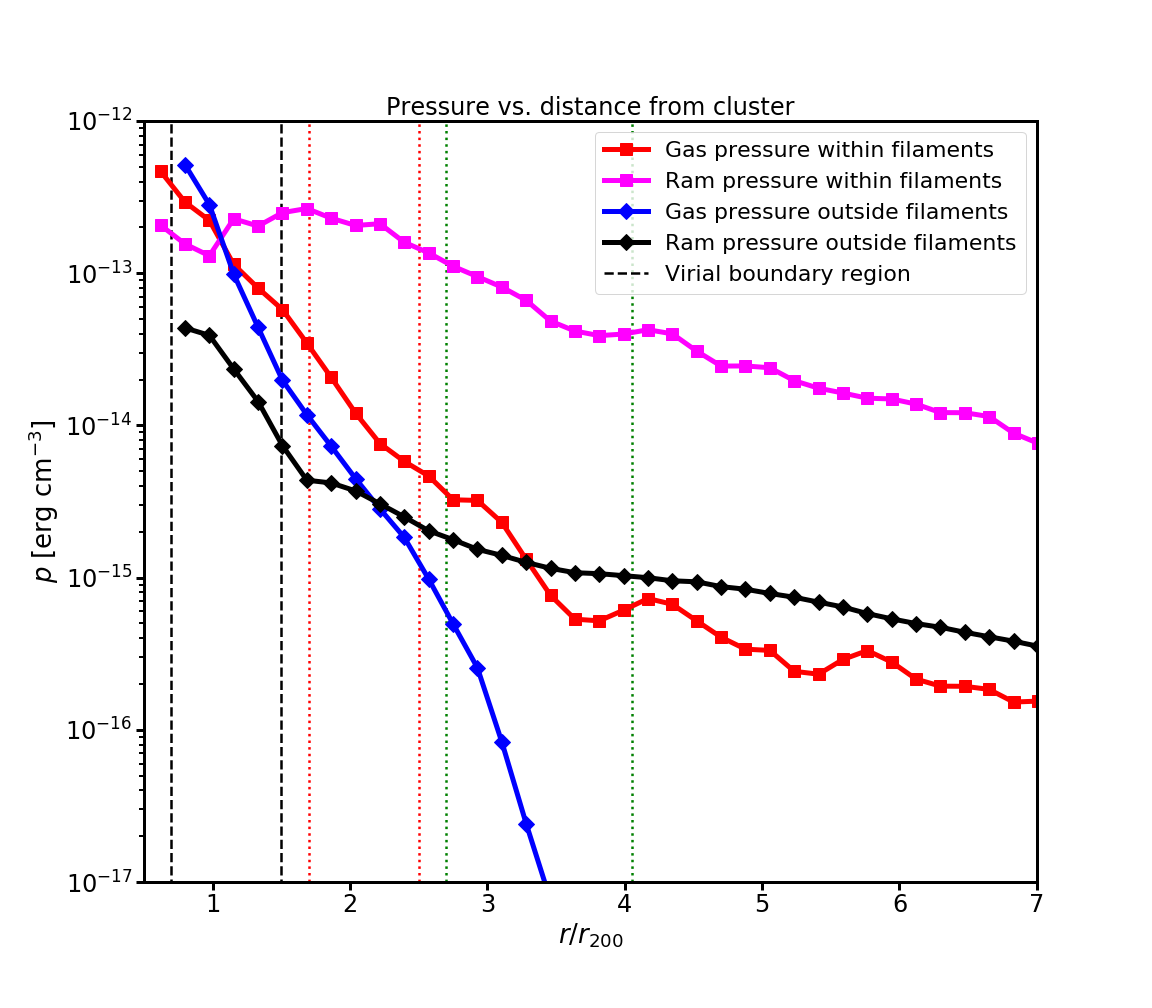}
        \caption{{\it Top left panel:} The median overdensity of the gas outside the virial boundary of the cluster as a function of the distance from the cluster centre inside (red symbols) and outside (green symbols) the filaments. The dashed blue (orange) line shows $\rho \propto r^{-1.6}$ ($\rho \propto r^{-4}$) approximation for the density profile inside (outside) filaments. The horizontal dash-dotted line shows the median overdensity within the filaments at radii $4r_{200} < r < 7r_{200}$.  The mean cosmic density is denoted with a horizontal dotted line. The approximate radial range of the elongated virial boundary is indicated with the vertical dashed lines. {\it Top right panel:} The same as the left panel, but for the median temperature. The dashed lines indicate the piece-wise power-law approximation to the temperature profile in different zones (IF1-3 and OF1-2, see Section \ref{zones}).
        {\it Bottom left panel:} The same as top panels, except for the median radial velocity towards the cluster. The blue curve indicates the free-fall velocity given by the mass of the cluster. {\it Bottom right panel:} The median gas and ram pressure outside the virial region of the cluster as a function of the distance from the cluster centre for the gas within (square symbols) and outside (diamond symbols) filaments.} 
       \label{fig:temperature:density:median}
   \end{figure*}

In this work, we are focusing on the potentially enhanced temperatures and densities in the cluster-filament interface due to interaction with the central cluster. As a baseline for comparison, we first considered the gas properties at the largest distances from the cluster where the cluster effects should be minimised. While the temperature and density profiles of the filament gas quite monotonically decrease with the distance from the cluster centre, they start to level off at $\sim 4 r_{200}$ (see Fig.~\ref{fig:temperature:density:median}, top panels). The median density of the filament gas beyond $4 r_{200}$ is $\sim 10$ times the cosmic mean baryon density. This is consistent with the general large scale structure formation scenario whereby the baryon overdensity in the Cosmic Web filaments is $10-100$ \citep[e.g.,][]{1999ApJ...514....1C, 2001ApJ...552..473D}. The median temperature at these radii is $\sim 10^{6}$~K, again consistent with the general predictions of WHIM temperatures of $T = 10^5-10^7$~K. Thus, the data implies that beyond $4 r_{200}$ the filaments are approaching the general Cosmic Web network.

\section{Characterising the cluster-filament interface}
\label{Overall}
We will now shift our focus to the main topic of interest in this paper, namely the outskirts of cluster CE-29, and will begin by characterising the gas properties within this region, both within filaments as well as outside. Due to the high elongation of the cluster (Section~\ref{C29}), we will not apply the formal $r_{200}$ to separate the cluster and the filaments. Instead, we study the gas in the cluster-filament interface which we define as the volume within the filaments, radially limited by being outside the 3D boundary of the virial region (determined in Section~\ref{C29}) but within 4 times the nominal $r_{200}$ from the cluster centre. 

The 3-dimensional thermodynamic structure of the cluster-filament interface region (see Fig.~\ref{fig:density_filaments}) indicates that the gas is indeed concentrated along the galactic filaments all the way to the cluster boundary. This lends confidence to our aim of observationally locating the infalling gas using the DisPerSE filament detection method acting on spectroscopic galaxy surveys. 

The temperature structure is more complex (see Fig.~\ref{fig:density_filaments}). The inverse correlation between infalling gas density and its susceptibility to shocks, as well as turbulent motions and outflows of hot gas driven by the cluster results in a hotter and somewhat chaotic environment in the cluster vicinity compared with larger distances, more readily affecting the more tenuous gas outside filaments; we will discuss this issue further in Sections~\ref{heatoutside} and \ref{heatinside}. Thus, the fraction of hot gas captured within filaments is lower ($\sim 50$\%, see Section~\ref{gassampling}) than that in the main Cosmic Web far from clusters \citep[$\sim 80$\%,][]{Tuominen2021}. However,
focusing on filaments with X-ray observations is preferable since, while being hot, they stand out from their surroundings with their higher gas density.

Using the procedure described in Section~\ref{gassampling}, we constructed profiles of the thermodynamic properties as a function of distance from the cluster centre, in the radial range extending from the virial boundary up to $4 r_{200}$. We first produced a single median profile considering the gas from all filaments, i.e., all such (0.2~Mpc)$^3$ boxes located within 1~Mpc distance from the filament spine points detected with DisPerSE. We will additionally perform a more detailed analysis of the individual profiles in Section~\ref{indiv}. For comparison, we also specified an ``outside filaments'' gas region which is located safely outside the filaments,  i.e., the boxes further than 2~Mpc from any filament spine. 
 
In the following, we will describe the main thermodynamic and kinematic properties of gas in the interface region, as derived from the above methods.

\subsection{Density}
 
The density of the gas inside filaments decreases from the cluster level of overdensity $\sim 100$ at the inner edge of the cluster-filament interface ($r \sim 0.7-1.5 r_{200}$) to the typical Cosmic Web filament level of overdensity $\sim 10$ at $r = 4 r_{200}$, following approximately  $\rho \propto r^{-1.6}$, providing a high-density path from the cluster to the Cosmic Web  (see Fig.~\ref{fig:temperature:density:median}, top left panel).

On the other hand, the density of the gas outside filaments drops much faster with distance, following  approximately $\rho \propto r^{-4}$. At $r \gtrsim 2.5 r_{200}$, the density outside filaments decreases to the cosmic mean baryon density level. At $r \gtrsim 4 r_{200}$, the gas outside filaments is underdense by a factor of $\sim 10$, i.e., it enters the void domain (Fig.~\ref{fig:temperature:density:median}, top left panel).

\begin{figure*}[bt]
 \includegraphics[width=0.33\textwidth]{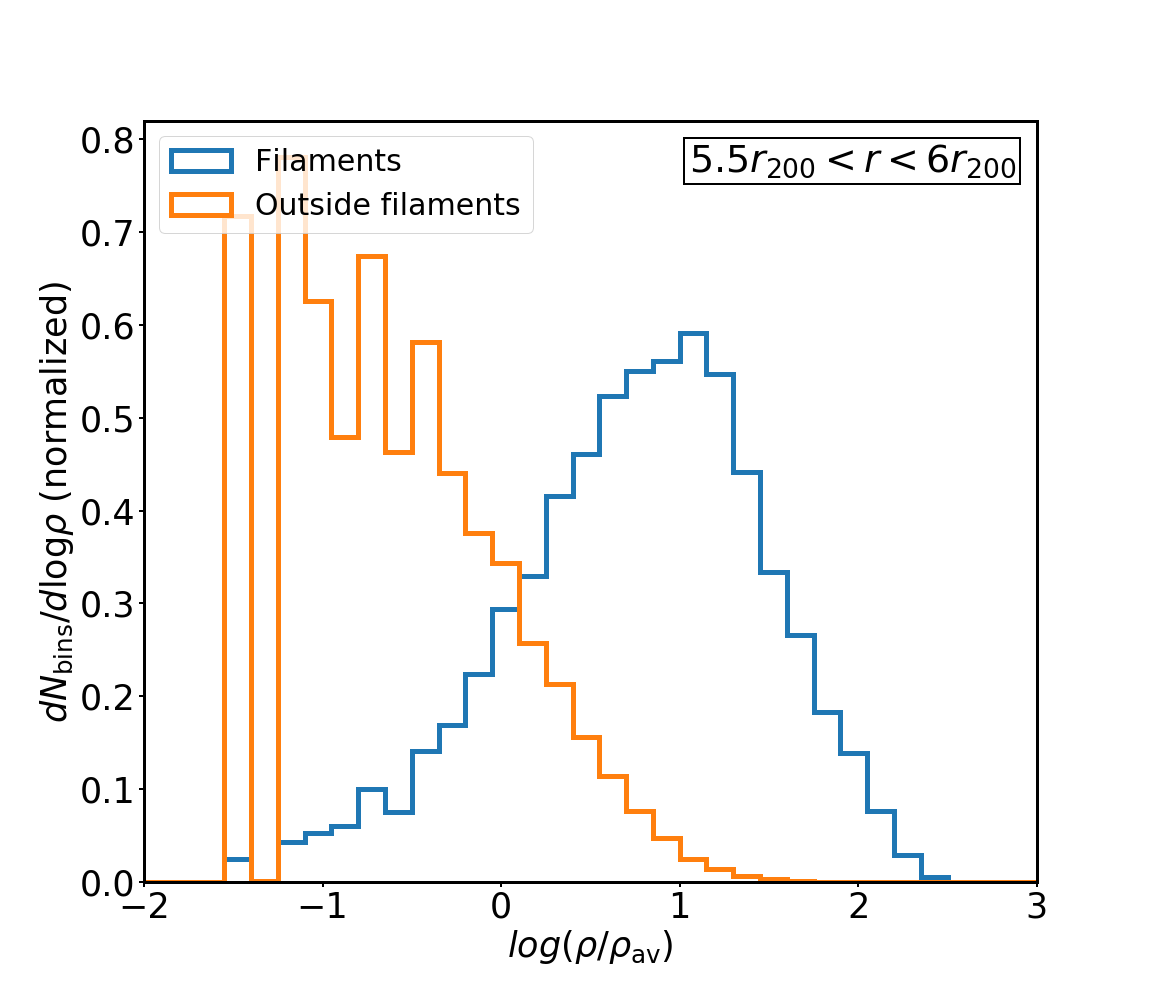}
\includegraphics[width=0.33\textwidth]{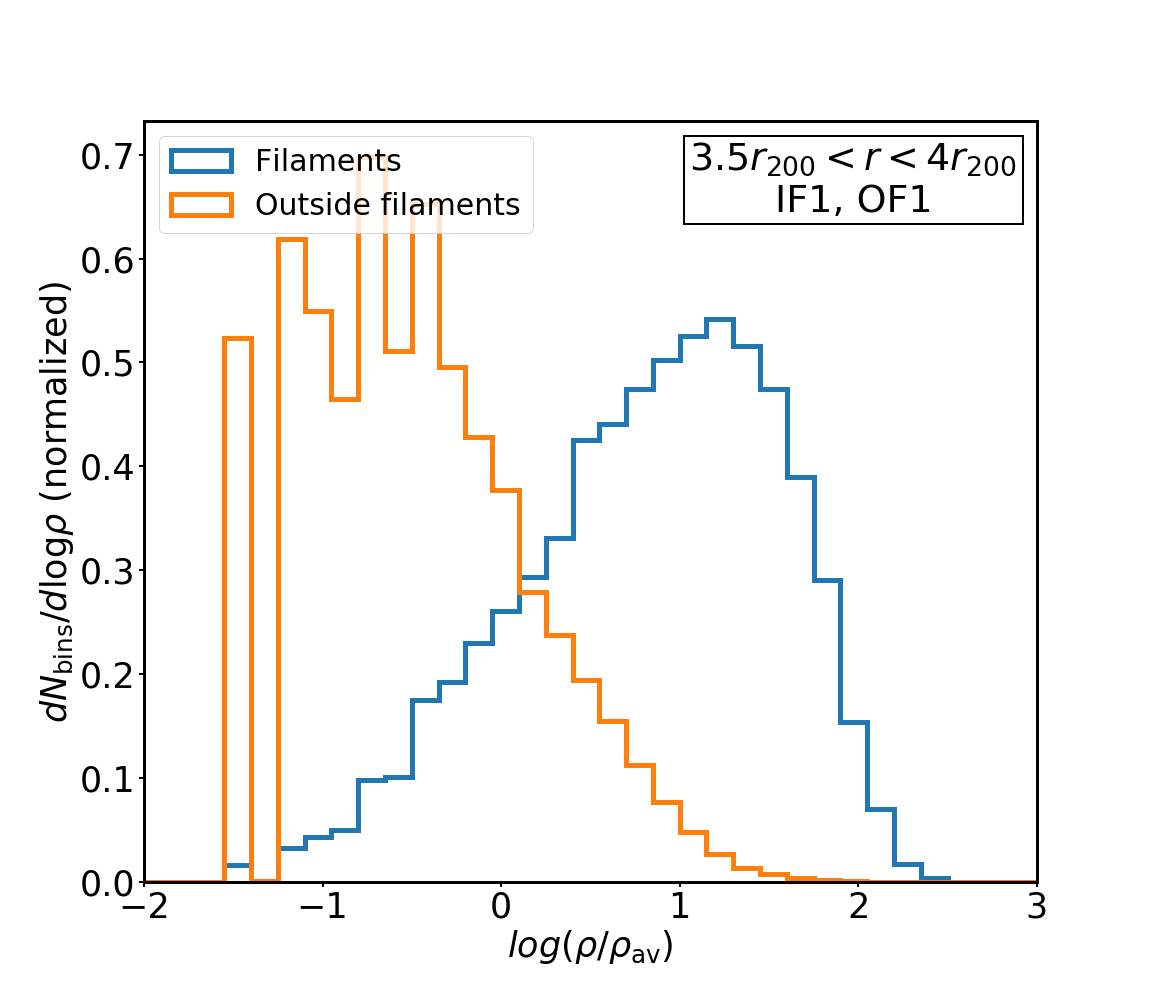}
\includegraphics[width=0.33\textwidth]{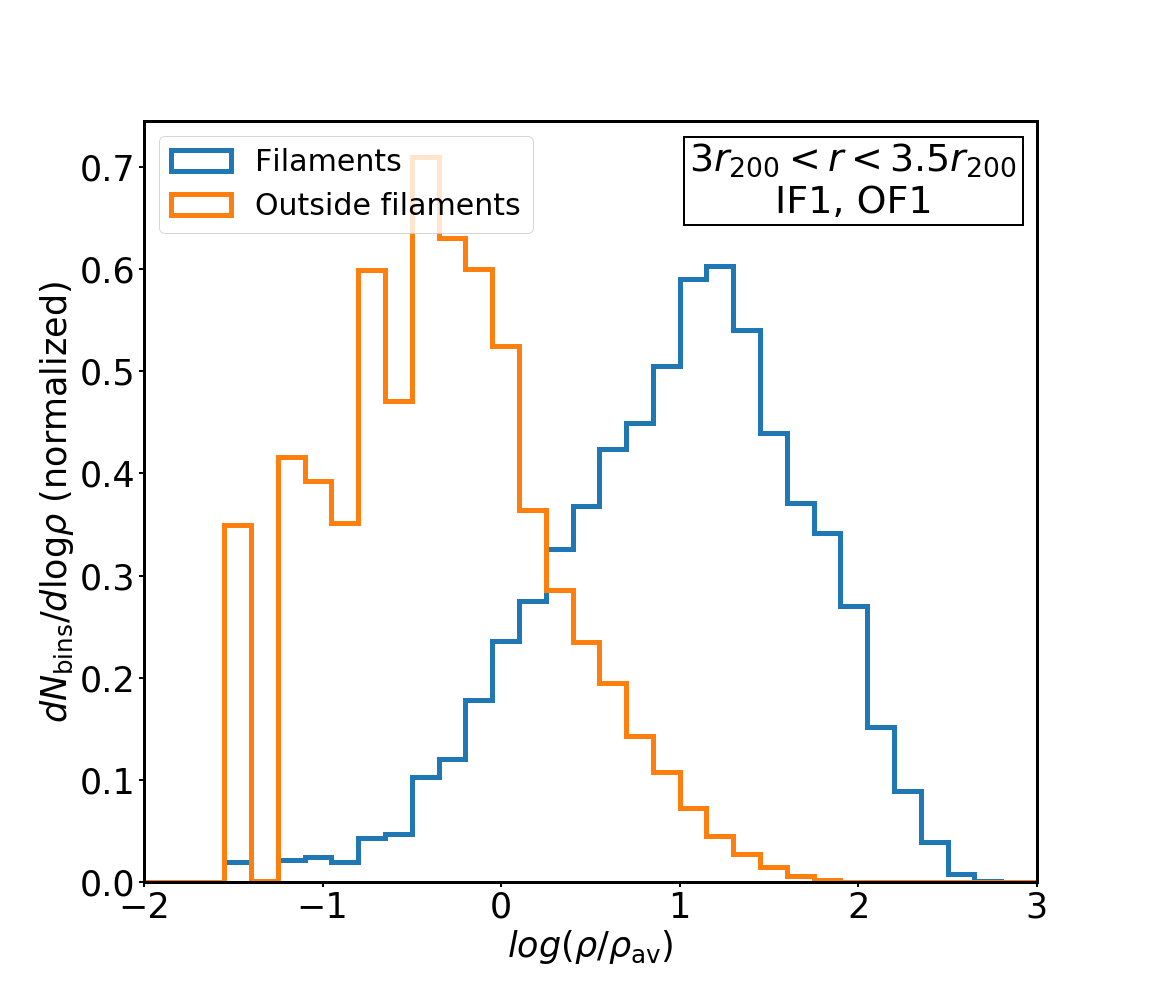}
\includegraphics[width=0.33\textwidth]{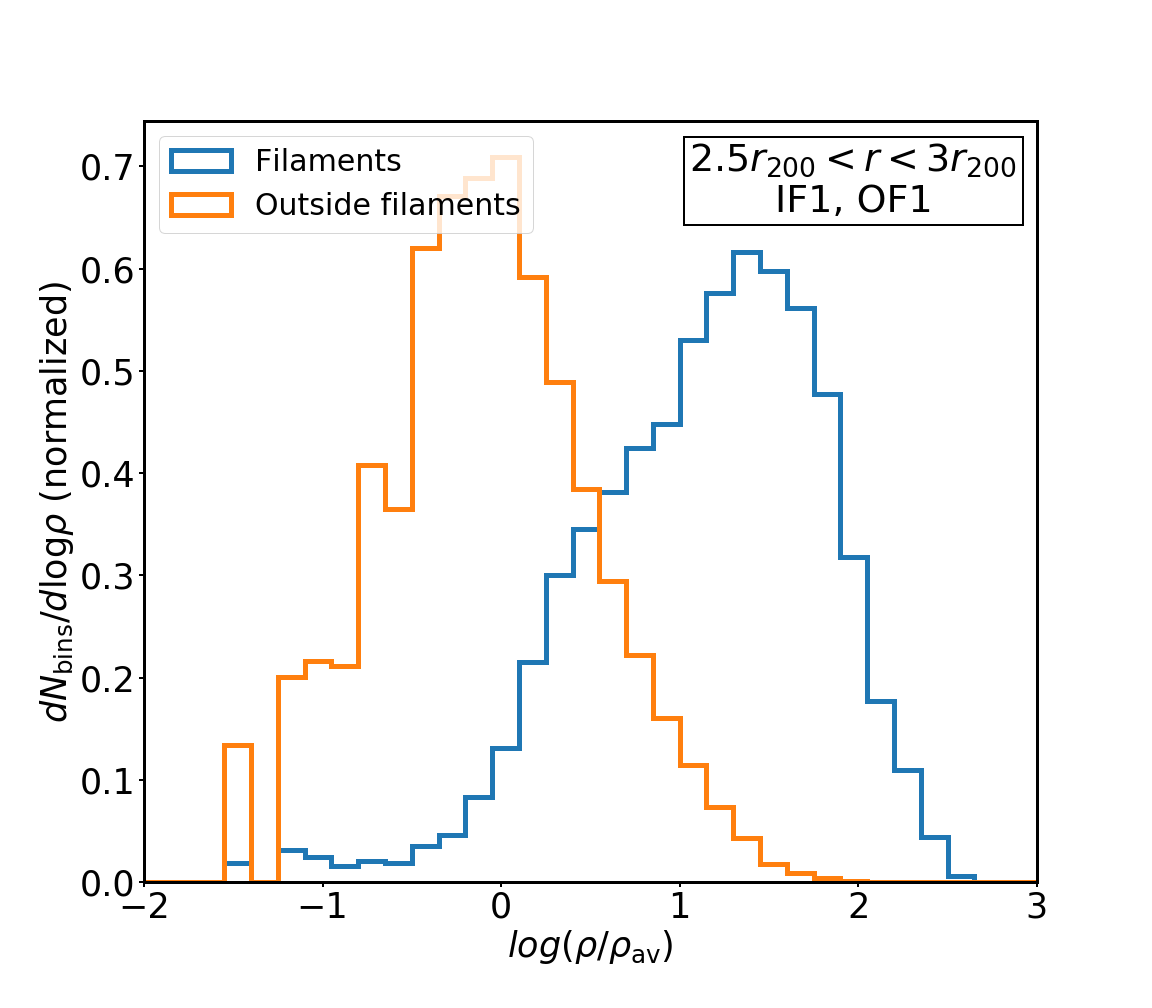}
\includegraphics[width=0.33\textwidth]{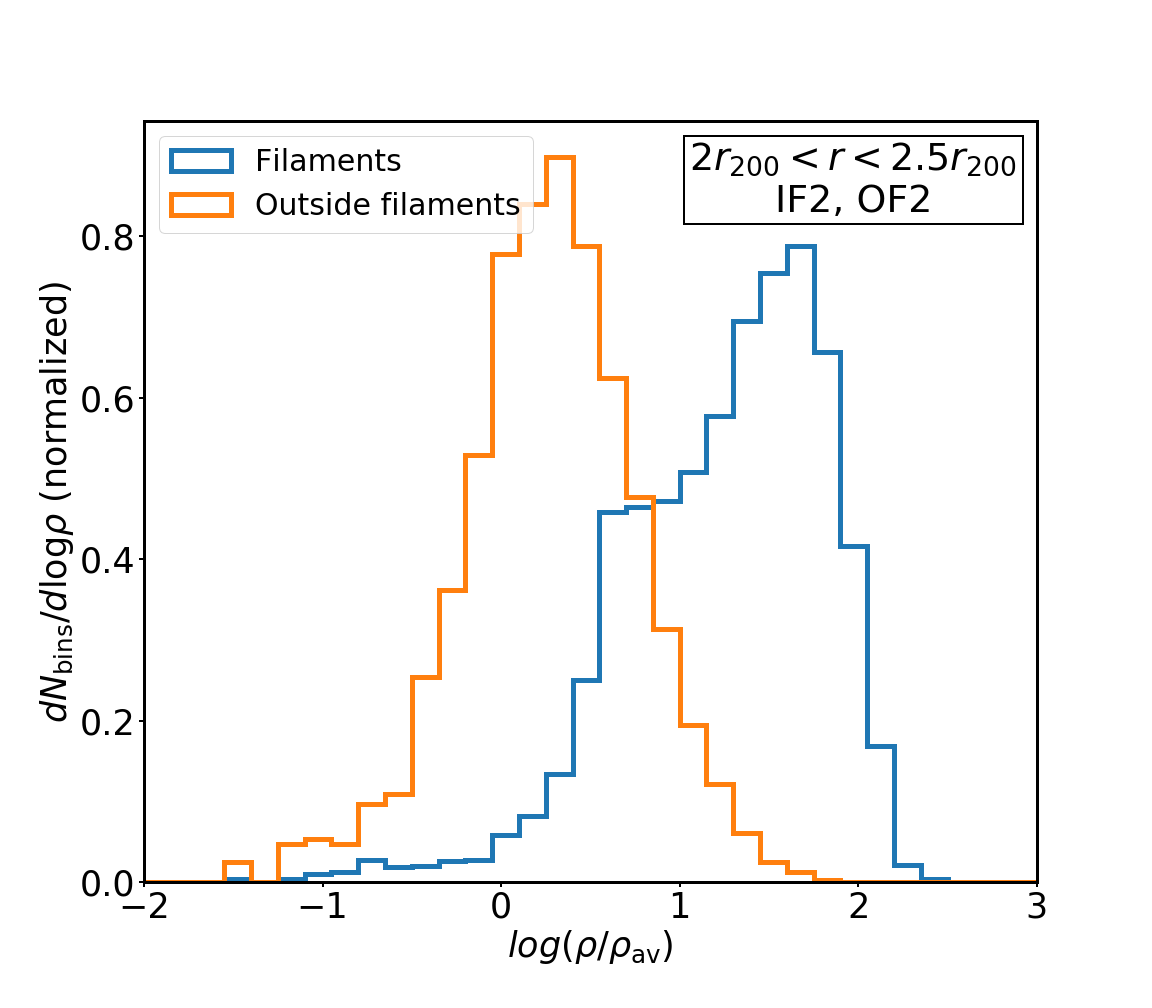}
\includegraphics[width=0.33\textwidth]{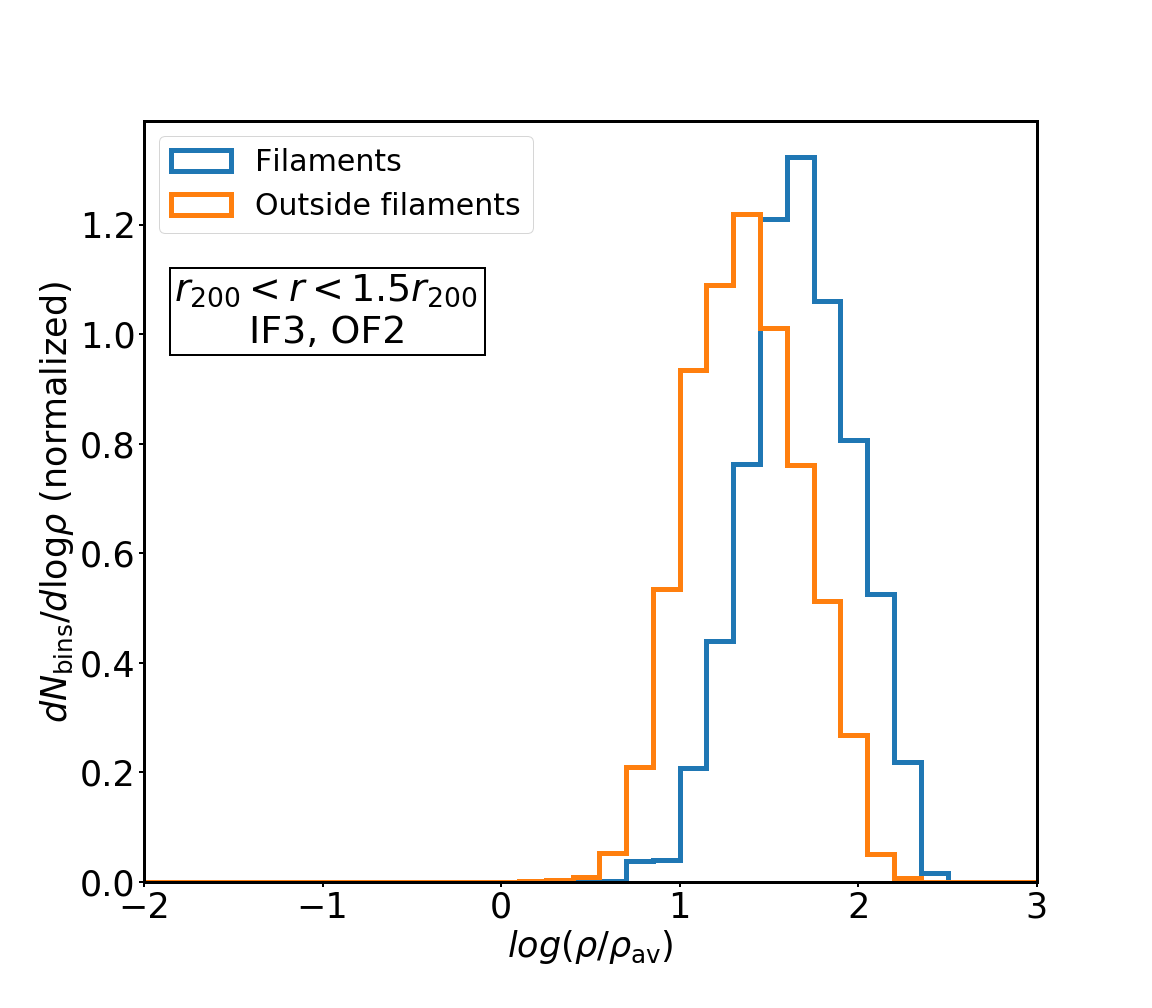}
\vspace{-0.5cm}
\caption{Overdensity distributions within (blue line) and outside (orange line) filaments, at different radial distances from the cluster centre. The distance is decreasing from left to right and top to bottom as indicated in the panels. Note that all distributions have been normalised independently for visual clarity.}
\label{fig:dens:hist}
\end{figure*}

Figure ~\ref{fig:dens:hist} shows the overdensity distributions in selected radial ranges, separately for gas within and outside filaments. At all radii, the distribution of the gas overdensity within filaments peaks at systematically higher values than outside filaments. Thus, when searching for the densest gas in the cluster outskirts in observations, it is beneficial to focus on the locations of the filaments which can be detected relatively easily using spectroscopic galaxy surveys such as Sloan Digital Sky Survey \citep[SDSS; e.g.,][]{2014MNRAS.438.3465T}.
 
\subsection{Temperature}

\label{zones}
Within the cluster-filament interface, the median temperature profile of the gas inside filaments suggests three different zones (see Fig. \ref{fig:temperature:density:median}, top right panel). When the gas flows from the Cosmic Web domain at $r > 4 r_{200}$ towards  $r \sim 2.5 r_{200}$ (Zone IF1), the temperature increases from the WHIM temperatures of $T \sim 10^5 - 10^6$~K, following $T \propto \sim r^{-2.6}$ down to $r \sim 2.5 r_{200}$, reaching a level of a few times $10^6$~K. Continuing inwards, in a relatively narrow radial range of $r \sim 1.7-2.5 r_{200}$ (Zone IF2) the gas temperature rapidly increases by almost an order of magnitude, exceeding $10^7$~K, following approximately $T \propto \sim r^{-5.1}$. Between $\sim 1.7 r_{200}$ and the virial boundary (Zone IF3) the temperature profile is flatter ($T \propto \sim r^{-0.9}$), reaching $\sim 5 \times 10^7$~K where the filaments enter the cluster.

Compared with the filaments, the gas outside filaments is significantly cooler near the outer edge of the interface region ($r \approx 4 r_{200}$), at the expected void level of $T = 10^3 - 10^4$~K \citep{Cautun2014, Tuominen2021}. However, it heats up rapidly between $\sim 4 - 2.7 r_{200}$ (Zone OF1), reaching $10^7$~K. Inwards from this distance, the temperature profile is flatter, following approximately $T \propto r^{-1.0}$ all the way towards the virial boundary (Zone OF2). As a result, between $2 - 3 r_{200}$, the gas outside the filaments is, on average, hotter than that within filaments.

At the radial range of $0.7 - 1.5 r_{200}$, the temperature profiles of the gas inside and outside filaments (IF3 and the inner part of OF2 in the top right panel of Fig.~\ref{fig:temperature:density:median}) are very similar to those inside the cluster (see Fig.~\ref{fig:cluster}, right panel); the temperature varies between $\sim 3-6 \times 10^7$~K. The similarity implies a causal connection between the gas in the different domains mentioned above. We will discuss this issue further in Section~\ref{IF3}.

\subsection{Radial velocity}
\begin{figure*}[bt]
   \centering
 \includegraphics[width=0.48\hsize]{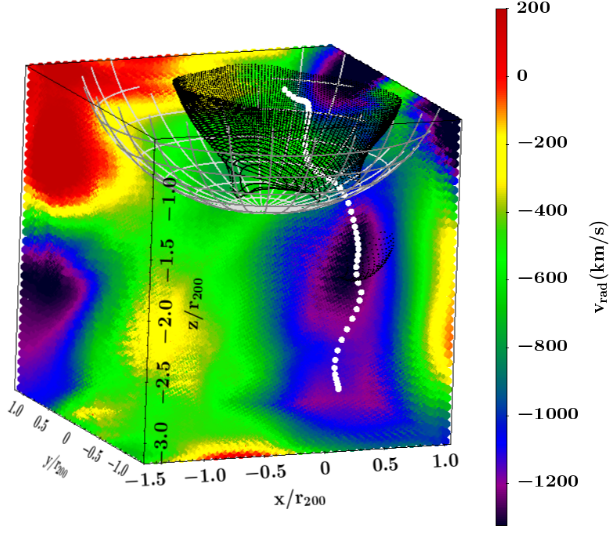}
  \includegraphics[width=0.48\hsize]{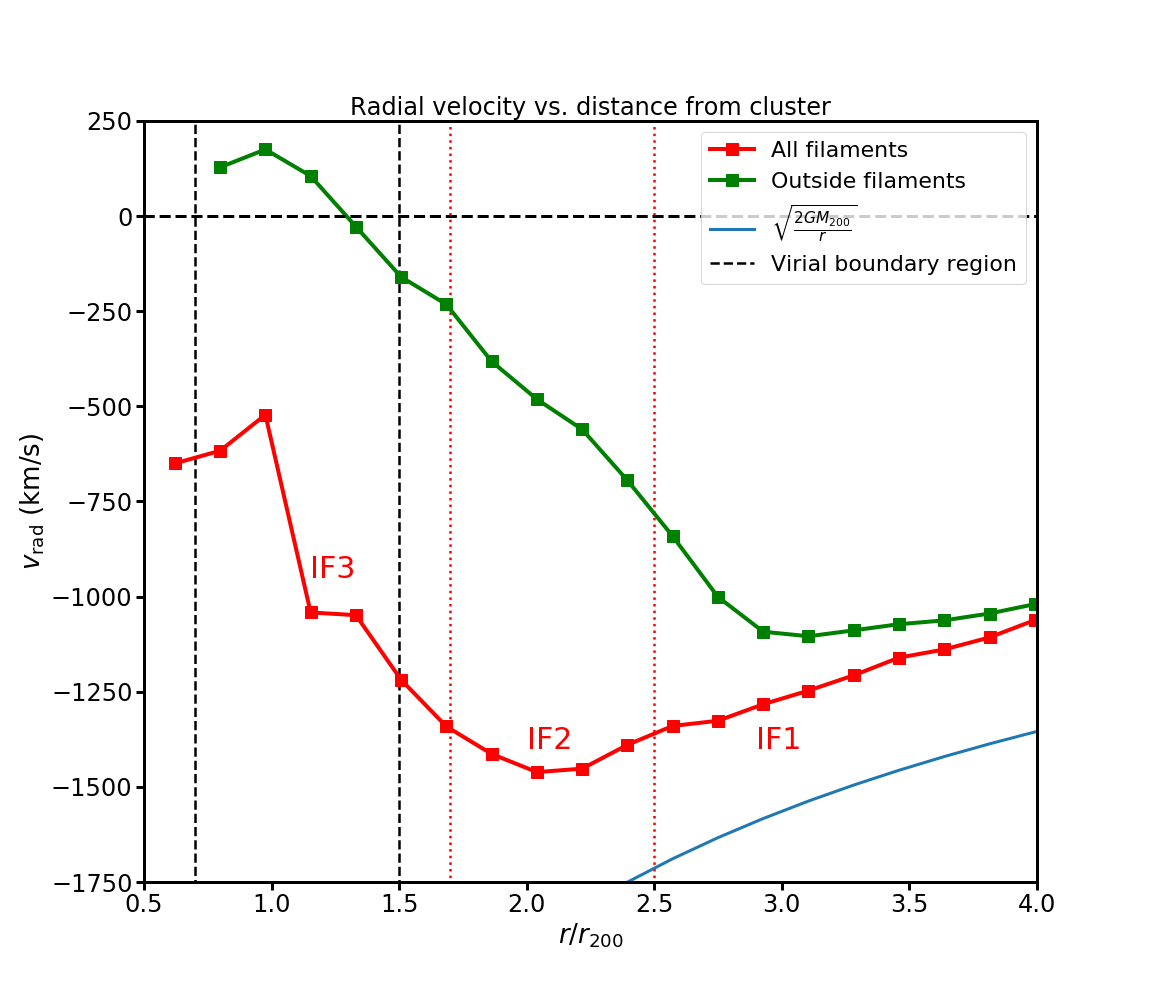}
  \includegraphics[trim=300 50 250 70 , clip, width=0.48\hsize]{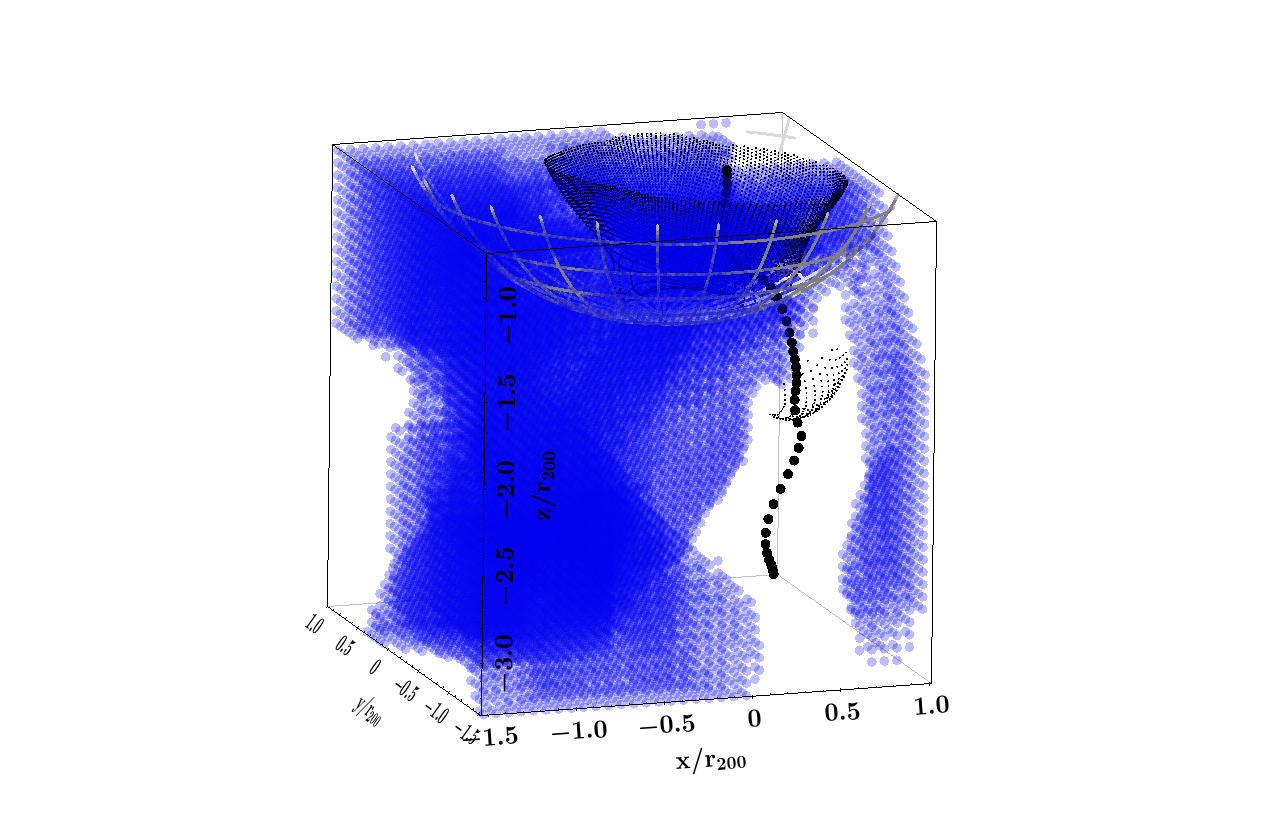}
\includegraphics[width=0.48\hsize]{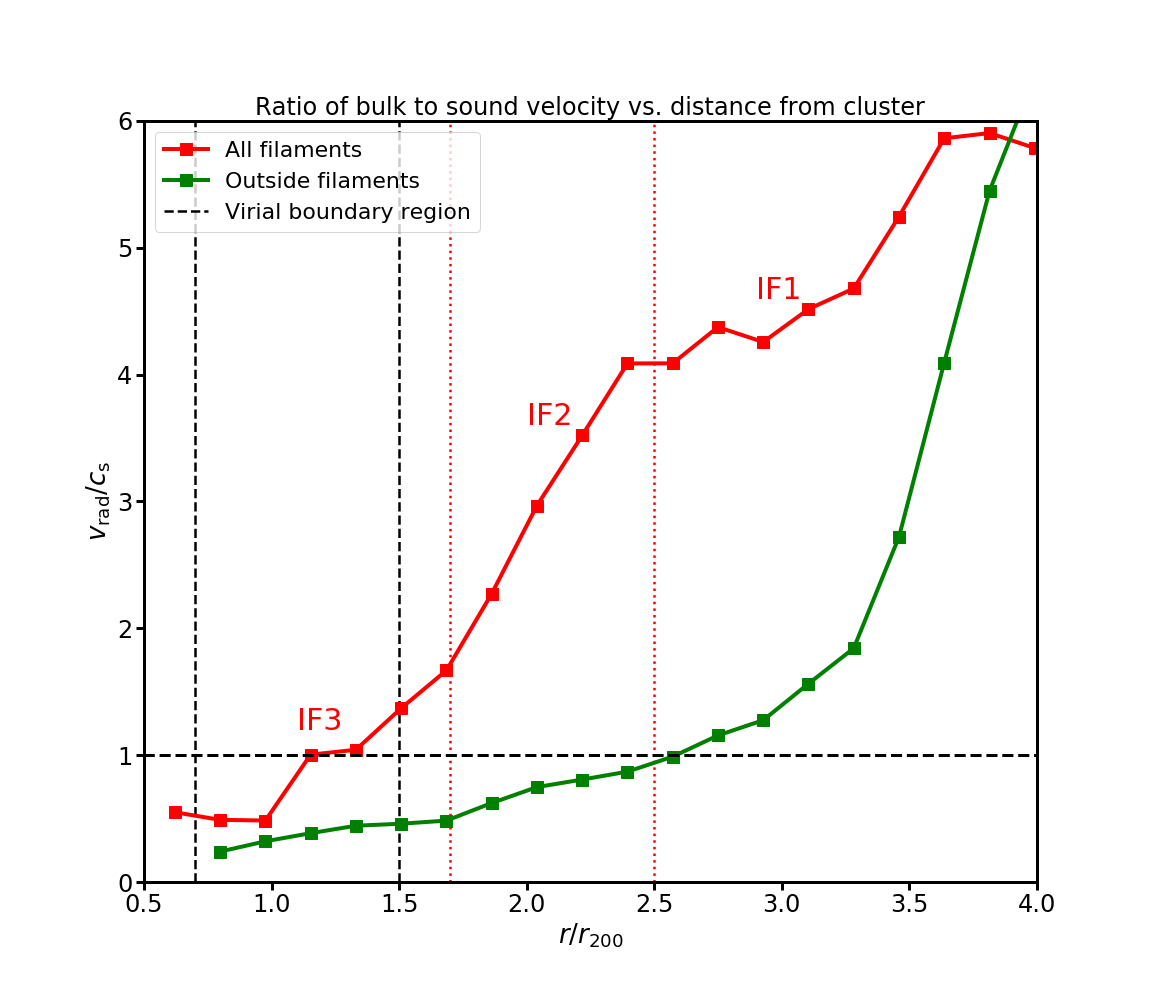}
    \caption{{\it Top left panel:} The radial velocity structure around filament F1 (white dots). Negative (positive) values correspond to gas inflow (outflow). The values are averages in each sight line and thus the image is illustrative rather than quantitatively accurate. The black symbols indicate the boundary of the virial region. The grey surface marks the radius $r = 1.5r_{200}$ from the cluster centre. {\it Top right panel:} radial velocity repeated from Fig.~\ref{fig:temperature:density:median}. {\it Bottom left panel:} The radial Mach number structure around filament F1 (black dots). The subsonic regions (values below 1) are shown in blue and supersonic regions in white, i.e. in the 2D projection white regions indicate the lack of subsonic gas along the line of sight. {\it Bottom right panel:} The median ratio of bulk to sound velocity (a proxy for Mach number) of the gas outside the virial boundary as a function of distance from the cluster centre.}
         \label{fig:3D:vrad:fil2}
   \end{figure*}

The 3D distribution and the median profile of the radial velocity (see Fig.~\ref{fig:3D:vrad:fil2}, top panels) indicate that at $3-4 r_{200}$ the gases inside and outside filaments fall with the same velocity ($\sim 1\,000$~km/s). From $r \sim 3 r_{200}$ inwards the behaviour of the gas in the two environments starts to differ. The gas inside the filaments has a higher velocity and keeps accelerating when falling closer to the cluster, as on a cosmic gas highway. In contrast, the gas outside filaments already starts to slow down at these radii. The deceleration of the filament gas begins closer to the cluster centre, at $\sim 2r_{200}$.
 
The gas outside the filaments stops around the virial boundary ($r\sim 0.7-1.5 r_{200}$), while the gas inside filaments still has a median velocity of $\sim 500-1\,000$~km/s at these radii. The latter is related to the fact that individual filaments can penetrate inside the virialised region (see Fig.~\ref{vir_fil.fig}).

\section{Physical interpretation}
\label{physics}
Having characterised the general radial behaviour of the infalling gas in the outskirts of CE-29 cluster, we now turn to more detailed analysis and physical interpretation of the kinematic and thermodynamical behaviour of gas in the interface region within and outside filaments. The following interpretations are specific to the CE-29 cluster. A rigorous statistical analysis of a full sample of C-EAGLE clusters is required for generalising the features reported here. While we stress the approximate nature of our spherically-averaged analysis, it will be seen below that the simplifications made nevertheless retain sufficient information about the gas thermodynamics for a coherent physical interpretation. To complement the spherically-averaged analysis and to inform on the inherent scatter in filament properties, we will also discuss the latter on an individual filament basis (Section~\ref{indiv}). Furthermore, two specific filaments with markedly distinct properties will be selected for a more detailed analysis. The 3-dimensional structure of gas in the cluster vicinity will also be discussed. This will yield a better understanding of the main features of the thermodynamic properties indicated by the overall profiles, i.e., the different temperature zones (see Fig.~\ref{fig:temperature:density:median}, top right panel) and the slowing down of the infalling gas (see Fig.~\ref{fig:temperature:density:median}, bottom left panel).

\subsection{Slowing down of the infalling gas}
\label{sec:press}

\begin{figure*}[h]
\includegraphics[width=0.33\textwidth]{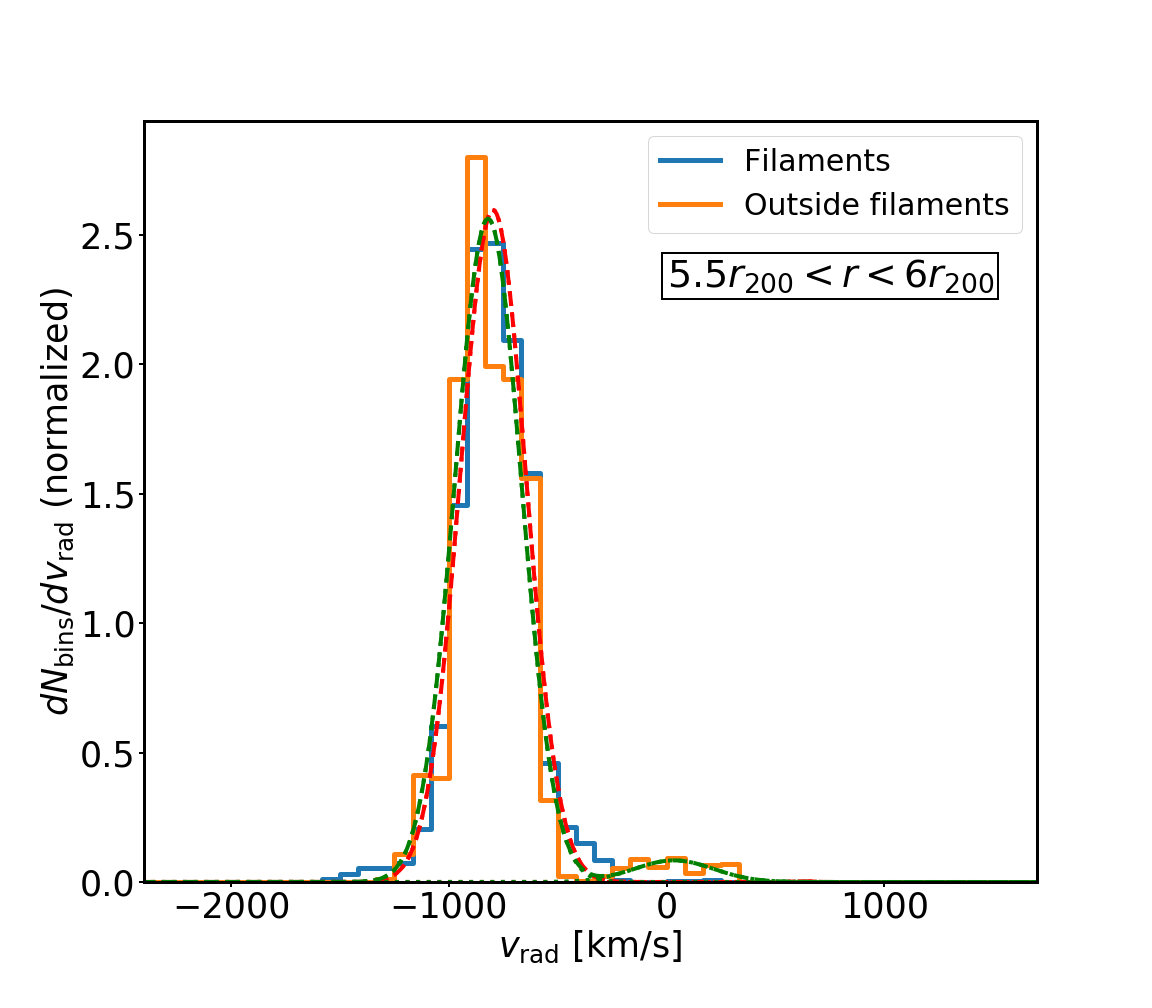}
\includegraphics[width=0.33\textwidth]{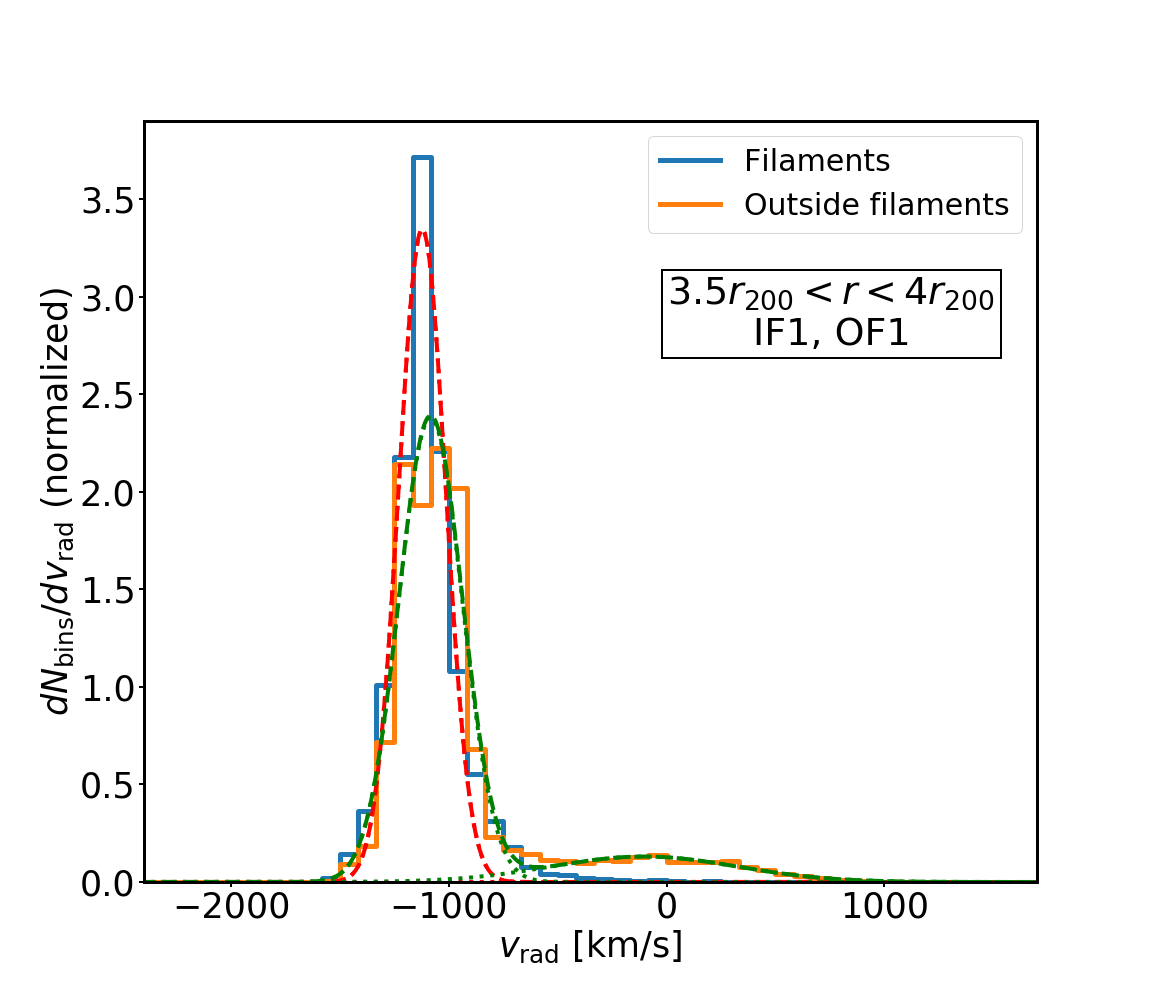}
\includegraphics[width=0.33\textwidth]{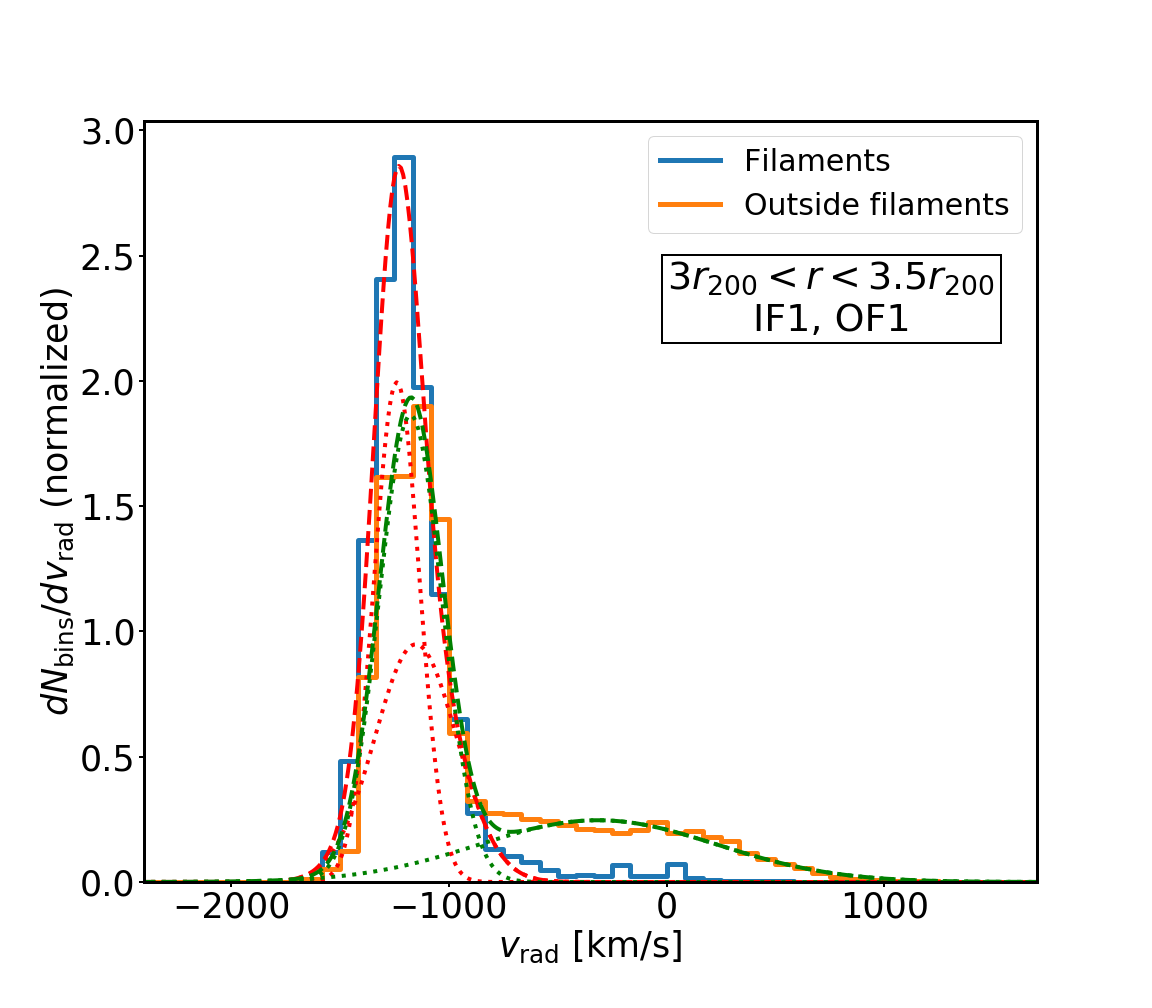}
\includegraphics[width=0.33\textwidth]{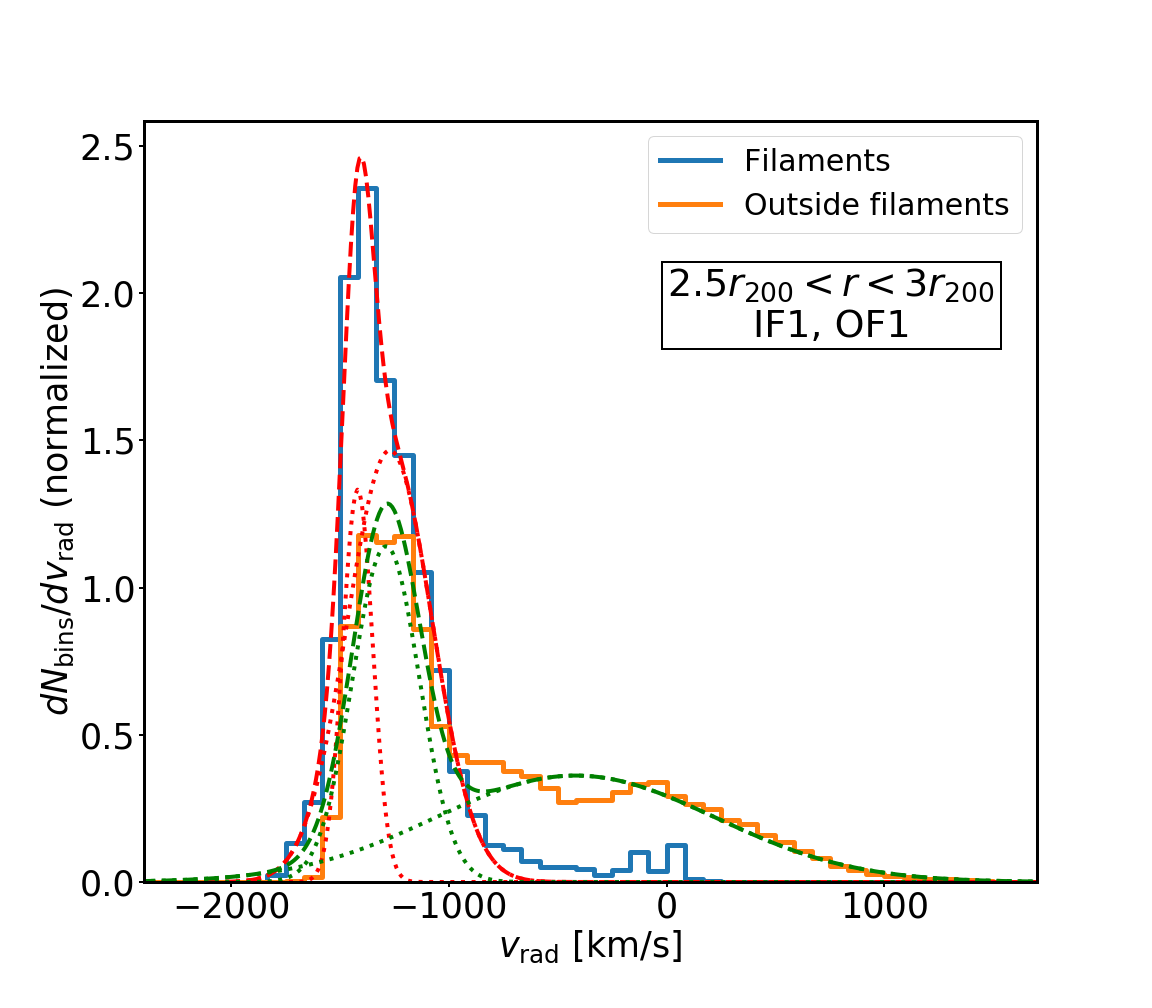}
\includegraphics[width=0.33\textwidth]{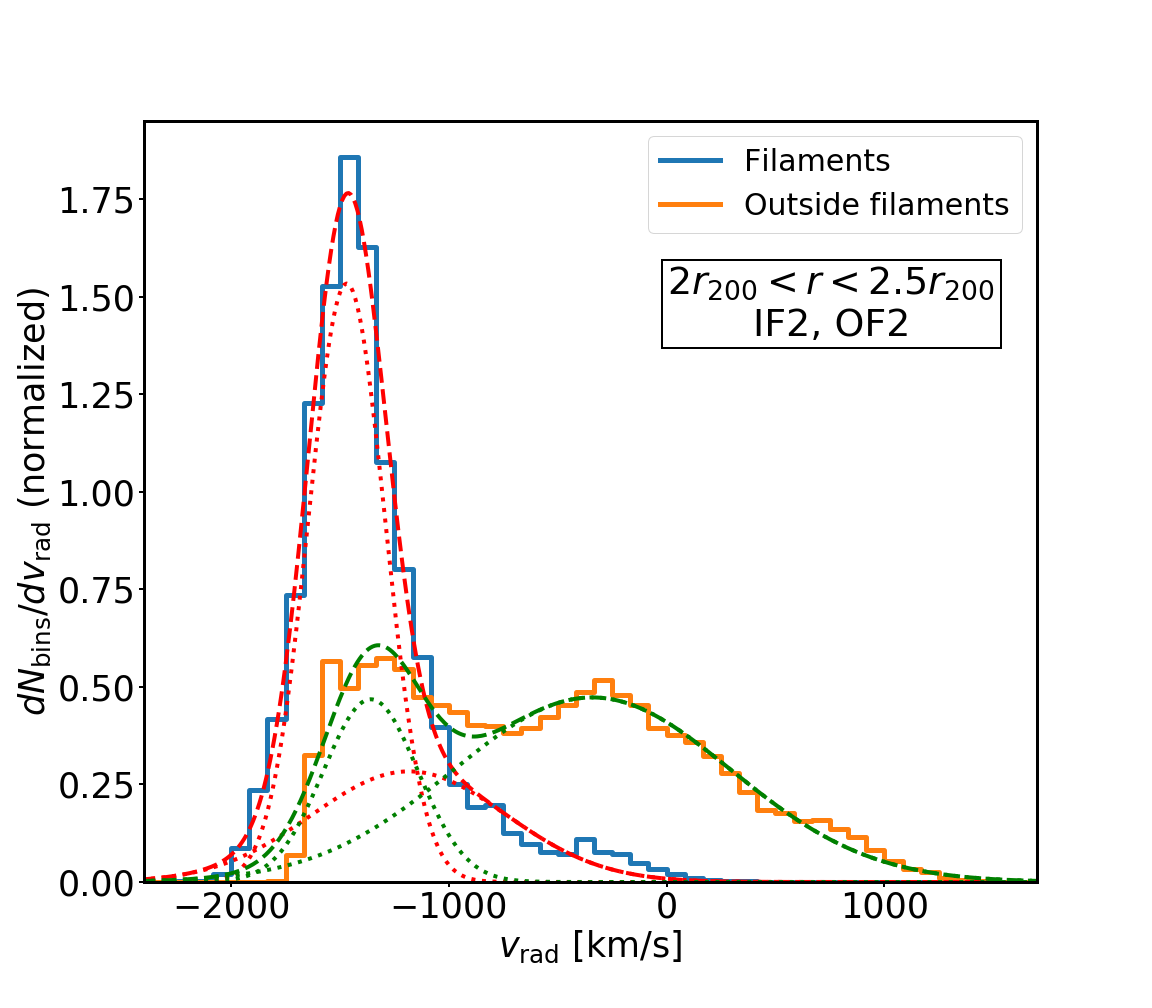}
\includegraphics[width=0.33\textwidth]{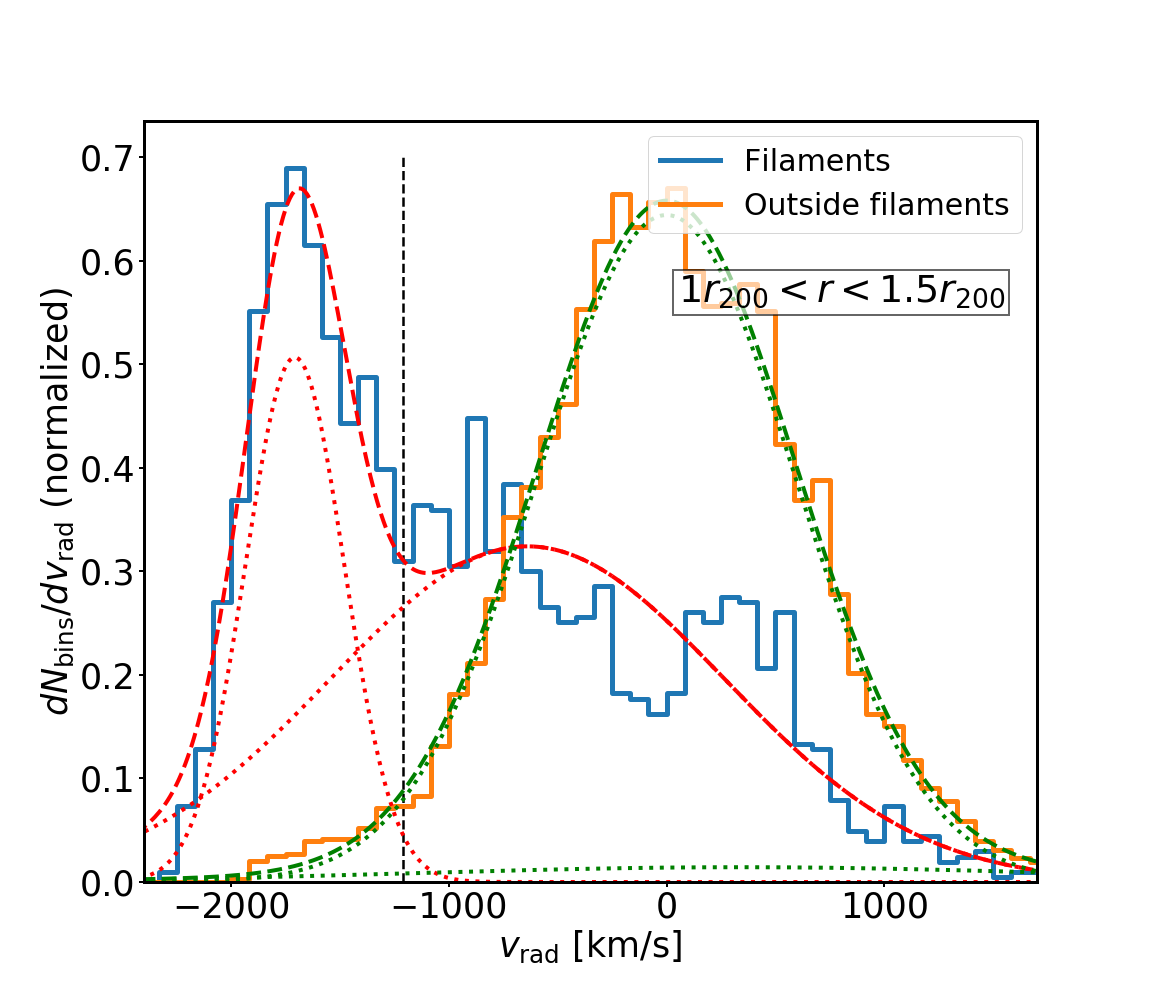}
\vspace{-0.5cm}
\caption{Radial velocity distributions within (blue line) and outside (orange line) filaments at different radial distances from the cluster centre. The distance is decreasing from left to right and top to bottom as indicated in the panels. The dashed red and green curves show double gaussian fits to the radial velocity distributions within and outside filaments, respectively. The individual gaussian components of the fits are shown by dotted lines. The vertical dashed black line on the bottom right panel shows $v_{\rm ff}/2$ at $1.25 r_{200}$; within filaments, approximately half of the gas (by mass) retains radial velocities $v_{\rm rad} < v_{\rm ff}/2$ at $r_{200} < r < 1.5r_{200}$.}
\label{fig:vrad:hist}
\end{figure*}

We first investigate how the infalling gas is slowed down depending on whether it approaches the cluster via filaments or from the outside. In Figure \ref{fig:vrad:hist}, we show the distributions of gas radial velocities within and outside filaments in selected radial ranges. Also shown are double gaussian fits for each distribution (see Appendix \ref{sec:app:signif} for more details), with the rationale that  the infalling gas both within and outside filaments can be broadly classified into two phases, i.e. the free-falling component and the decelerated/shocked one. By tentatively identifying the above components with the two gaussians, the relative magnitudes of the latter
inform about the fraction of decelerated vs. free-falling gas in the two environments.

The ability of the accreting material to penetrate into the cluster is determined mainly by its ram pressure, defined for an approximately radial infall as $p_{\rm ram} = \rho v_{\rm rad}^2$.  As long as $p_{\rm ram}$ significantly exceeds the thermal pressure $p_{\rm gas} = 2\rho k_{\rm B} T/m_{\rm p}$ ahead of it, the gas remains in free-fall. At the largest radii ($r > 3r_{200}$), the ambient gas pressure around the cluster CE-29 is very low compared to the ram pressure of the infalling gas (see Fig.~\ref{fig:temperature:density:median}, bottom right panel). Thus, the gas both inside and outside filaments is presumably dynamically affected mainly by the gravitational pull of the cluster. The velocity distributions inside and outside the filaments are consequently relatively narrow and almost identical (see Fig.~\ref{fig:vrad:hist}) and comparable to the free-fall velocity derived from the cluster mass (see Fig.~\ref{fig:temperature:density:median}, bottom left panel).

\subsubsection{Inside filaments}

The ambient gas pressure becomes comparable to the ram pressure of the gas within filaments only once the latter has reached the vicinity of the virialized region of the cluster (see Fig.~\ref{fig:temperature:density:median}, bottom left panel). At this stage, the filament gas begins to slow down while dissipating its kinetic energy, most likely via shocks starting at $\sim 2.0 r_{200}$. Yet, by mass, approximately half of the gas inside filaments remains within $\sim 50\%$ of the free-fall velocity ($v_{\rm rad} < v_{\rm ff}/2$) down to the virial region (bottom right panel in Fig.~\ref{fig:vrad:hist}) and therefore penetrates into the cluster (for comparison, the free-falling gaussian component accounts for 28\% of the filament by volume). The persistence of a significant free-falling component is in agreement with the results of \citet{Zinger2016,Zinger2018}, who found that gas streams associated with filaments can penetrate deep into the cluster in approximately half of the simulated clusters that they studied, and that the penetration was more efficient in dynamically unrelaxed clusters.

The dissipation associated with slowing gas down heats it up (see Fig.~\ref{fig:temperature:density:median}, top right panel) and its thermal pressure increases above the ram pressure, indicating that a substantial fraction of the infalling material has become subsonic and dominated by thermal (rather than kinetic) energy (see Section~\ref{shock} for further discussion about shocks). On average, the supersonic ($v_{\rm rad}/c_{\rm s} > 1$) to subsonic ($v_{\rm rad}/c_{\rm s} < 1$) transition takes place at $r \sim 1.2 r_{200}$, i.e., within the radial limits of the elongated cluster virial boundary (see Fig.~\ref{fig:3D:vrad:fil2}, bottom right panel).

\subsubsection{Outside filaments}

In contrast, outside of the virial region the density of the gas outside filaments is more than an order of magnitude below that of the filament gas. Consequently, its ram pressure is lower and its infall starts to be affected by the gas pressure ahead already at $r\sim 3 r_{200}$. This is manifested via the appearance and subsequent growth of a separate low-velocity component (in terms of the absolute value) in the distribution, indicating that an increasing fraction of gas is being slowed down (Figure~\ref{fig:vrad:hist}, top right and bottom left panels). By $r \approx 2 - 2.5r_{200}$, the low-velocity component has increased sufficiently so that the overall distribution is clearly bimodal, supporting our two-phase interpretation. By that stage, much of the gas outside filaments has been slowed down; this corresponds to the radius where gas and ram (median) pressures have become comparable (Fig.~\ref{fig:temperature:density:median}).

As within filaments, the deceleration is accompanied by heating of the gas and transition into the subsonic domain (see Fig.~\ref{fig:3D:vrad:fil2}, bottom right panel). However, unlike filaments, the gas outside filaments does not retain a free-falling component near the virial radius; its radial velocity distribution can be well fitted by a single gaussian with mean velocity $\overline{v}_{\rm rad} \approx 0$ (bottom right panel on Fig.~\ref{fig:vrad:hist}), suggesting that essentially all of the gas accreting in this mode has been processed through shocks by this stage.

\begin{figure*}[bt]
 \includegraphics[width=0.33\textwidth]{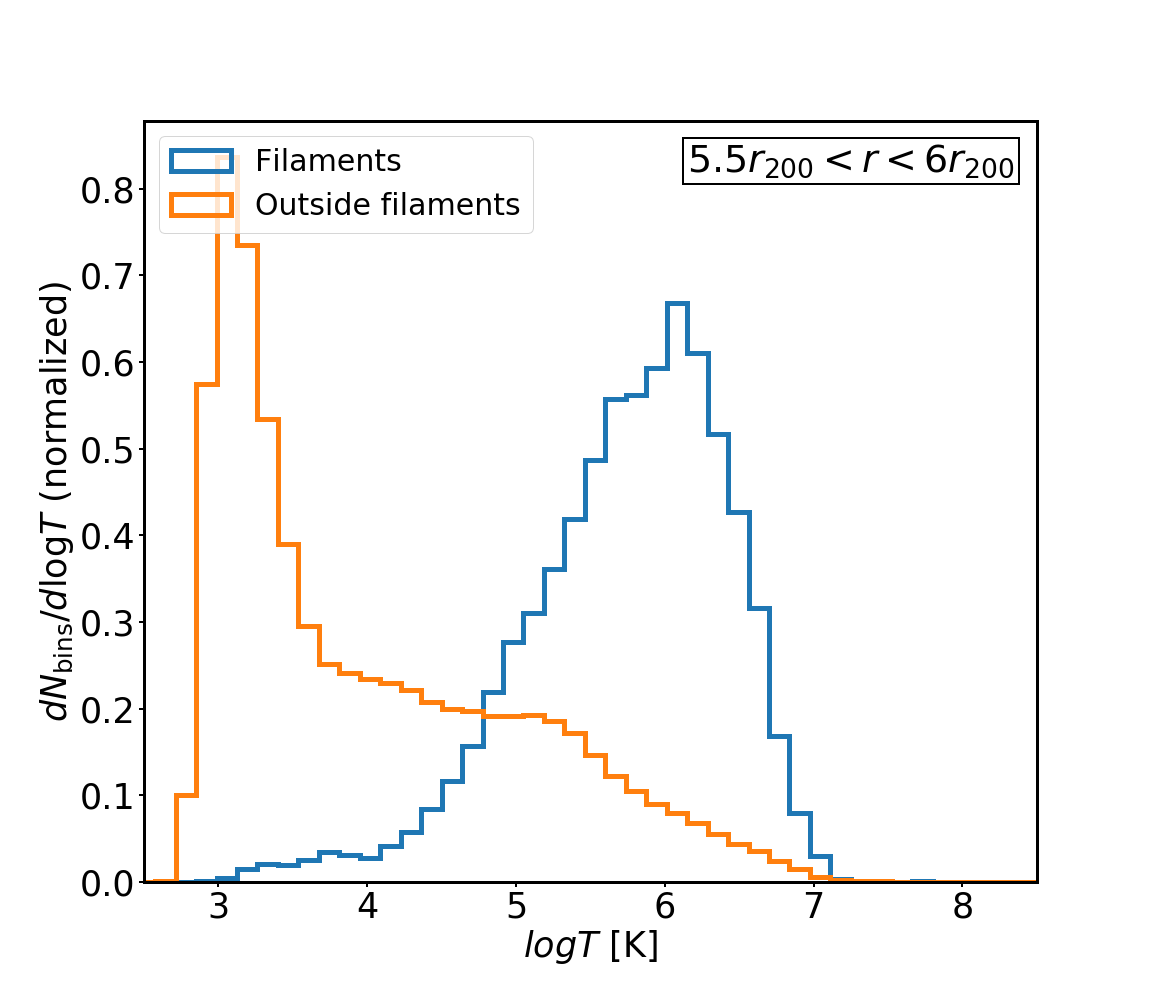}
\includegraphics[width=0.33\textwidth]{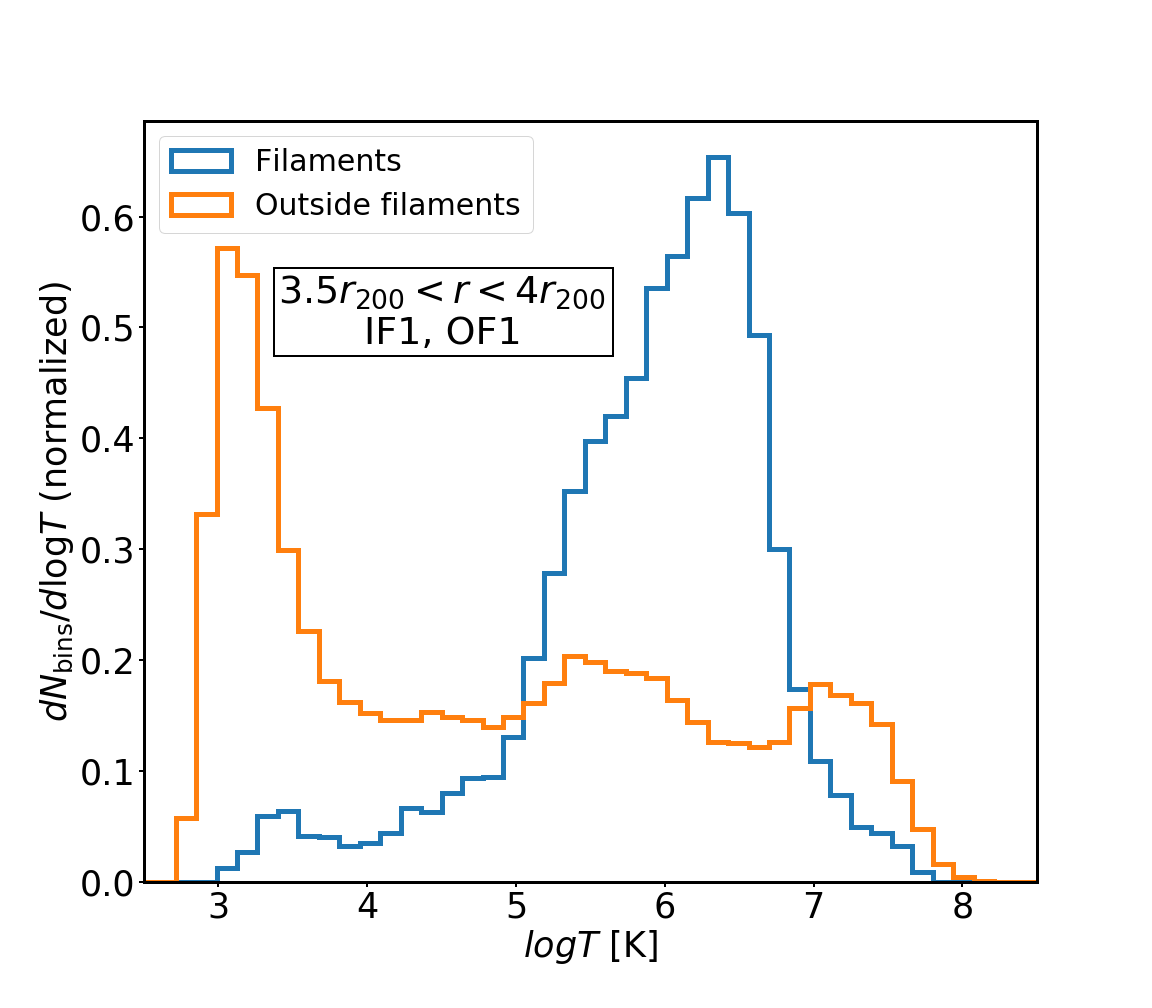}
\includegraphics[width=0.33\textwidth]{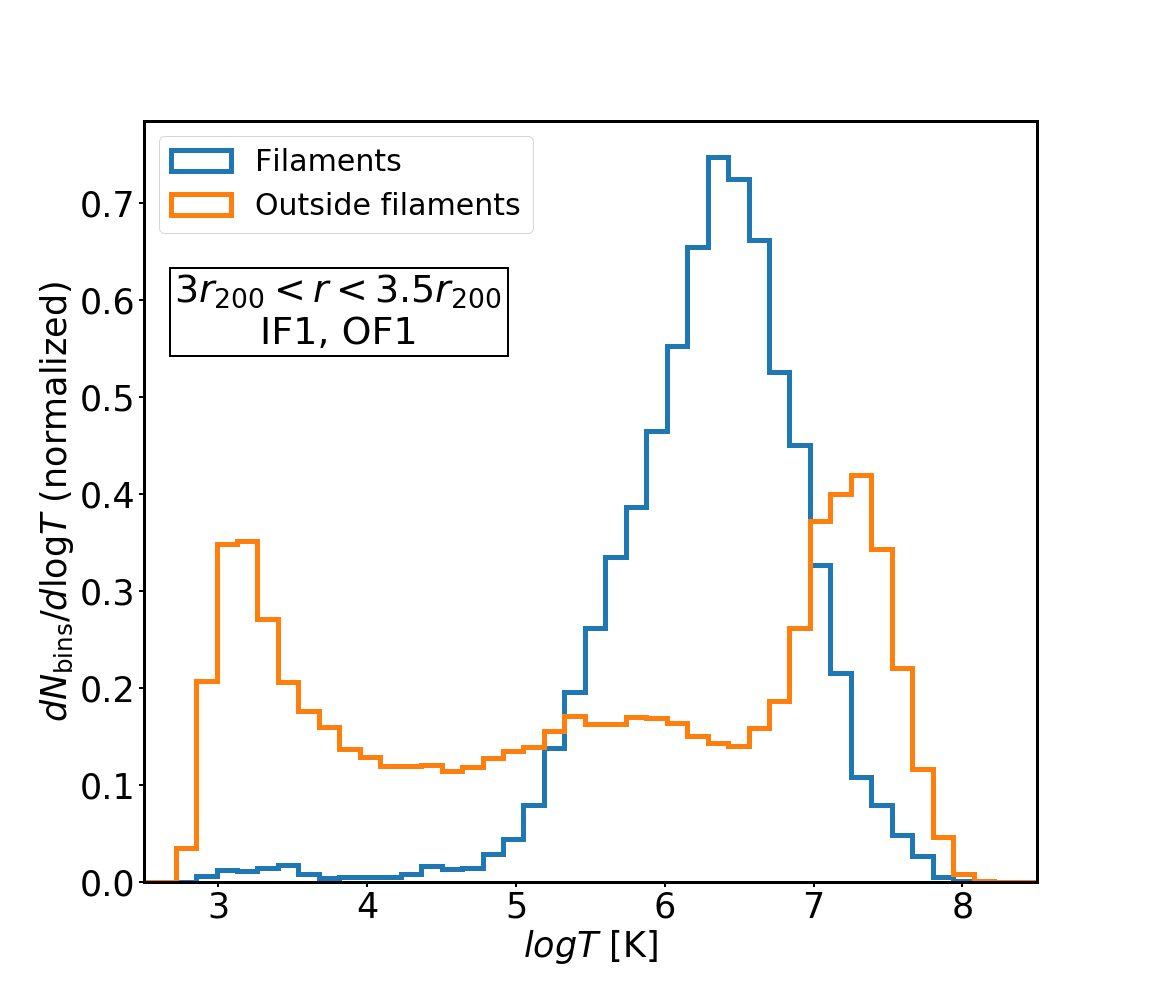}
\includegraphics[width=0.33\textwidth]{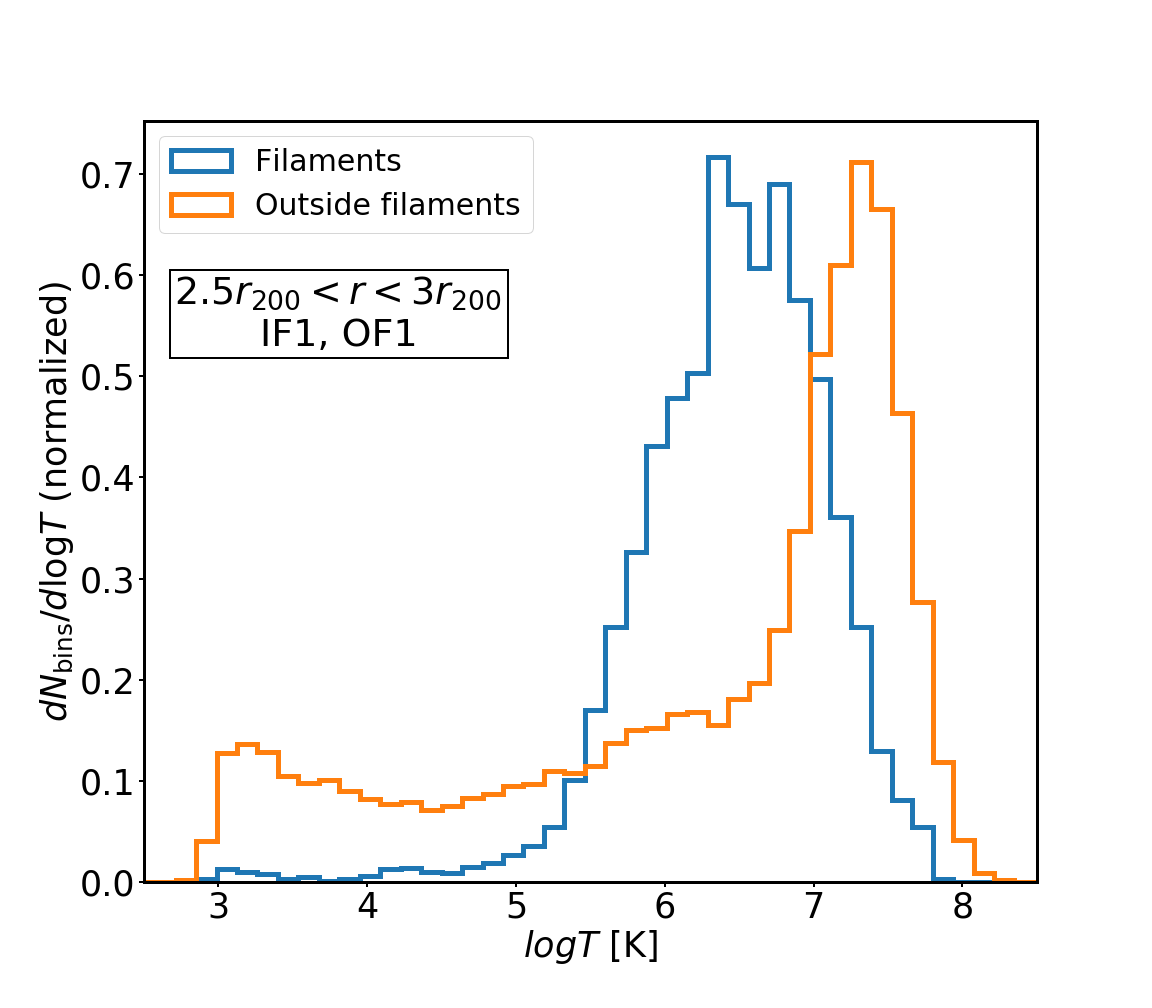}
\includegraphics[width=0.33\textwidth]{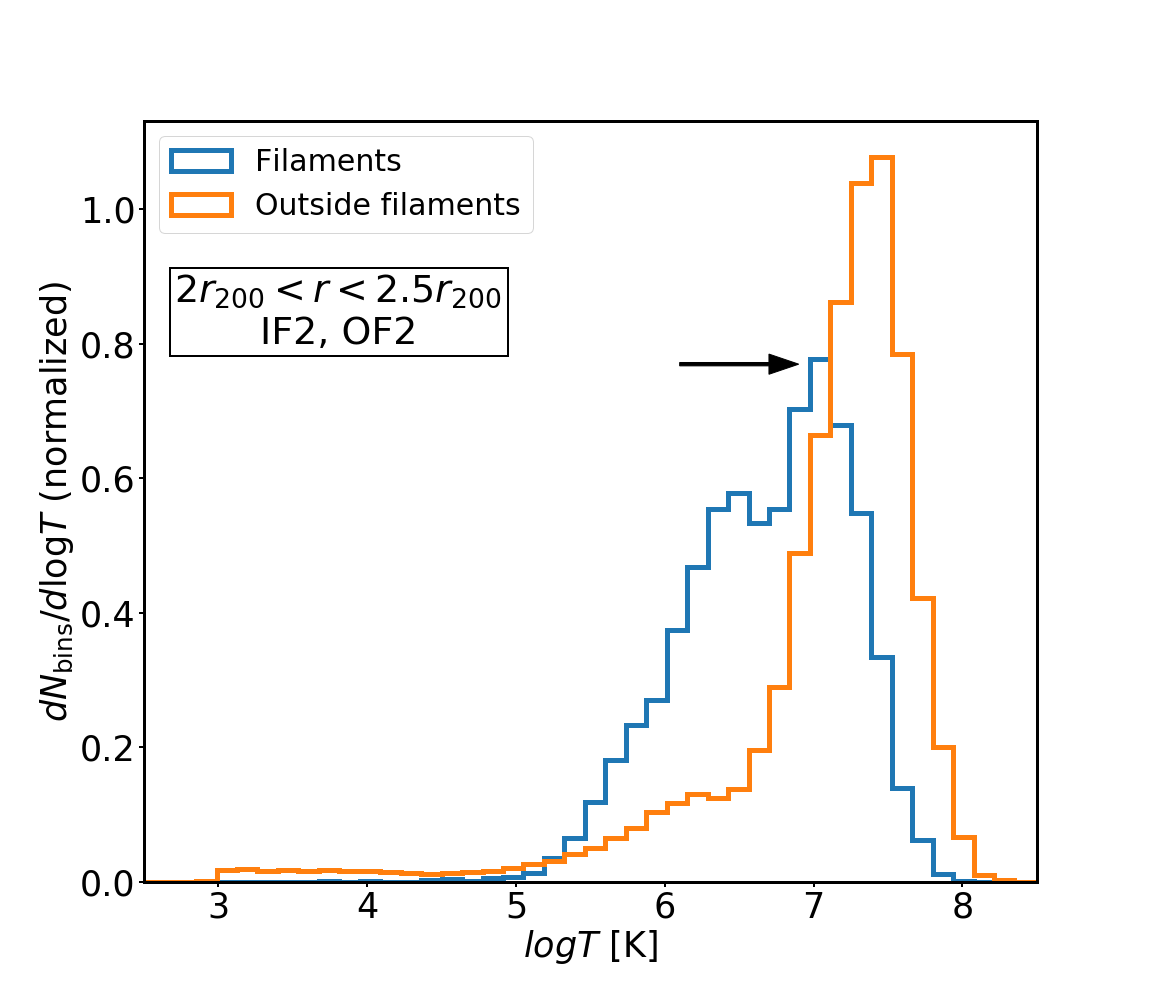}
\includegraphics[width=0.33\textwidth]{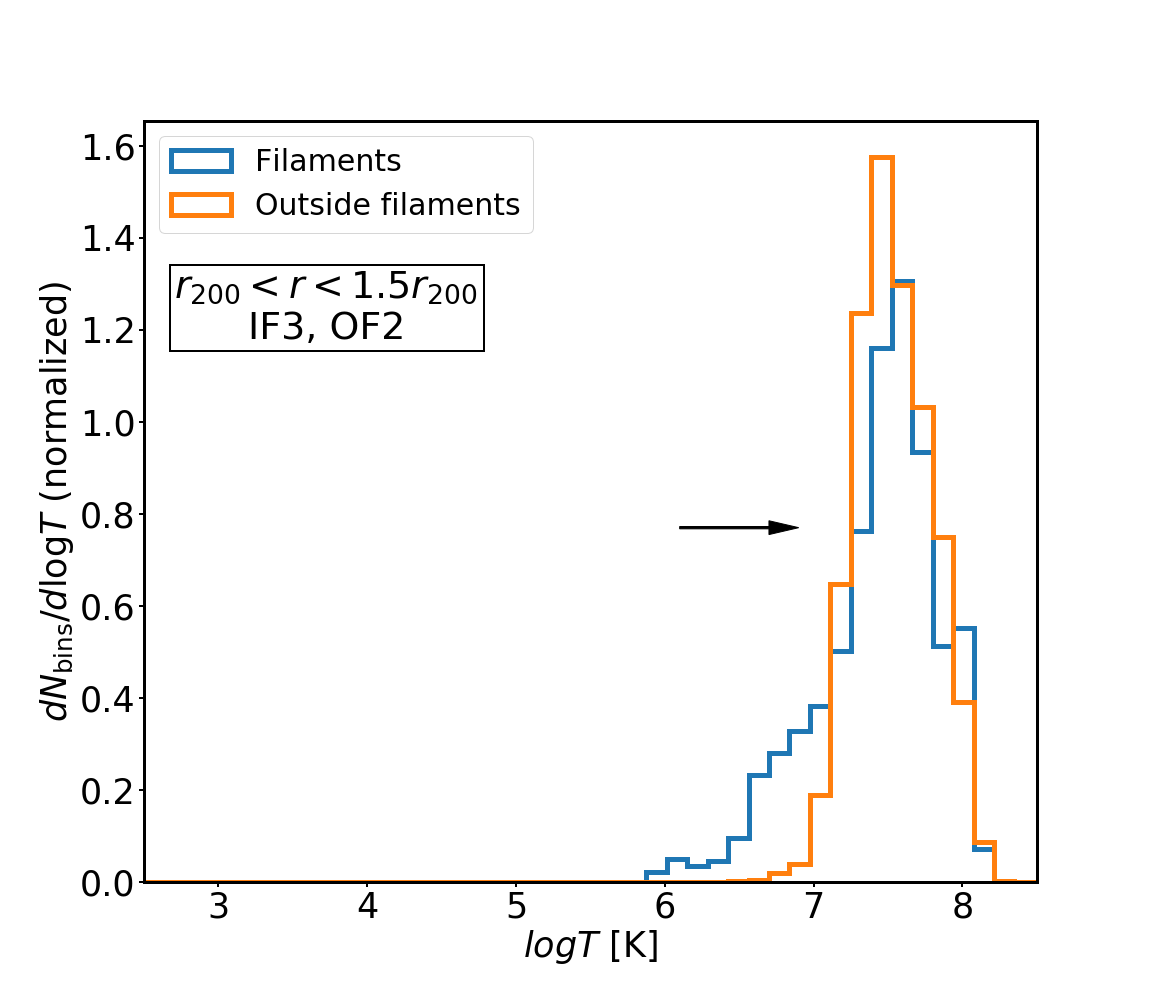}
\vspace{-0.5cm}
\caption{Temperature distributions within (blue line) and outside (orange line) filaments at different radial distances from the cluster centre. The distance is decreasing from left to right and top to bottom as indicated in the panels. The secondary peak discussed in the text is indicated with a black arrow in the bottom middle panel.}
\label{fig:temp:hist}
\end{figure*}

\begin{figure*}[bt]
\includegraphics[width=0.5\textwidth]{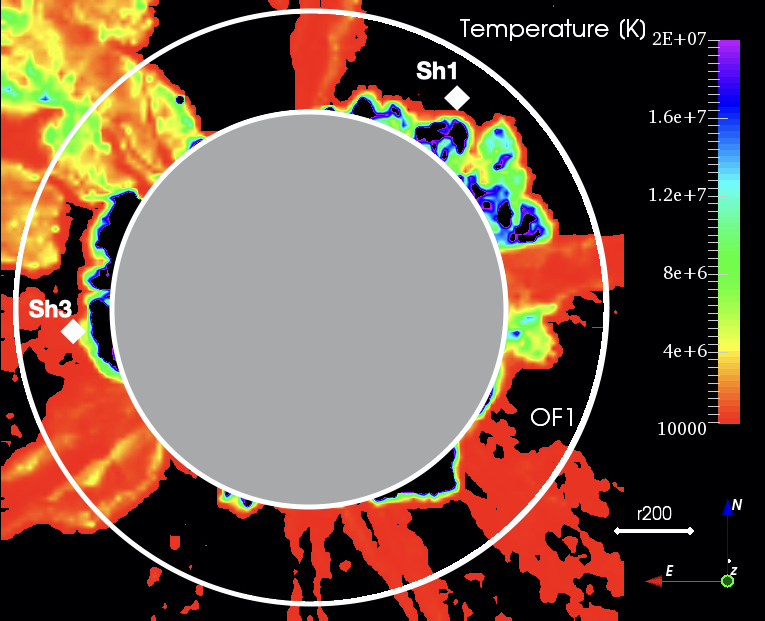}
\includegraphics[width=0.5\textwidth]{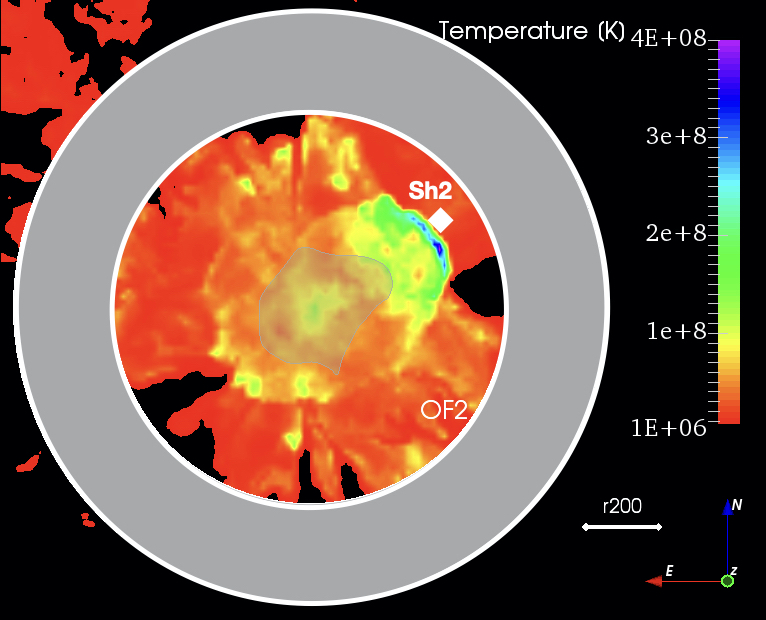}
\vspace{-0.0cm}
\caption{
The temperature of the gas in the OF1 (\emph{left}; black region corresponds to $T < 10^4$~K) and OF2 (\emph{right}; black region corresponds to $T < 10^6$~K) zones in a planar cross-section of size $(24~{\rm Mpc})^2 \approx (8.5 r_{200})^2$, crossing through the cluster centre. Note that the temperature scale is different for the two panels, in order to maintain the details in the two regions with very different properties. The white diamonds labelled Sh1--Sh3 mark the locations of heating events (possibly shocks) in gas outside filaments, for referencing in the text. The shaded area in the middle of the right panel shows the virial boundary of the cluster.}
\label{fig:T_only:2D}
\end{figure*}

\begin{figure*}[p]
\includegraphics[trim=60 10 0 0 0, clip, width=0.5\textwidth]{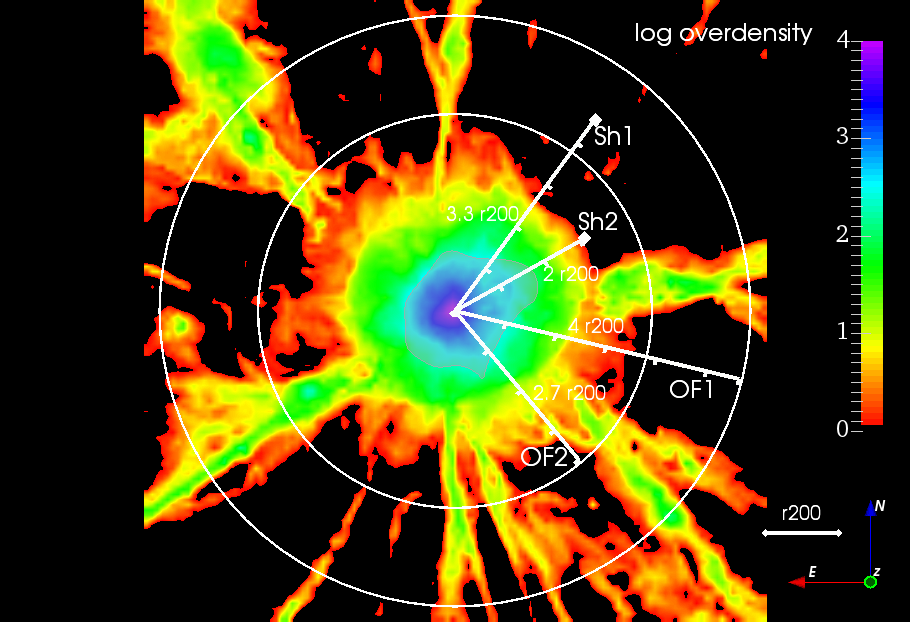}
\includegraphics[trim=60 10 0 0 0, clip, width=0.5\textwidth]{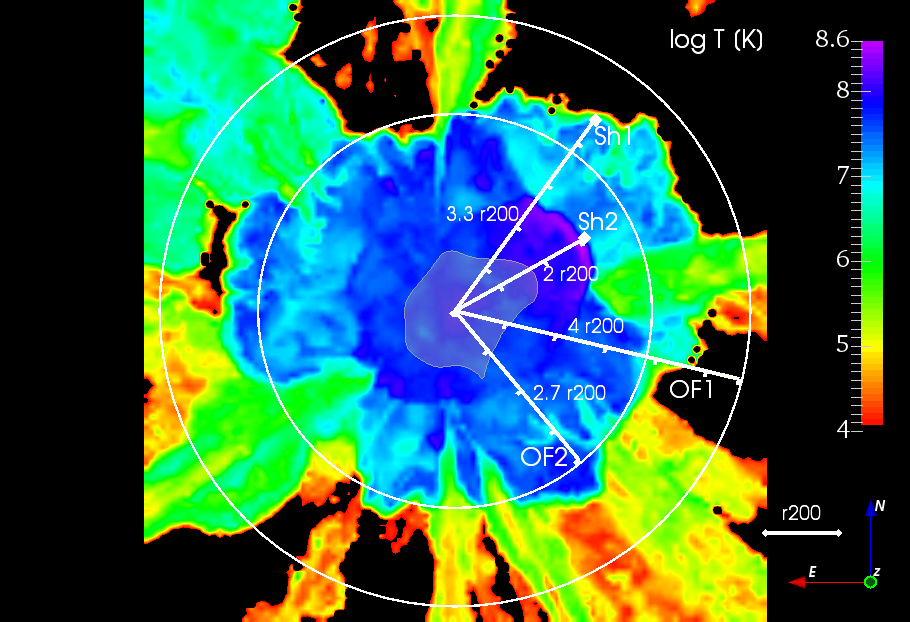}
\includegraphics[trim=60 10 0 0 0, clip, width=0.5\textwidth]{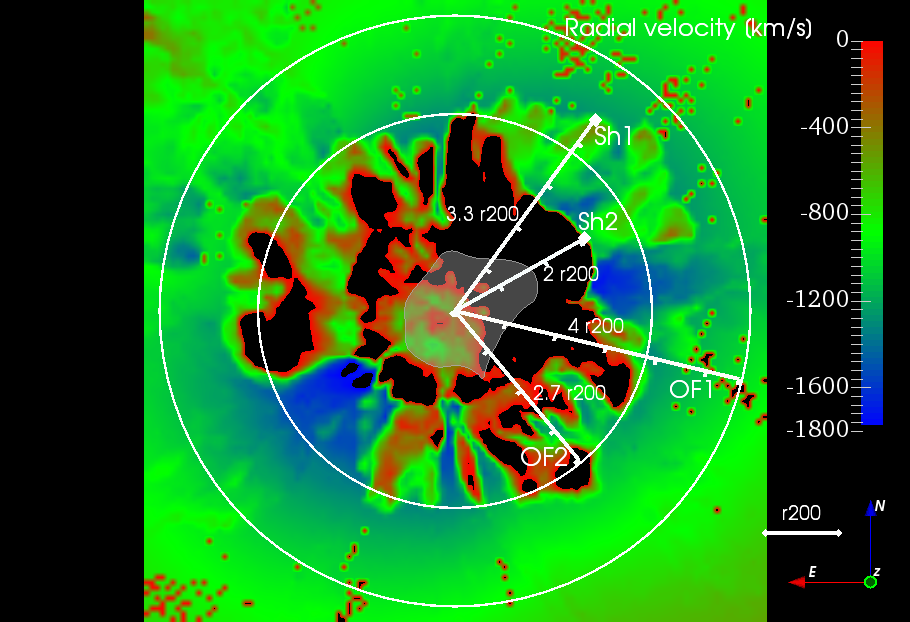}
\includegraphics[trim=60 10 0 0 0, clip, width=0.5\textwidth]{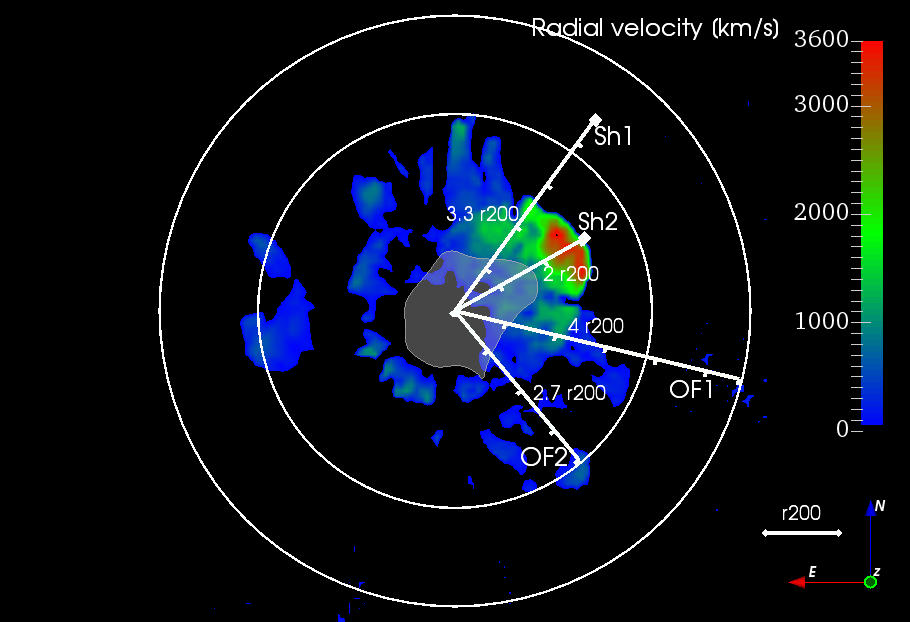}
\includegraphics[trim=60 10 0 0 0, clip, width=0.5\textwidth]{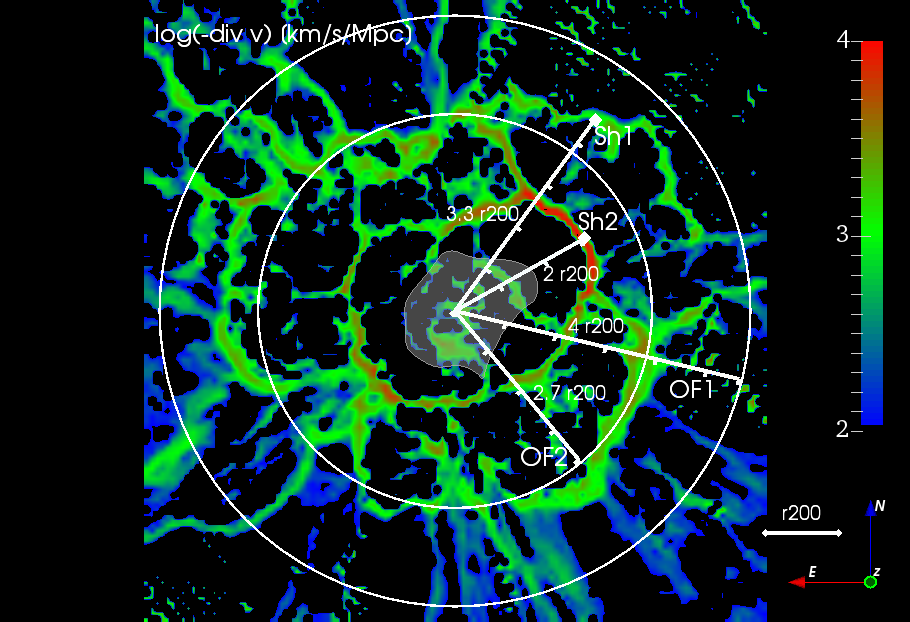}
\includegraphics[trim=60 10 0 0 0, clip, width=0.5\textwidth]{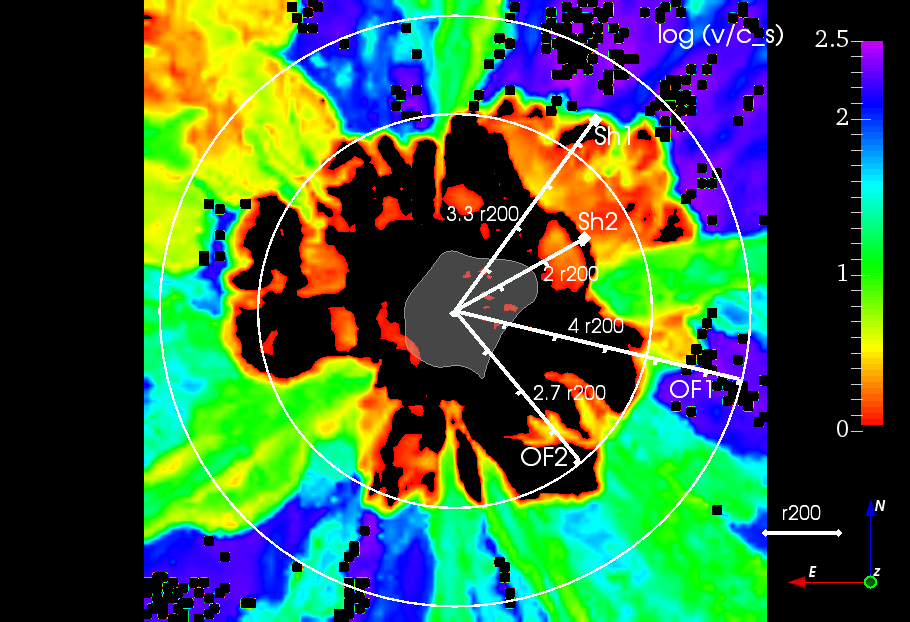}
\vspace{-0.0cm}
\caption{
Different gas properties in the same region as in Fig.~\ref{fig:T_only:2D}. The panels show gas density (\emph{top left}), temperature (\emph{top right}), radial velocity separately for inflows (\emph{middle left};  black region corresponds to outflow) and outflows (\emph{middle right};  black region corresponds to inflow), negative gas velocity divergence (\emph{bottom left}; high values correspond to strong compression), and the ratio of bulk speed to sound speed (\emph{bottom right}). The shaded area in the centre indicates the virial region of the cluster. The white circles delineate the radial regions marked OF1 and OF2 in Figure \ref{fig:temperature:density:median}. White diamonds labelled Sh1 - Sh2 mark the locations of heating events (possibly shocks) in gas outside filaments, for referencing in the text; white lines indicate the cluster-centric distances of the heating events.
}
\label{fig:out:2D}
\end{figure*}

\subsection{Gas heating outside filaments}
\label{heatoutside}

The next goal of our analysis is to gain a better understanding of the mechanisms responsible for gas heating within the interface region, and the related temperature behaviour in different zones delineated in Fig.~\ref{fig:temperature:density:median} (top right panel). We will first consider regions unassociated with filaments.

\subsubsection{Zone OF1}
\label{sec:OF1}

Far from the cluster (beyond $4r_{200}$), most of the volume outside filaments is filled by cold gas with temperatures at the expected void level of $T = 10^3-10^4$~K (see Fig.~\ref{fig:temperature:density:median}, top right panel and Fig.~\ref{fig:temp:hist}, top left panel). The free-falling gas has a low sound speed and is consequently highly supersonic (see Fig.~ \ref{fig:3D:vrad:fil2}, bottom panels). The accretion of such gas is known to create ``external shocks'' surrounding clusters, filaments, and sheets, with very high Mach numbers ($\gtrsim 10$) \citep[e.g.,][]{Miniati2000, Ryu2003}. We assume that due to this mechanism the temperature distribution outside filaments in our data begins to acquire a new high-temperature component with $T \sim 10^7$~K (Fig.~\ref{fig:temp:hist}, upper middle panel) near the outer boundary of Zone OF1 ($r\sim 4r_{200}$).

In order to learn more about the structure of this region, we employed SlicerAstro software \citep{2017A&C....19...45P} to produce thin "slices" of the full simulation box. The 3D structure reveals that within Zone OF1 ($r = 4.0 - 2.7 r_{200}$), where the overall temperature profile undergoes a steep rise from $10^4$~K to $10^7$~K outside the filaments (Fig.~\ref{fig:temperature:density:median}, top right panel), the infalling gas undergoes abrupt heating episodes (see Figs.~\ref{fig:T_only:2D} and \ref{fig:out:2D} for an example of a thin slice through the simulation box). The locations of the corresponding temperature jumps within OF1 (e.g., at Sh1 in Figs.~\ref{fig:T_only:2D} and \ref{fig:out:2D}) match with the local minima on the velocity divergence $\nabla \cdot \vec{v}$ map (bottom left panel of Fig.~\ref{fig:out:2D}), both of which are indicative of a shock \citep{Ryu2003,Schaye2015}.

Due to the very inhomogeneous spatial distribution of the heating episodes, the spherically averaged temperature profile does not exhibit a clear-cut temperature break but instead a rather smooth variation. The steepness of the temperature profile of the gas outside filaments at these radii reflects the extreme temperature gradient across the shock front, where the $\sim 10^{3}$~K gas entering from the voids is heated by several orders of magnitude to $T\sim 10^7$~K.

\subsubsection{Zone OF2}
The 3D velocity structure indicates that during the initial shock heating episode in Zone OF1, most of the accreting gas loses only a fraction of its bulk kinetic energy, continuing its infall towards the cluster (e.g., Sh1, top right panel of Fig.~\ref{fig:out:2D}) into Zone OF2. The gas becomes subsonic in this episode (Fig.~\ref{fig:out:2D}, bottom right panel), which places it in causal contact with the cluster, in a sense that its subsequent evolution depends on the physical conditions in the circumcluster region, which in turn are influenced by the cluster properties and evolutionary history. For example, in the case of CE-29 the infalling gas encounters an outflow of cluster gas in Zone OF2 at velocities $\sim 2\,000-3\,000$~km/s (peaking at location Sh2 in Fig.~\ref{fig:out:2D}, right middle panel) ejected at an earlier epoch. The outflow may be due to AGN feedback or pressure effects related to the abovementioned pre-merger state of CE-29 (Section~\ref{C29}, see also \citealt{2022arXiv220112370B}). At this location, the infalling and already heated gas undergoes another shock against the outflowing material, yielding very high temperatures ($\sim 10^8$~K). This heating episode and the corresponding temperature discontinuity are spatially coincident with the global minimum of the velocity divergence (Fig.~\ref{fig:out:2D}, bottom left panel), reinforcing the shock interpretation.

The outflow-induced reheated gas structure around Sh2 is very hot ($\sim 10^{8}$~K), i.e.,  hotter than that inside the virial boundary of the cluster, and its spatial extent appears comparable to the virial region of the cluster (see Fig.~\ref{fig:out:2D}). Such a hot spot might be too hot to be detected via X-ray line emission since oxygen, the most common metal, is fully ionised at these temperatures. However, the detection might be feasible via the Sunyaev-Zeldovich (SZ) effect which scales linearly with the temperature. Depending on the properties (phase, mass of the sub-clusters etc.), there may be such cluster mergers where the merger outflow-induced hot spots have less extreme temperatures and thus would be useful targets for X-ray spectroscopy.

\subsection{Gas heating inside filaments}
\label{heatinside}

Next, we analyse the temperature behaviour and gas heating inside filaments, with a goal to obtain a physical explanation for the different temperature zones (see Fig.~\ref{fig:temperature:density:median}, top right panel), as well as to contrast the behaviour and properties of gas within and outside filaments in the interface region.

\subsubsection{Zone IF1}
The temperature distribution of the gas inside filaments behaves very differently from the gas outside filaments (see Fig.~\ref{fig:temp:hist}):  it has a well-defined single peak that gradually shifts from $T \sim 10^{6}$~K at the largest radii to $T \approx 3 - 4 \times 10^{6}$~K at the inner boundary of Zone IF1 ($\sim 2.5r_{200}$). This corresponds to the rising temperature profile in Zone IF1 (see Fig. \ref{fig:temperature:density:median}, top right panel). Due to its higher density, the gas inside filaments is less sensitive to shocks (as pointed out by \citealt{Rost2021}) than that outside and retains the bulk of its kinetic energy until it approaches the cluster virial boundary. Consequently, the temperature of the gas inside filaments towards the inner boundary of IF1 is smaller than that outside filaments.

The gradual increase of filament gas temperature towards smaller radii within Zone IF1 can arise from a few different effects related to the increasing influence of the cluster. First, it is well established that the filament gas within the Cosmic Web is heated in the first place by the ram, turbulent and thermal pressure of the ambient gas accreting onto the filament from the sides due to the gravitational pull of the latter. As the filament approaches the cluster, the ambient conditions begin to change which also alters the physical properties of the accreting gas (in other words, the outer boundary conditions of the accretion problem change). The ambient gas also begins to slow down at larger radii from the cluster compared to the filaments (Fig.~\ref{fig:temperature:density:median}, bottom left panel), creating shearing motions near the filament-environment interface. The latter can give rise to instabilities and oblique shocks that heat the filament by dissipating a fraction of its kinetic energy without disrupting the filament altogether. Finally, visual inspection reveals locations where the gas structure deviates from our definition of a filament volume, i.e. a cylinder with a fixed radius of 1 Mpc around the spines detected with DisPerSE. This leads to some degree of artificial mixing of gas identified as residing inside or outside filaments; since the gas temperature outside filaments rises rapidly with decreasing radius, such mixing may also contribute to the temperature rise in filaments within IF1.

\subsubsection{Zone IF2}

At $r < 2.5r_{200}$, a possibly separate higher-temperature component starts to appear at $T \sim 10^7$~K (see Fig.~\ref{fig:temp:hist}, bottom middle panel), which could indicate that filaments are starting to undergo termination shocks against the cluster material (see Section~\ref{shock} for a case study). The typical temperatures and sound speeds of the filaments are already sufficiently high that the termination shocks have relatively modest Mach numbers, with a median ${\cal M} \sim 3 - 4$ (see Fig.~\ref{fig:3D:vrad:fil2}, bottom panels). The latter corresponds to a temperature jump of $\lesssim 10$ at the shock according to the Rankine-Hugoniot conditions (see Eq.~(\ref{eq:RH:temp}) below), hence the putative higher-temperature (shocked) component is not as clearly separable from the overall distribution as in the outside filament gas (see Section~\ref{sec:OF1}).

The onset of the cluster-induced shock heating is reflected in the steepening of the temperature profile in Zone IF2 (Fig.~\ref{fig:temperature:density:median}, top right panel), while the shock interpretation is reinforced by the filament gas starting to slow down at these radii (see discussion in Section~\ref{sec:press} and Fig.~\ref{fig:temperature:density:median}). Thus the infalling filament gas undergoes qualitatively similar evolution as the gas outside filaments, while the main difference is in the distance from the central cluster where the bulk of the shock heating takes place (Zones IF2 and OF1, respectively), as well as in the strength of the respective shocks.

\subsubsection{Zone IF3}
\label{IF3}

The hot component in the temperature distribution becomes dominant and shifts to $T \sim 3\times 10^7$~K by $r < 2r_{200}$. At the same radii (IF3) the filament gas slows down quite rapidly and dissipates most of its kinetic energy (see Fig.~\ref{fig:temperature:density:median}), again consistent with the shock scenario.

Note, however, that this hot component cannot be unambiguously associated with that of the shocked filament gas --- inwards from the shock, the filament gas flow may be disrupted and can mix with 1) the infalling gas outside filaments and 2) the gas associated with the cluster which has been heated by the shocks due to the gravitational collapse during the cluster formation (as in the case of filament F1 below). This is indicated by the similarity of the temperature profiles of the gas inside the cluster, inside filaments and outside filaments at similar distances from the cluster centre at $r < 2r_{200}$ (see Figs.~\ref{fig:cluster} and \ref{fig:temperature:density:median}). Due to this complexity, delineating and interpreting the thermal state separately for the gas inside the cluster and of that infalling through filaments or outside is not straightforward at these radii.

\subsection{Individual filaments}
\label{indiv}
 
Given the above understanding of the physical processes taking place at the cluster-filament interface, we proceed on a more detailed level and investigate the filaments individually in order to avoid various averaging effects in our analysis. It is likely that in order to detect the hot infalling gas in observations, one needs to stack many filaments (see Section \ref{Introduction}). Thus, the details of individual filaments may not be very important in this context, and the overall profile derived above could be sufficient for estimating the detection feasibility. However, in this work, our goal is to gain a better understanding of the underlying physical mechanisms, which manifest most clearly when considering filaments on an individual basis. Thus, we first derived the radial profiles of density and temperature for individual filaments, in order to evaluate the scatter of the thermodynamic properties between the filaments and to derive some insight into the cause of the scatter.

As discussed in Section~\ref{Filaments}, the individual filaments F1--9 are defined between the virial boundary and the first splitting point (up to $\sim r_{200}$ distance from the virial boundary). Since the detectable X-rays probably originate from the close proximity of the cluster virial boundary, this radial limitation is acceptable. In order to extract the profiles of the relevant quantities, we applied a modified version of the procedure we used for the median profile described in Section~\ref{gassampling}. Namely, we are interested here in the local individual phenomena and conditions, primarily driven by the matter density of the cluster. Thus, we construct the individual profiles of the thermodynamical properties of the gas as a function of the radial distance from where the individual filament spine crosses the cluster virial boundary (rather than from the cluster centre), defined in Section~\ref{C29}.

\begin{figure*}
\includegraphics[width=0.48\hsize]{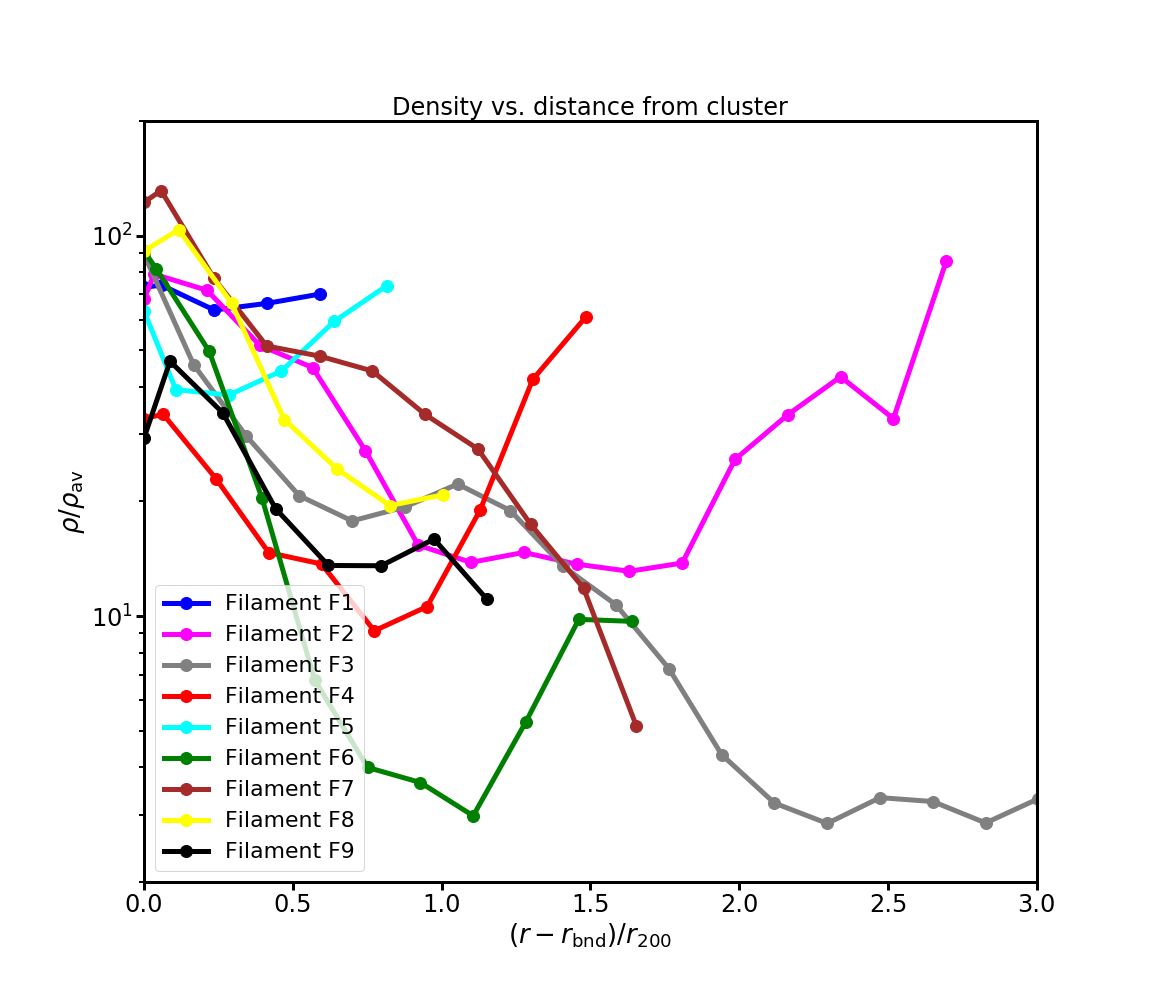}
\includegraphics[width=0.48\hsize]{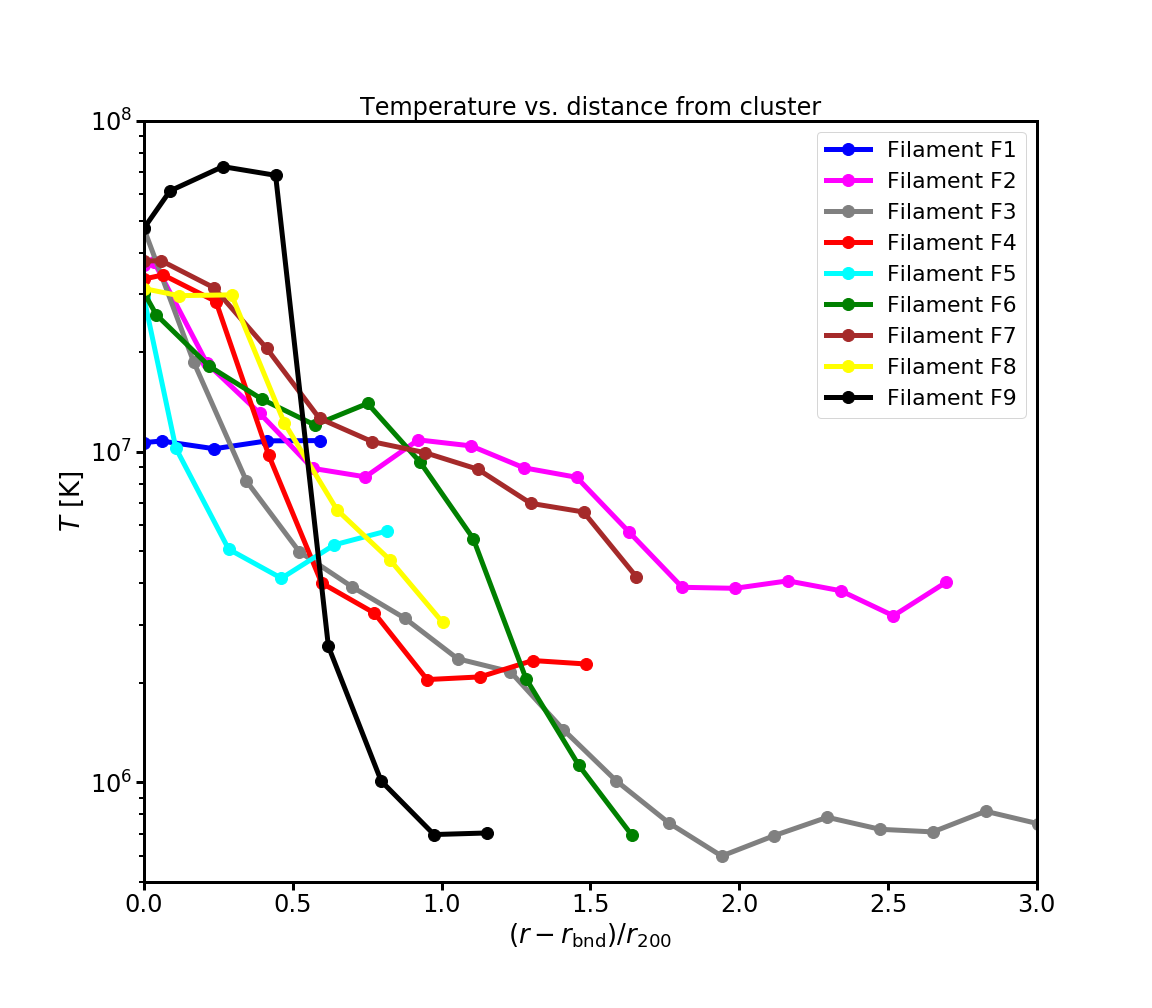}
\caption{The median gas overdensity (\emph{left}) and temperature (\emph{right}) within individual filaments, as a function of radial distance from the local cluster boundary (normalised to the nominal $r_{200}$).}
\label{fig:temperature:density:indiv}
\end{figure*}

At a distance further than $\sim 0.5-1.0 r_{200}$ from the cluster boundary, the gas in most filaments is approaching the cluster relatively undisturbed and nearly in free fall along the cosmic gas highway, while the temperature and density vary by almost an order of magnitude between filaments (Fig.~\ref{fig:temperature:density:indiv}). Closer to the cluster, the temperature of the gas in most filaments exhibits behaviour similar to the overall median profile for all filaments taken together (Fig.~\ref{fig:temperature:density:median}, top right panel), i.e., it increases rapidly. The location of the highest temperature gradient approximately marks the location of the termination shock for each filament --- its location varies between filaments as it depends on the individual filament properties (e.g., ram pressure), as well as the local ambient conditions. The corresponding temperature jumps in each filament are smeared out over a distance comparable to the filament radius (1~Mpc), which is to be expected given that neither the filament nor the shock normal is perfectly radial (see Fig.~\ref{fig:temperature:density:indiv}, right panel). After the termination shock, the scatter in temperature and density becomes markedly smaller between filaments (roughly a factor 3) and probably reflects more the properties of the cluster that the shocked filaments have joined/mixed with, rather than the individual filament properties ahead of the shock.

The large scatter in temperature and density before the termination shock is likely caused by the complexity of the environmental and evolutionary factors,  which are difficult to disentangle. For example, for a subset of filaments the higher ambient density and proximity of prominent clumps seem to correlate with higher filament temperatures; however, this is not the case for all filaments. A significantly larger sample of filaments is required to put such a tentative correlation on a firmer footing and to investigate how the ambient gas influences the filament not just through its density, but also through pressure, turbulent motions, etc. Finally, the snapshot data analysed in this paper cannot account for evolutionary effects, and including an explicit temporal dimension in the analysis is beyond the scope of this work.
 
Given the above issues we limit our study of the individual filaments to examining filaments F1 and F9 (see Figs.~\ref{fig:F1:2D} and \ref{fig:fil5:2D}), which bracket the variation of the gas temperature and density close to the virial boundary of the cluster (see Fig.~\ref{fig:temperature:density:indiv}).

\subsubsection{Filament F1}
\begin{figure*}
\includegraphics[trim=0 10 0 0 0, clip, width=0.5\textwidth]{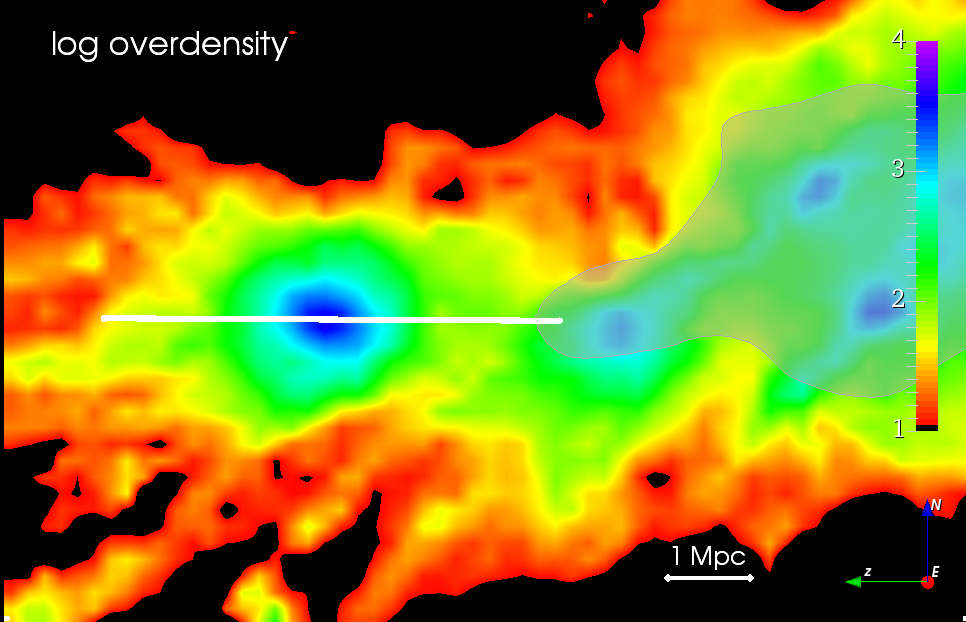}
\includegraphics[trim=0 10 0 0 0, clip, width=0.5\textwidth]{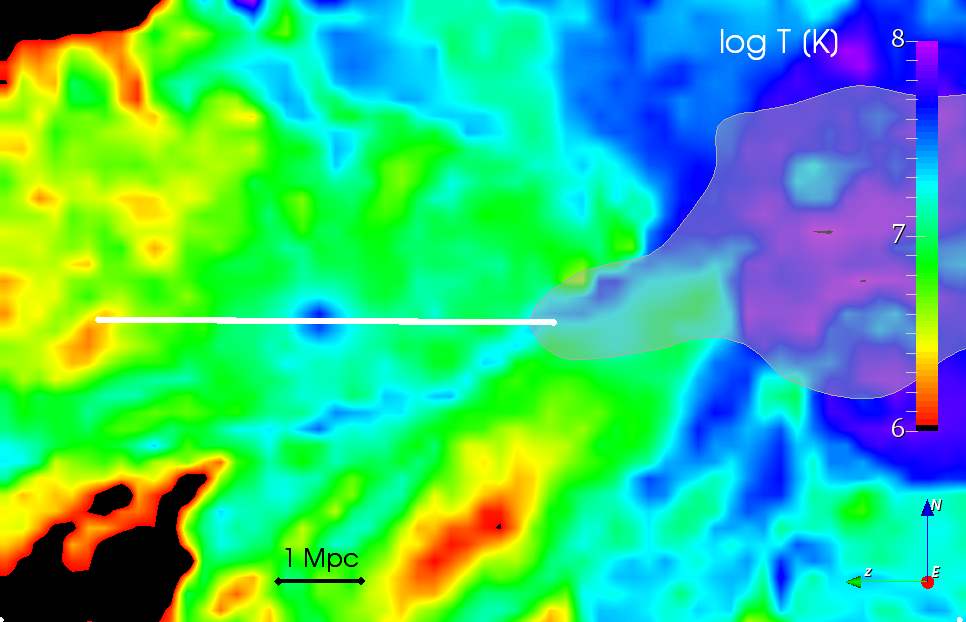}
\includegraphics[trim=0 10 0 0 0, clip, width=0.5\textwidth]{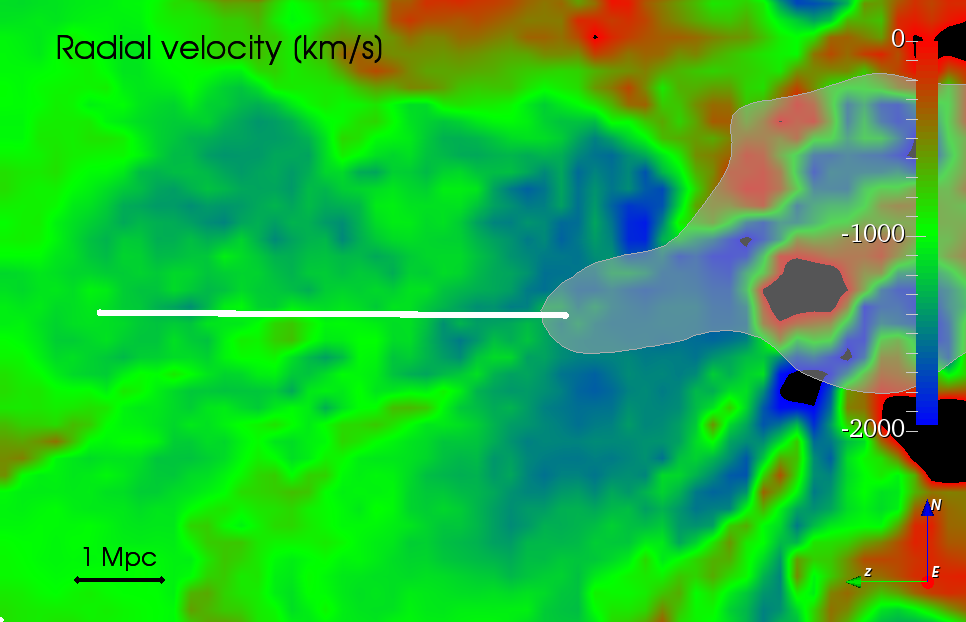}
\includegraphics[trim=0 10 0 0 0, clip, width=0.5\textwidth]{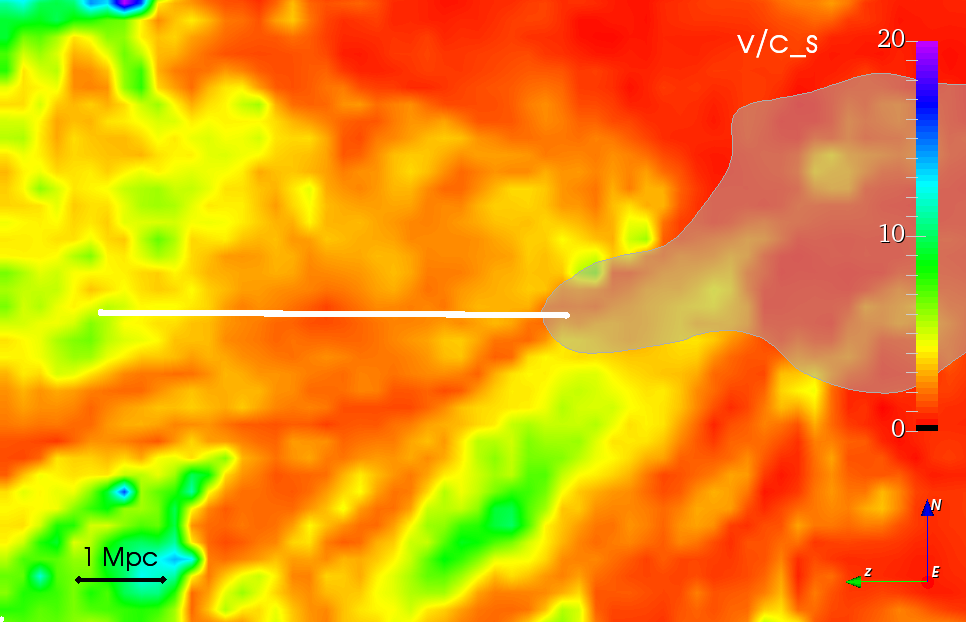}
\vspace{-0.0cm}
\caption{The overdensity (\emph{top left}), temperature (\emph{top right}), radial velocity (\emph{bottom left}), and the ratio of bulk speed to sound speed (\emph{lower right}) of the gas around filaments F1 in a planar cross-section of size $(24~{\rm Mpc})^2 \approx (8.5 r_{200})^2$, crossing through the cluster centre. The grey area on the right-hand side of the figures marks regions within the cluster boundary. The white line marks the direction approximately parallel to the shock normal. }
\label{fig:F1:2D}
\end{figure*}

Filament F1  appears ``disturbed'': it approaches the cluster through a denser environment than other filaments (see Fig.~\ref{fig:F1:2D}, top left panel), contains a prominent halo at $3r_{200}$, and connects to another halo just at the cluster boundary (Fig.~\ref{fig:F1:2D}). Accordingly, it is significantly thicker and denser ($r_{\perp} \sim 2$~Mpc, $\rho_{\rm gas} \sim 100 \rho_{\rm av}$) than most other filaments.

The denser filament has comparatively higher penetrating power compared with the rest and is, therefore, still in free fall (i.e., on a gas highway) when it crosses the cluster boundary. This is supported by Fig.~\ref{fig:F1:2D} (bottom left), which shows that the filament has not decelerated (i.e., undergone a termination shock) until well inside the cluster boundary. This also accounts for its lower temperature ($\sim 10^7$~K) near the boundary compared with other filaments ($\sim 3\times 10^7$~K; Fig.~\ref{fig:temperature:density:indiv}), because the others have already been heated up by their respective termination shocks by that stage. In contrast, if one compares the filaments on their respective ``gas highways'' (rather than at a fixed distance from the local cluster boundary), i.e., ahead of their termination shocks, then the ``disturbed'' filament F1 is comparatively hotter ($T\sim 10^7$~K) than the rest ($\sim 10^6 - 10^7$~K), which is not unexpected given its more turbulent environment and higher density.

\subsubsection{Filament F9}

In contrast, filament F9 is one of the more isolated ones, approaching the cluster through a relatively low-density region (see Figs.~\ref{fig:temperature:density:indiv} and \ref{fig:fil5:2D}). Hence, it is comparatively undisturbed and cool --- i.e., it approximately retains the properties characteristic of filaments within the Cosmic Web domain (overdensity $\sim 10$, $T \sim 10^5-10^6$~K) all the way to the clearly identifiable termination shock, where it exhibits an abrupt temperature jump by almost two orders of magnitude. Hence filament F9 serves as the ``cleanest" representative of general scenario presented in this paper, showing a clear cosmic highway region, penetration into the interface region, and the final termination shock against the cluster material. For these reasons, we chose the shock in filament F9 for a more detailed analysis below.

\subsubsection{Shock heating in filament F9}
\label{shock}

\begin{figure*}
\includegraphics[trim=0 160 0 0 0, clip, width=0.5\textwidth]{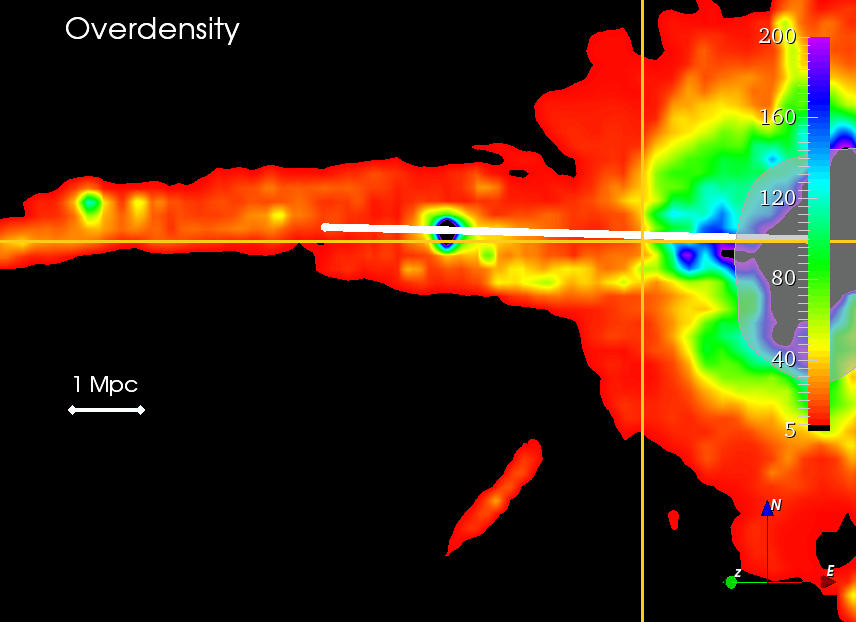}
\includegraphics[trim=0 160 0 0 0, clip, width=0.5\textwidth]{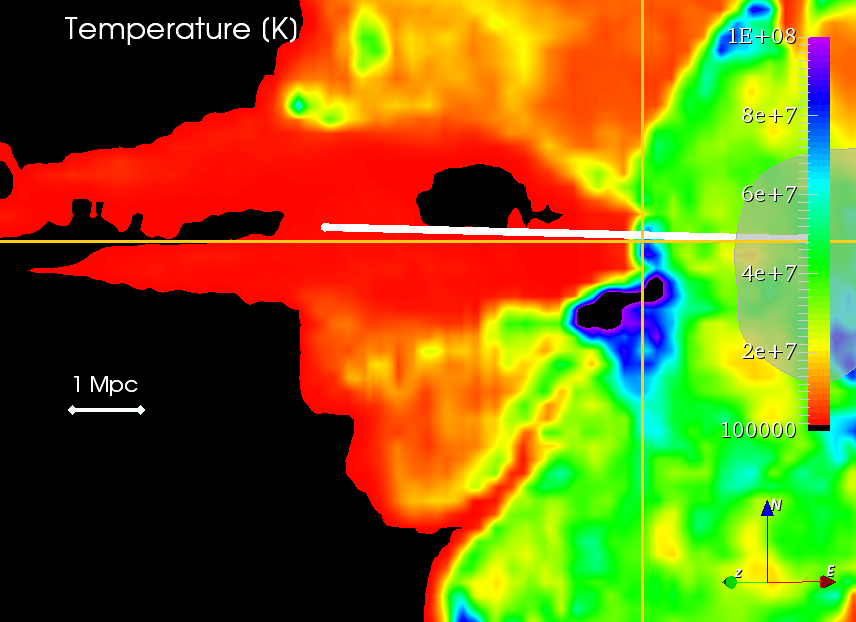}
\includegraphics[trim=0 160 0 0 0, clip, width=0.5\textwidth]{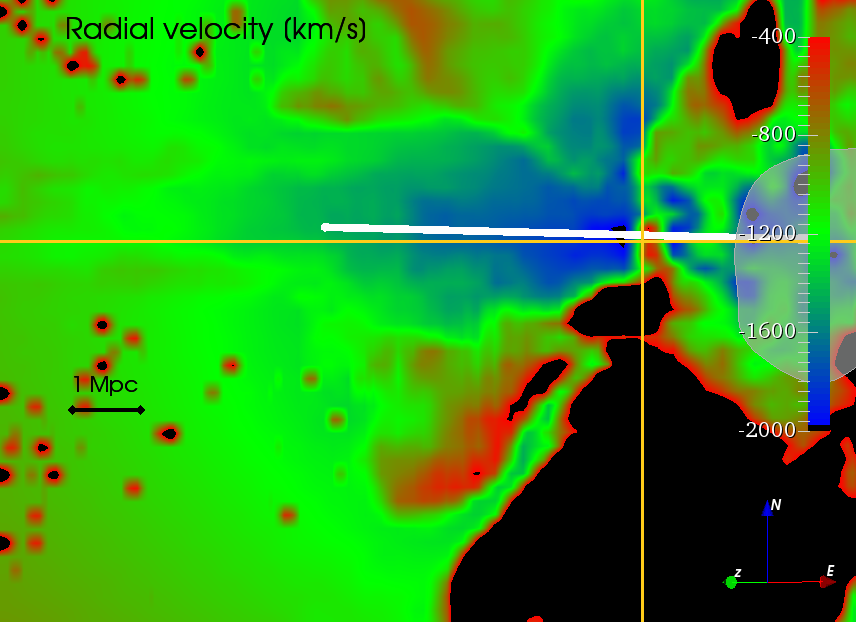}
\includegraphics[trim=0 160 0 0 0, clip, width=0.5\textwidth]{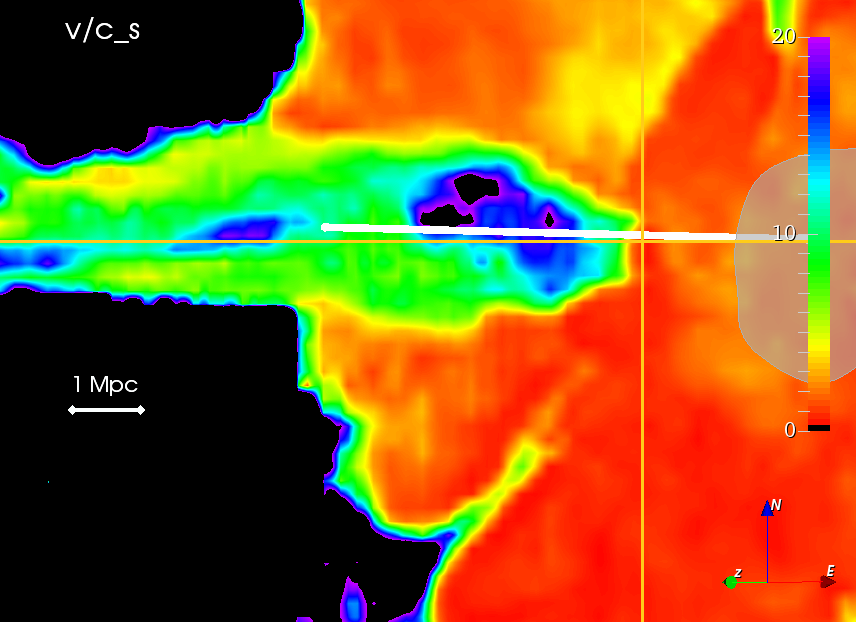}
\vspace{-0.0cm}
\caption{As Fig.~\ref{fig:F1:2D} but for filament F9. The yellow cross indicates the accretion shock location.
 }
\label{fig:fil5:2D}
\end{figure*}

\begin{figure*}
\includegraphics[width=0.5\textwidth]{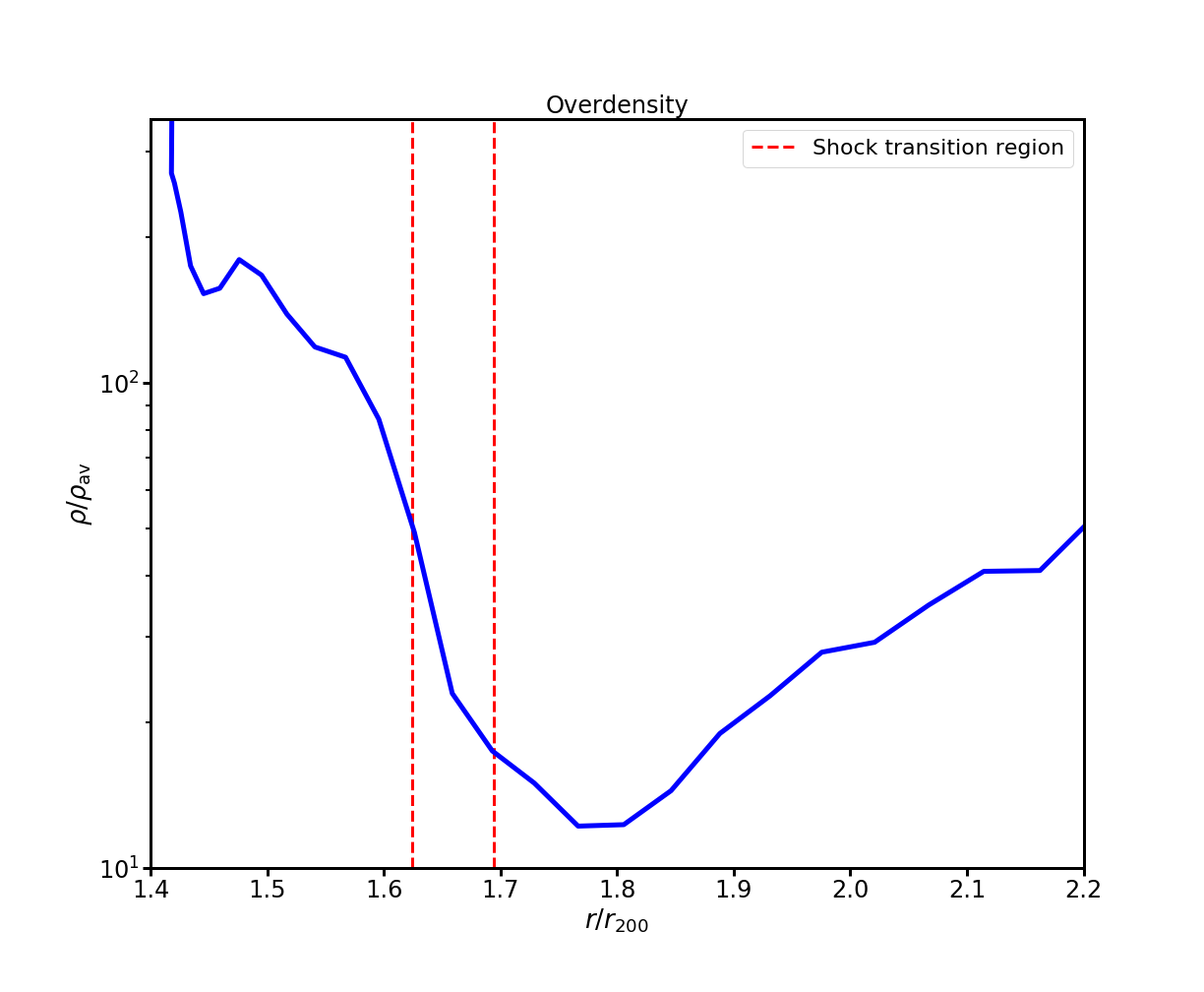}
\includegraphics[width=0.5\textwidth]{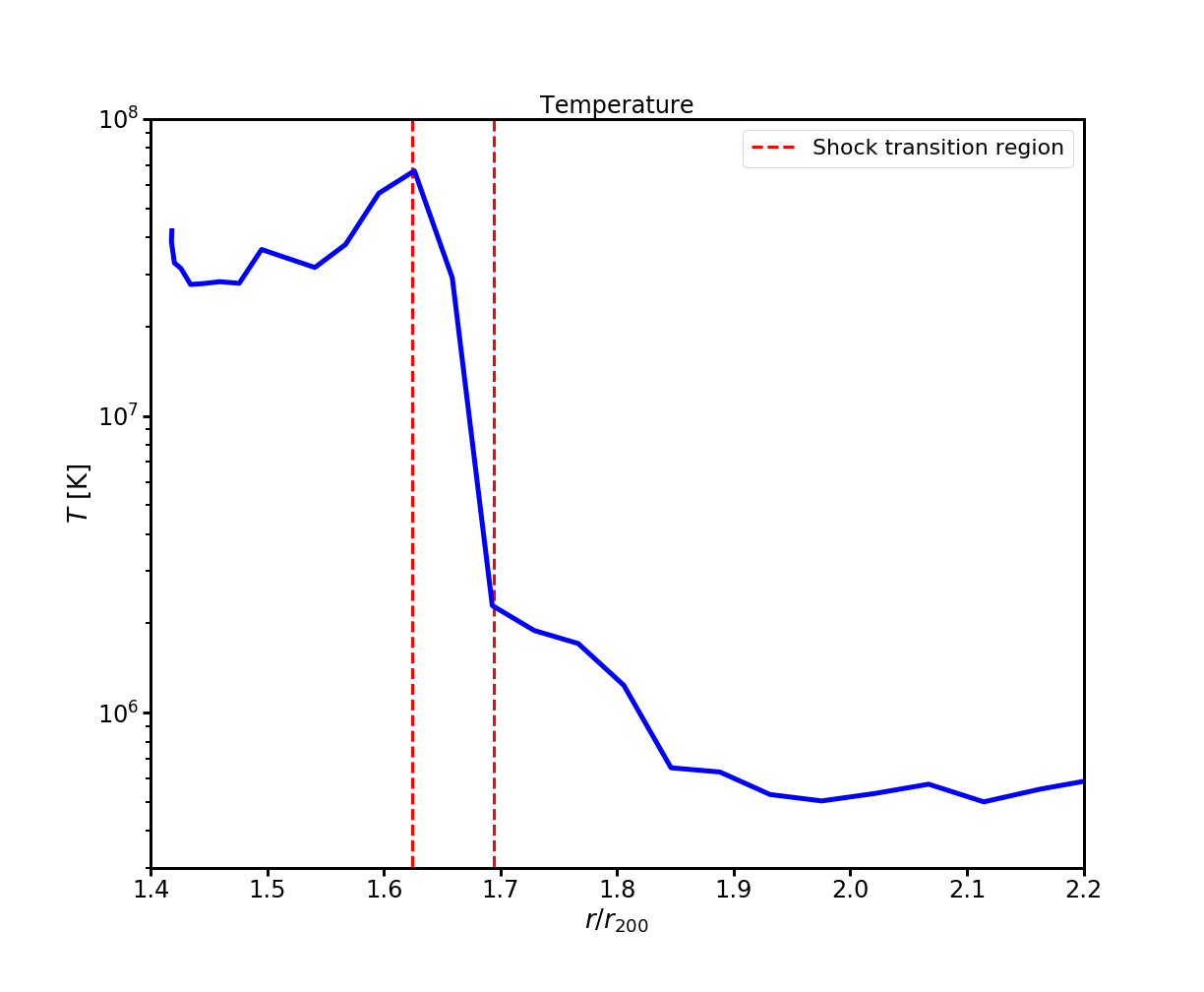}
\includegraphics[width=0.5\textwidth]{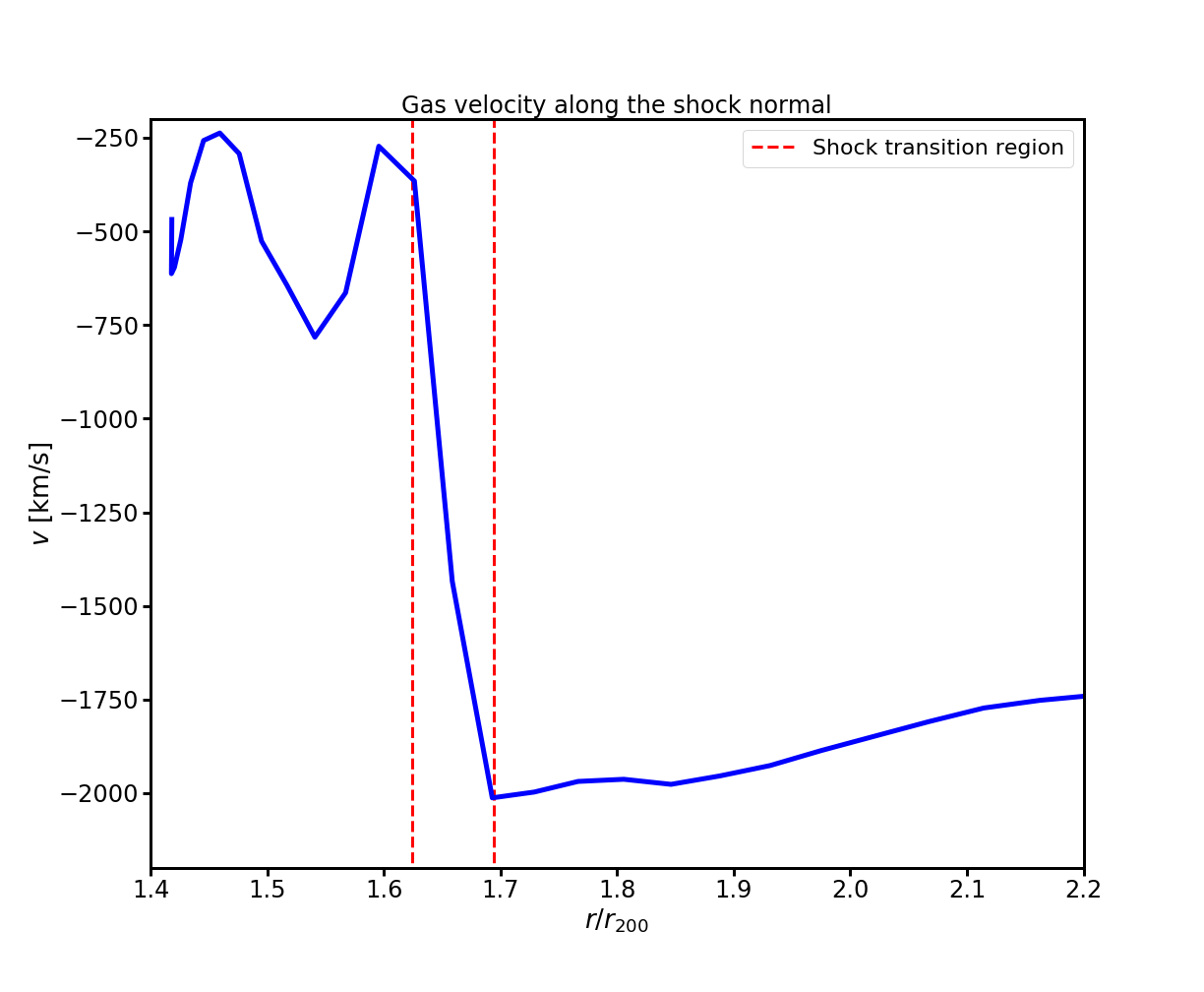}
\includegraphics[width=0.5\textwidth]{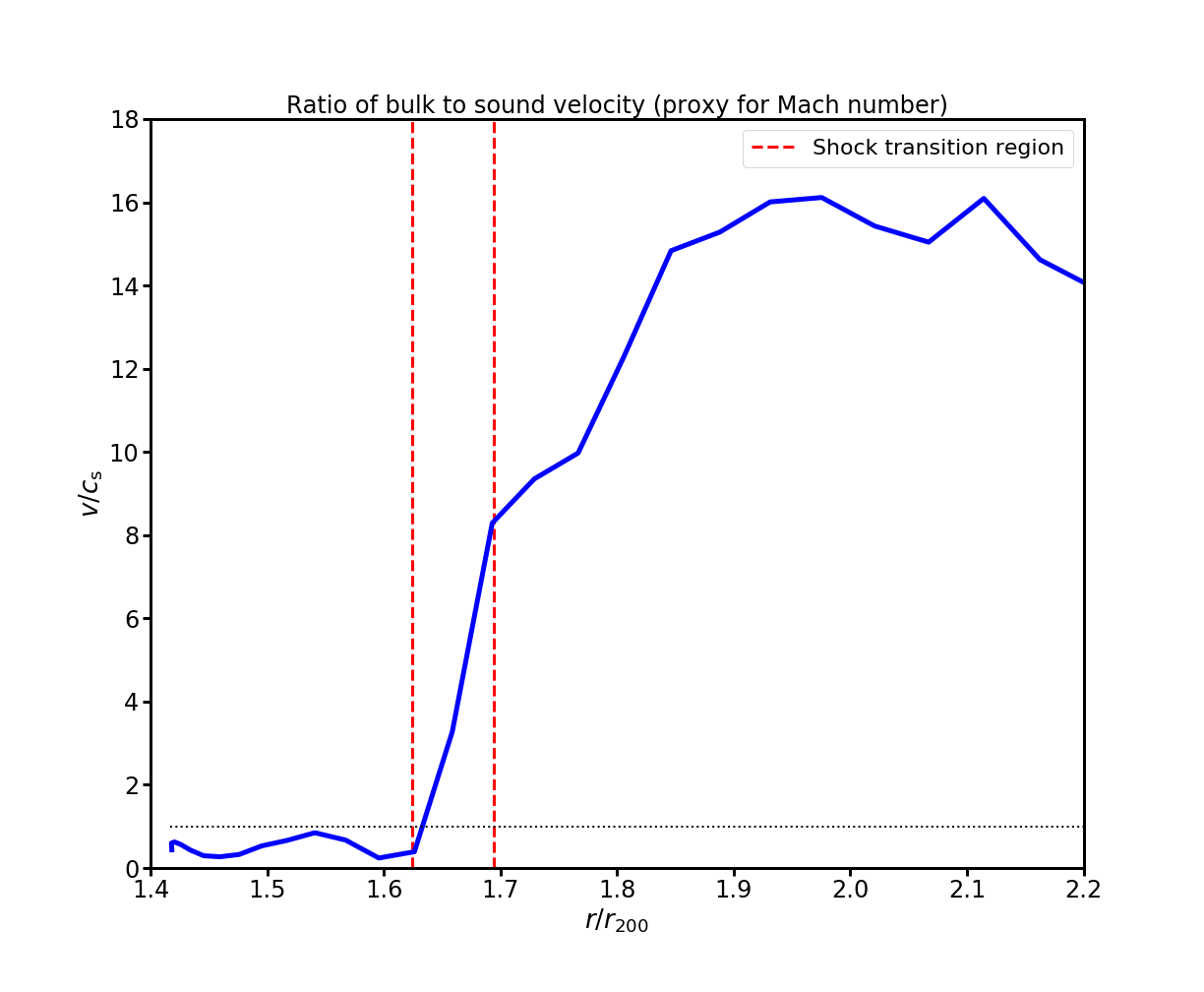}
\includegraphics[width=0.5\textwidth]{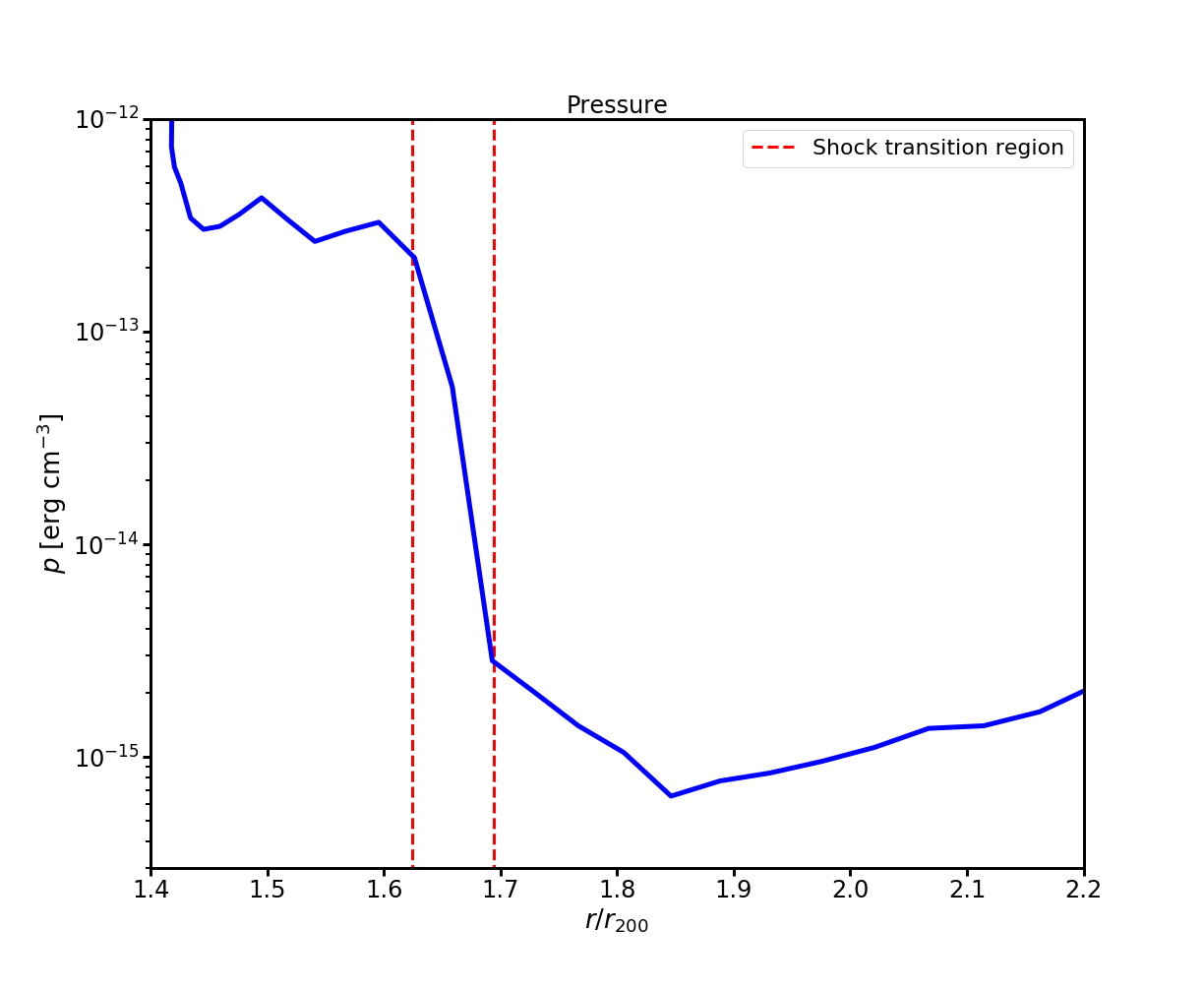}
\includegraphics[width=0.5\textwidth]{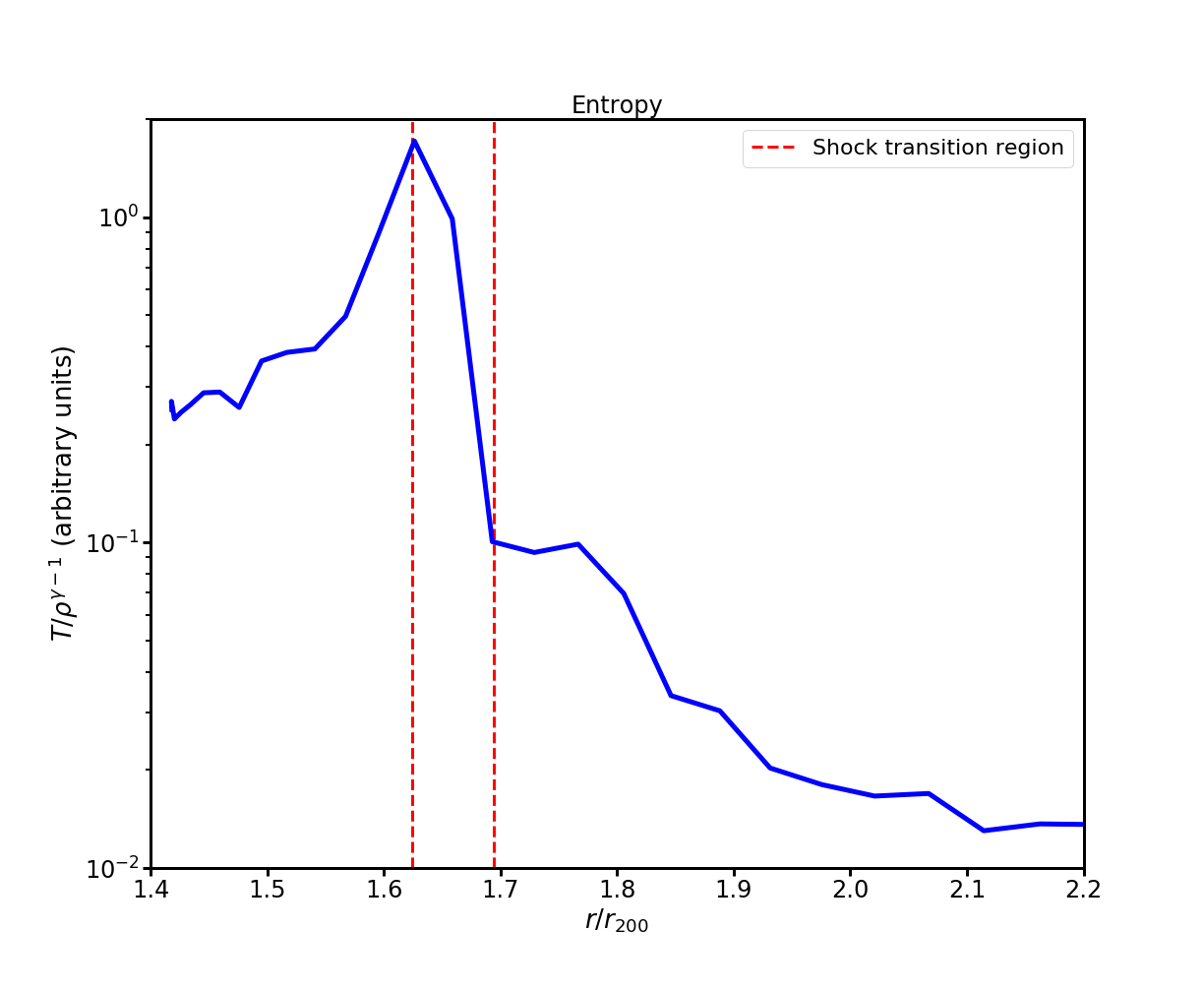}
\vspace{-0.0cm}
\caption{Profiles of gas properties in the shock vicinity in filament F9, along the shock normal (indicated by a white line in Fig.~\ref{fig:fil5:2D}). Radii above $1.7 r_{200}$ (below $1.6 r_{200}$) correspond to the upstream (downstream) region.}
\label{fig:fil5:sh}
\end{figure*}

The average thermodynamic behaviour of the full filament population in the cluster vicinity (discussed in Section~\ref{Overall}) is well understood in terms of the ram pressure and shock heating arguments. While a systematic study of shocks taking place within individual filaments is beyond the scope of this paper, we present here, as proof of concept, a specific example (filament F9) exhibiting a particularly prominent accretion shock. We identified a location of discontinuity of thermodynamic properties of the gas in this filament at a distance of $\approx 1.7 r_{200}$ from the cluster centre (see Figs.~\ref{fig:fil5:2D} and \ref{fig:fil5:sh}). The Rankine-Hugoniot jump conditions for density, velocity (in the shock frame), pressure and temperature in a perpendicular plane-parallel shock are
\begin{align}
    \frac{\rho_2}{\rho_1} &= \frac{v_{1, {\rm sh}}}{v_{2, {\rm sh}}} = \frac{(\gamma+1) {\cal M}_1^2}{(\gamma - 1){\cal M}_1^2 + 2}
    \label{eq:RH:dens} \\
    \frac{P_2}{P_1} &= \frac{2\gamma {\cal M}_1^2 - (\gamma - 1)}{\gamma + 1}
    \label{eq:RH:press} \\
    \frac{T_2}{T_1} &= \frac{[(\gamma - 1){\cal M}_1^2 + 2][2\gamma{\cal M}_1^2 - (\gamma -1)]}{(\gamma + 1)^2 {\cal M}_1^2}~,
    \label{eq:RH:temp}
\end{align}
where ${\cal M}_1$ is the Mach number upstream of the shock, $\gamma$ is the ratio of specific heats ($\gamma = 5/3$ for mono-atomic ideal gas), and subscripts 1 and 2 refer to quantities in the immediate upstream and downstream of the shock front, respectively.

The measured velocity jump in the cluster frame is $v_1/v_2 \approx 5.5$ (upstream velocity $v_1 \approx -2\,000$~km/s in the cluster frame). This is inconsistent with a stationary shock front (since in the shock frame $v_{1, {\rm sh}}/v_{2, {\rm sh}} \le 4$, where the limiting value corresponds to ${\cal M} \rightarrow \infty$) and suggests that the shock front is propagating in the direction opposite to the flow velocity. The observed ratio of the upstream bulk to sound velocities is $|v_1|/c_{{\rm s}, 1} \approx 8$; with the oppositely propagating shock, it follows that $|v_{1, {\rm sh}}| > |v_1|$ and that the upstream Mach number ${\cal M}_1 = |v_{1, {\rm sh}}|/c_{{\rm s}, 1} > 8$. From eq.~(\ref{eq:RH:dens}) one then obtains $3.8 \lesssim v_{1, {\rm sh}}/v_{2, {\rm sh}} \le 4$, which limits the shock velocity to a relatively narrow range around $v_{\rm sh} \sim 200$~km/s in the cluster frame. This yields ${\cal M}_1= |v_1 - v_{\rm sh}|/c_{\rm s}\approx 9$, where $v_{1, {\rm sh}} = v_1 - v_{\rm sh}$ is the upstream velocity in the shock frame.

Using the above value for the Mach number for a planar shock, the Rankine-Hugoniot conditions (Eqs.~\ref{eq:RH:dens}--\ref{eq:RH:temp}) yield $\rho_2/\rho_1 = v_{1, {\rm sh}}/v_{2, {\rm sh}} \approx 3.9$, $P_2/P_1 \approx 100$, and $T_2/T_1 \approx 26$. The measured discontinuities across the shock along its normal (white line in Fig.~\ref{fig:fil5:2D}) are $\rho_2/\rho_1 \approx 2.8$, $P_2/P_1 \approx 80$ and $T_2/T_1 \approx 30$. The theoretical and measured values are consistent within $\sim 20 - 40 \%$. The differences can be ascribed to the non-planarity of the shock front, as well as the multi-dimensional nature of the gas flow in the shock vicinity (i.e., that we use values in a thin slice to represent the full system). Thus, accounting for the above uncertainties, the measured discontinuities are consistent with an accretion shock taking place in filament F9.

\section{Implications for X-ray detection of the infalling gas}\label{Xray}

We will continue by discussing some of the general implications of the above analysis of a simulated Coma-like C-EAGLE cluster CE-29 for the X-ray emission and absorption line detection of the infalling hot intergalactic gas in the cluster-filament interface with future instruments onboard ATHENA and XRISM missions.

The gas density at the virial boundary ($r \sim 0.7-1.5 r_{200}$) of the cluster CE-29 is $\sim 100$ times the mean baryon density. To look for the observational signals from the infalling gas beyond the virial radius it is beneficial to follow the filaments, the Cosmic Gas Highways, which accumulate the gas over the cosmic history into the cluster-filament interface. The radial decrease of the gas density along filaments is less rapid (approximately $\rho \propto r^{-1.6}$) than outside  (approximately $\rho \propto r^{-4}$) and maintains an overdensity above 10 all the way to the Cosmic Web. 

By applying the DisPerSE filament detection algorithm on the galaxy distribution in the cluster outskirts we captured $\sim 50$\% of the mass of the intergalactic X-ray ($T > 10^{5.5}$~K) gas between the cluster boundary and a radial distance of $6 r_{200}$ from the cluster centre within the filaments. Thus, a practical way of locating the infalling X-ray gas in observations is to apply filament detection methods to the spectroscopic galaxy surveys. In future work, we will investigate the effects of applying magnitude cuts mimicking the sensitivity limits of specific surveys like SDSS and 4MOST in order to obtain a more realistic efficiency of capturing the hot WHIM.

The structure of the high-temperature gas close to the virial boundary of the cluster is not well spatially correlated with the filaments, due to the chaotic nature of the cluster-related shock heating episodes induced by the infalling and outflowing gases (see Fig.~\ref{fig:density_filaments}). This is probably the reason why our capture fraction of the hot WHIM is smaller than that in the main EAGLE Cosmic Web filaments  far from clusters \citep[$\sim 80$\%,][]{Tuominen2021}.
However, while the filaments close to the cluster boundary have temperatures  comparable to the hot ambient gas, they are preferable for the X-ray follow-up since they stand out from their surroundings with their higher gas densities.

Most of the filament gas mass (73\%) up to $2 r_{200}$ distance from the cluster CE-29 is hotter than $10^7$~K. At such high temperatures most of the common X-ray elements, in particular oxygen, are fully ionised. This implies (based on the temperature only) that at these distances in the case of very massive clusters, the infalling gas may be better detectable via \ion{Fe}{XXV-XXVI} rather than oxygen lines. This capability is provided by XRISM. Further than $2 r_{200}$ from CE-29 the median temperature in filaments decreases to levels suitable for \ion{O}{VII-VIII} production and the corresponding X-ray line wavelengths will be covered by ATHENA/X-IFU. A quantitative study of the X-ray detectability of a single massive cluster such as the one studied in this paper will be the subject of follow-up work, where we will consider in detail the distribution of metal ions in the abovementioned regions and their relevant emission/absorption properties. Future work will also extend the above study to lower mass clusters, which will dominate the observational samples. Their lower gravitational potentials and virial temperatures are expected to increase the importance of the ions of the most common metal, i.e., oxygen in the cluster-filament interfaces.

\section{Summary and conclusions}\label{Summary}
We summarise here the main conclusions of our analysis of the intergalactic gas inside and outside the filaments in the outskirts of the most massive ($M_{200} = 2.4\times 10^{15}~{\rm M}_{\odot}$) cluster (CE-29) in C-EAGLE simulations.

 \begin{itemize}
   
    \item 
     Using DisPerSE filament finder we detected the complex filamentary network of galaxies around the two merging subclusters. There are nine locations where the network crosses the virial boundary of the cluster.

     \item 
     The volumes within 1~Mpc distance from the filament spines (excluding those within $r_{200}$ of halos)
     capture a large fraction ($\sim 50$\%) of the hot ($T > 10^{5.5}$~K)  intergalactic gas falling towards cluster in the cluster-filament interface. This, along with the systematically higher gas densities within filaments, holds promise for locating the hot WHIM observationally.
    
    \item  
     The gas inside filaments is in free-fall all the way down to the radial distance of $\sim 2 r_{200}$ from the cluster, beyond which it starts to slow down. The ordered (bulk) motion of $\sim 50$\% of the filament gas persists into the virial region. On the other hand, the gas outside filaments starts decelerating at larger radii and stops near the virial boundary.
    
    \item 
    The slowing down of the infalling gas starts when its ram pressure becomes comparable to the thermal and/or turbulent pressure of the ambient gas in the cluster vicinity. Owing to its higher density, the ram pressure of the free-falling filament gas is higher than that outside filaments. Hence the filaments are able to penetrate deeper, i.e. almost all the way to the cluster virial boundary, before being decelerated by the increasing cluster gas pressure. They therefore serve as cosmic gas highways transporting matter from the Cosmic Web domain into the cluster.
    
    \item 
    When the gas slows down, a fraction of the energy associated with ordered gas motion along the filament is converted to thermal energy via shocks. Consequently, the gas is rapidly heated close to the virial boundary of the cluster. 
 
    \item
    The median temperature of the filament gas increases from the typical Cosmic Web WHIM level of $T = 10^5-10^6$~K at $r = 4 r_{200}$ to higher than $T = 10^7$~K at the virial boundary of the studied massive cluster.
   
    \item
    The rising filament temperature closer to the cluster is accompanied by a gradual increase of the filament gas density from the overdensity of $\sim 10$ at $r = 4 r_{200}$ to $\sim 100$ in the cluster boundary, following
    $\rho \propto r^{-1.6}$.

 \end{itemize}

\begin{acknowledgements}
The authors thank Toni Tuominen for useful discussions on EAGLE data analysis and Andrea Negri for providing the galaxy data for C-EAGLE simulations. We thank Rain Kipper for help with the analysis of our results. We acknowledge the support by the Estonian Research Council grants IUT40-2, PRG1006, and by the European Regional
Development Fund (TK133). S.E.H. was supported by the project 우주거대구조를 이용한 암흑우주 연구 (``Understanding Dark Universe Using Large Scale Structure of the Universe''), funded by the Ministry of Science. CDV acknowledges support by the Spanish Ministry of Science and Innovation (MICIU/FEDER) through research grant PID2021-122603NB-C22.
\end{acknowledgements}

\bibliographystyle{aa}
\bibliography{reference}

\begin{appendix}

\section{Statistical significance of the identification of free-falling and decelerated gas phases within and outside filaments}

\label{sec:app:signif}

\begin{figure}[h]
\includegraphics[width=0.48\textwidth]{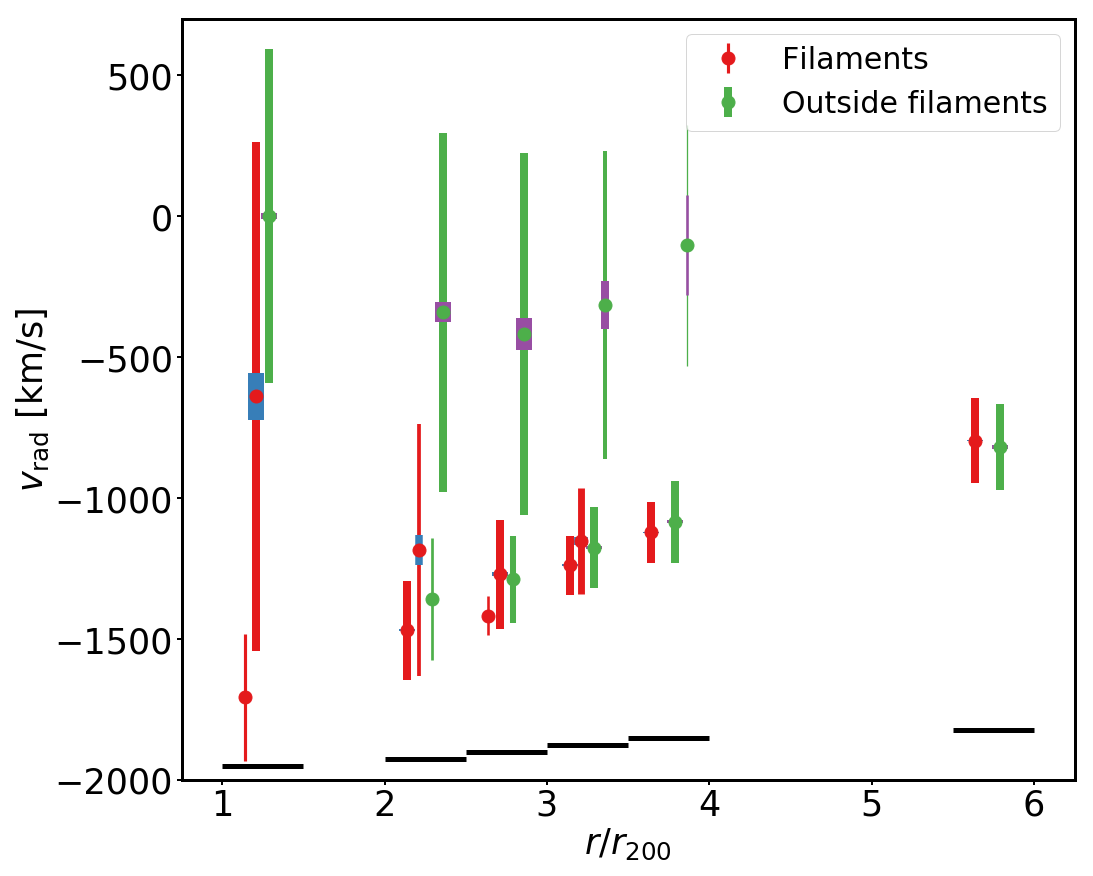}
\caption{
Results of double gaussian fitting of the radial velocity distributions, corresponding to the radial regions shown on the different panels in Figure \ref{fig:vrad:hist}. The red and green bars show the $1\sigma$ range of the fitted gaussian components within and outside filaments, respectively. The overlaid blue and purple bars show the standard errors of the fitted gaussian mean values. The relative line widths of the bars within a given radial range are proportional to the areas under the two fitted gaussian components (if the ratio is $<0.1$ in a given range, only the dominant component is shown). The black horizontal bars mark the radial ranges where the fits are performed.
}
\label{fig:vrad:prof}
\end{figure}

Figure \ref{fig:vrad:prof} shows the mean values of the double gaussian components fitted to the radial velocity distributions in Figure \ref{fig:vrad:hist}, along with the widths of the gaussians and the standard errors of the fitted means. At large radii, the velocity distributions both within and outside filaments are well fitted by a single gaussian, with mean values consistent with each other within errors. At $r\lesssim 4r_{200}$, a decelerated/shocked component appears in the outside filament gas, which is separated by more than $1\sigma$ in velocity space from the free-falling phase. At the same radii, the filament gas distribution is still consistent with a single (free-falling) component. Importantly, the standard errors of the fitted gaussian mean values (show by thicker bars overlaid on top of the longer bars that characterize the gaussian widths) show that the mean velocities of the free-falling and decelerated components are separated by $\gg 3\sigma$ whenever both components are present (note that some cases the above error bars are smaller than the symbol of the relevant data point and not visible in the figure), clearly indicating that the two components/distributions are statistically distinct.

Outside filaments, the decelerated component has become the dominant one by $r\sim 2.5r_{200}$, while the bulk of the filament gas still resides in the free-falling phase; the radial velocities of these components are separated by $> 1\sigma$ in terms of the gaussian widths and by $>10 \sigma$ in terms of the standard errors of the fitted gaussian means. The latter in particular shows that the radial velocities within and outside filaments are drawn from distinct underlying distributions,
supporting the case for an environment-dependent accretion history.

\end{appendix}

\end{document}